\documentclass[preprintnumbers,amsmath,amssymb,floatfix,10pt,prd,onecolumn,
superscriptaddress,nofootinbib]{revtex4}
\usepackage{latexsym}
\usepackage{epsfig}
\usepackage{epstopdf}
\usepackage{graphicx}
\usepackage{amssymb}
\usepackage{amsmath}
\usepackage{dcolumn}
\usepackage{bm}
\usepackage{color}
\usepackage{comment}
\usepackage[utf8]{inputenc}

\begin{document}

\title{\bf Reconstruction and Stability Analysis of Some Cosmological Bouncing Solutions in $F(\mathcal{R},T)$ Theory}

\author{M. Zubair}
\email{mzubairkk@gmail.com;drmzubair@ciitlahore.edu.pk}\affiliation{Department of Mathematics, COMSATS University Islamabad, Lahore Campus, Lahore-Pakistan}

\author{Quratulien Muneer}
\email{anie.muneer@gmail.com}\affiliation{Department of Mathematics, COMSATS University Islamabad, Lahore Campus, Lahore-Pakistan}

\author{Saira Waheed}
\email{swaheed@pmu.edu.sa}\affiliation{Prince Mohammad Bin Fahd University, Al Khobar, 31952 Kingdom of Saudi Arabia}

%

\begin{abstract}

The present article investigates the possibility of reconstruction of the generic function in $F(\mathcal{R},T)$ gravitational
theory by considering some well-known cosmological bouncing models namely exponential evaluation, oscillatory, power law and matter
bounce model, where $\mathcal{R}$ and $T$ are Ricci scalar and trace of energy-momentum tensor, respectively. Due to the complexity
of dynamical field equations, we propose some ansatz forms of function $F(\mathcal{R},T)$ in perspective models and examine that
which type of Lagrangian is capable to reproduce bouncing solution via analytical expression. It is seen that for some cases of
exponential, oscillatory and matter bounce models, it is possible to get analytical solution while in other cases, it is not possible
to achieve exact solutions so only complementary solutions can be discussed. However, for power law model, all forms of generic function
can be reconstructed analytically. Further, we analyze the energy conditions and stability of these reconstructed cosmological bouncing
models which have analytical forms. It is found that these models are stable for linear forms of Lagrangian only but the reconstructed
solutions for power law are unstable for some non-linear forms of Lagrangian.\\


\end{abstract}

\maketitle

\date{\today}

\section{Introduction}

In cosmological theories, the bouncing cosmology is one of the most interesting scenario which deals with
those solutions having initial singularity problems \cite{1*,2*,3*,4*}. The primary objective of such solutions
is to resolve the initial big bang singularity which is regarded as one of the biggest issues in cosmology \cite{5*}.
According to bouncing cosmology, initially our universe contracts up to a minimal radius, it bounces off at that point
and than it starts to expand. Thus our universe can never collapse to a singular point and therefore initial singularity
can be avoided. Additionally, cosmological bouncing is considered as a fascinating alternative to standard inflationary
cosmology and have been appeared in loop quantum cosmology \cite{6*}, matter bounce \cite{7*} and scalar field theories
\cite{1*,2*,3*,4*,7*,8*}. Actually, inflationary scenario solve some early universe issues like flatness, horizon, initial
singularity and baryon asymmetry issues and provides a complete picture of structure formation \cite{7*,9*,10*,11*}.
Although, inflationary model faces both the trans-Plankian as well as singularity issues during fluctuations, however,
before inflation when our universe undergoes the exponential expansion, singularity produces and than inflationary
scenario cannot describe the complete picture of universe. At this stage, matter bouncing scenario challenges these
issues which are faced by the inflationary models \cite{12*}.

In Einstein's theory and its modifications, bouncing scenarios have grasped attentions of researchers and in literature,
much work have been done on this subject. In \cite{13*}, Myrzakulov and Sebastiani presented the bouncing solutions using
viscous fluid in FRW flat space time by taking different scale factors (exponential and power law) along with some kinds
of fluid into account. They discussed a relationship between finite singularity and bounce and extended their work to
$f(R)$ gravity. Cheung et al. \cite{14*} proposed a new scenario for the production of dark matter regarding bouncing
cosmology where dark matter has been produced from plasma during contracting and expanding cosmic phases. In another study
\cite{15*}, authors discussed the $\Lambda$CDM bouncing scenario in which they studied that contracting universe contained
radiation as well as cold dark matter with positive cosmological constant. They assumed that loop quantum cosmology captured
the spacetime in the context of high curvature dynamics which guaranteed the removal and replacement of big bang singularity
by bounce. Further they used bounce to calculate the perturbations and examined that these are nearly scale invariant.
Bandyopadhyay and Debnath \cite{16*} discussed the cosmological bouncing in entropy corrected models namely logarithmic and
power law corresponding to Horava-Lifshitz gravity as well as fractal universe and also analyzed the null energy condition
near the bounce.

During the last few decades, substantial attempts have been made to develop the non-standard theories of gravity \cite{17*,18*,19*,20*}
by introducing some modifications in the action of Einstein gravity. Some noteworthy examples of such modified formulations include
$F(\mathcal{R})$ \cite{21*}, $F(\mathcal{R},T)$ (where $T$ represents the trace of $T_{ij}$) \cite{18}, $F(\mathcal{T})$,
where $\mathcal{T}$ represents torsion \cite{22*}, Gauss-Bonnet gravity \cite{23*,24*,25*} and Brans-Dicke theory \cite{26*} etc.
The bouncing cosmology has been explained in such modified gravitational frameworks by numerous researchers, for reference one
can see \cite{27*}. In this respect, Oikonomou and Odintsov \cite{28*} explained the bouncing scenario along with type $IV$
singularity at bouncing point in Gauss Bonnet $F(G)$ and $F(\mathcal{R})$ gravity. They also analyzed the stability of solutions
corresponding to $F(G)$ and $F(\mathcal{R})$ theories. In another paper \cite{28**}, same authors discussed $\Lambda$CDM bouncing
model in $F(R)$ theory and checked stability properties as well as gravitational particle production which are essential
features for checking viability of the $\Lambda$CDM bounce. Further, some bouncing scenarios namely matter bounce, super bounce,
singular bounce and symmetric bounce in unimodular $F(\mathcal{R})$ gravity have been explored by researchers and they presented the
behavior of Hubble radius for all bouncing scenarios \cite{29*}. Elizalde et al. \cite{30*} discussed the matter bounce in extended form by
considering the framework of ghost free $F(\mathcal{R},G)$ gravity, while Boisseau et al. \cite{31*} produced the bouncing universe
in scalar-tensor gravity having negative scalar field potential.

Modified gravitational frameworks (geometrical modification of GR action) are considered as viable dark energy candidates
and predict the late time cosmic acceleration successfully. In this context, Harko et al. \cite{18}, have proposed the
$F(\mathcal{R},T)$ gravity (with $\mathcal{R}$ and $T$ as Ricci scalar and energy-momentum tensor trace) in which they have
suggested different forms of $F(\mathcal{R},T)$ such as $F(\mathcal{R},T)=\mathcal{R}+2F(T)$, $F(\mathcal{R},T)= F(\mathcal{R})+G(T)$
and $F(\mathcal{R},T)=F_1(\mathcal{R})+F_2(\mathcal{R})F_3(T)$, where $F(\mathcal{R})$, $G(T)$, $F_1(\mathcal{R})$, $F_2(\mathcal{R})$
and $F_3(T)$ are some arbitrary functions of $\mathcal{R}$ and $T$. Many researchers have utilized these forms of $F(\mathcal{R},T)$
function and addressed the phenomenon of late time cosmic acceleration \cite{34*}. In this regard, Singh et al. \cite{32*} explained
the bouncing cosmology to obtain a specific form of Hubble parameter in $F(\mathcal{R},T)$ theory. Shabani and  Ziaie \cite{33*}
studied the cosmological bouncing solutions using perfect fluid in $F(\mathcal{R},T)$ gravity. They explored the properties of
such bouncing solutions, checked the stability and examined the validity of energy constraints near bouncing point. They obtained those
cosmological scenarios which can exhibit the non-singular bounce before and after de-Sitter cosmic phase and also proposed the
general solution for matter bounce. In the present work, we shall also adopt the same form of Lagrangian function $F(\mathcal{R},T)$
to reconstruct bouncing models.

In modified theories, reconstruction of cosmological models is an open problem. Many reconstruction schemes have been proposed
in literature \cite{24,24**,25**} which are substantially used to understand the inter-conversion of matter dominated and DE cosmic
phases in $F(\mathcal{R})$ gravity. By assuming some known cosmic evolution, the corresponding form of Lagrangian can be
calculated which can reproduce the same background evolution. In literature \cite{19}-\cite{21}, researchers have discussed
various cosmological techniques of reconstruction in $F(\mathcal{R},T)$ gravity. Bamba et al. \cite{35*} explained the
reconstruction scenario for two bouncing (exponential and power law) solutions in $F(\mathcal{R})$ and $F(G)$ gravity theories and
examined the stability of reconstructed bouncing models in both theories. Further, these works are extended by Bamba et al. \cite{32}
where they discussed bouncing cosmology in a gravitational framework involving interaction of Gauss-Bonnet invariant and dynamical
scalar field and reconstructed the forms of potential and Gauss-Bonnet coupling function of scalar field using hyperbolic and
exponential forms of scale factor. By using conformal transformation, they also explained the link between the bouncing behaviors
in Jordan and Einstein frames. Odintsov et al. \cite{36*} studied the bouncing cosmology in the framework of $F(R)$ gravity theory
via reconstruction scheme, and confronted their model with the recent observations. They solved dynamical equation numerically and
discussed different qualitative features and observable measures of the proposed model. In \cite{22}, Nojiri et al. executed such
reconstruction technique to obtain the $F(\mathcal{R})$ gravity model, which were adopted in $F(\mathcal{R},G)$ and Gauss-Bonnet
theories \cite{23}. Cruz-Dombriz et al. \cite{9} investigated the existence of analytical bouncing solutions in extended
teleparallel gravity namely $f(T,T_G)$ theory using FRW model for symmetric, oscillatory, superbounce, matter bounce, and singular
bounce. Caruana et al. \cite{33} explored possibility of reconstruction of analytical form of Lagrangian function using some well-known
bouncing models like symmetric, oscillatory, superbounce, matter bounce, and singular bounce for flat FRW geometry in $f(T,B)$ theory
and obtained significant solutions.

Zubair and Kousar \cite{26} explained the reconstruction scenario in $F(\mathcal{R}, \mathcal{R}_{\alpha\beta}\mathcal{R}^{\alpha\beta},\phi)$
theory and discussed the validity of all energy bounds graphically. Mishra et al. \cite{28} also investigated different scenarios
of anisotropic cosmological reconstruction in $F(\mathcal{R},T)$ gravity and explored the effect of coupling constant and anisotropy
on the cosmic dynamics using more general approach. In a recent paper \cite{33**}, Shamir studied some bouncing models in $F(G,T)$ gravity
(with $G$ as the Gauss-Bonnet term and $T$ as the trace of energy-momentum tensor) by taking well-famed EoS parameter along with two
forms of generic functions $F(G,T)$ involving logarithmic and linear trace terms and concluded that their discussed bouncing solutions
are significant. In the context of modified theories, the stability of cosmological solutions is regarded as a captivating issue and numerous
studies are already available on this topic. In stability analysis, both Hubble parameter $H$ and energy density $\rho$ are
perturbed by introducing the isotropic and homogeneous perturbations. In this respect, stability of de-Sitter, power law, phantom and
non-phantom matter fluid solutions have been analyzed in $F(G,T)$ gravity \cite{29a}. Salako et al. \cite{29b} investigated
the reconstruction, thermodynamics and stability of $\Lambda$CDM model in the framework of teleparallel theory. In another
paper \cite{29c}, authors performed the stability analysis for cosmological models using dynamical system analysis (fixed point theory)
in $F(\mathcal{R})$ gravity, while Shabani et al. \cite{29d} examined the stability of Einstein static universe in
Einstein-Carton-Brans Dicke gravity. In another work \cite{29e}, authors presented bouncing scenario in $F(\mathcal{R},T)$ gravity
and explored the energy conditions near the bouncing point and discussed the stability of obtained solutions. Godani and Samanta \cite{29f}
estimated the cosmological parameters (Hubble and deceleration) to analyze the stability and energy conditions in a modified
gravity.

In our work, we shall employ reconstruction scenario in $F(\mathcal{R},T)$ theory to analyze some bouncing cosmological
models namely exponential, oscillatory, power law and matter bounce models. In all models, we shall choose minimal and
non-minimal coupling of $\mathcal{R}$ and $T$ in Lagrangian and reconstruct their solutions. In three models,
only complementary solutions exist for non-minimal coupling and for power law model, all forms of Lagrangian have
analytical solution. Further, we shall analyze the stability for these reconstructed bouncing solutions. This paper is organized
in this pattern. Section $\textrm{II}$ provides the set of field equations for FRW metric along with some basic assumptions
used for this work. Section $\textrm{III}$ comprises the reconstruction of generic function $F(R,T)$ by using
exponential, oscillatory, power law and matter bounce models with different ansatz of this function. Section $\textrm{IV}$
deals with the validity of energy conditions for these reconstructed models. Further, we analyze the stability of these
reconstructed solutions against perturbations of FRW universe model in Section $\textrm{V}$. Final section will conclude
and present the obtained results briefly.

\section{Basic Mathematical Structure of $F(\mathcal{R},T)$ Gravity and its Field Equations}

Here, we shall present a brief review of the $F(\mathcal{R},T)$ gravity, its field equations along with some necessary
assumptions used for this work. Let us start from the action \cite{18} of this theory defined as follows
\begin{equation}\label{1}
S=\frac{1}{2k^2}\int F(\mathcal{R},T)\sqrt{-g}d^{4}x+\int L_{m}\sqrt{-g}d^{4}x,
\end{equation}
where $L_{m}$ stands for the ordinary matter Lagrangian density and $g$ represents the determinant of $g_{ij}$. By
varying the above action with respect to metric tensor, the following set of field equations can be obtained:
\begin{eqnarray}\label{2}
8\pi T_{ij}-F_{T}(\mathcal{R},T)T_{ij}-F_{T}(\mathcal{R},T)\Theta_{ij}=F_{\mathcal{R}}(\mathcal{R},T)\mathcal{R}_{ij}
-\frac{1}{2}F(\mathcal{R},T)g_{ij}+(g_{ij}\Box-\nabla_{i}\nabla_{j})F_{\mathcal{R}}(\mathcal{R},T),
\end{eqnarray}
where $F_{\mathcal{R}}(\mathcal{R},T)=\frac{\partial F(\mathcal{R},T)}{\partial \mathcal{R}},~
F_{T}(\mathcal{R},T)=\frac{\partial F(\mathcal{R},T)}{\partial T}$, while $\square=g^{i j}\nabla_{i}\nabla_{j}$. Also,
$\nabla_{i}$ symbolizes the covariant derivative and the mathematical expression of $\Theta_{ij}$ is defined by
\begin{equation}\nonumber
\Theta_{ij}=\frac{g^{\alpha\beta}\delta T_{ij}}{\delta
g^{ij}}=-2T_{ij}+g_{ij}L_{m}-2g^{\alpha\beta}\frac{\partial^{2}L_{m}}{\partial
g^{ij}\partial g^{\alpha\beta}}.
\end{equation}
It is worthy to mention here that the energy-momentum tensor in $F(\mathcal{R},T)$ gravity is not conserved and yields
the following relation:
\begin{eqnarray}\label{3}
\bigtriangledown^{i}T_{ij}=\frac{F_T}{K^{2}-F_T}\bigg[(T_{ij}+\Theta_{ij})\bigtriangledown^i \ln F_T+\bigtriangledown^i\Theta_{ij}
-\frac{1}{2}g_{ij}\bigtriangledown^i T\bigg].
\end{eqnarray}
In this work, we assume the matter Lagrangian as $L_{m}=-p$, and consequently, $\Theta_{ij}$ takes the form as
\begin{equation}\label{4}
\Theta_{ij}=-2T_{ij}-p g_{ij}
\end{equation}
and then field equations and energy-momentum conservation take the following form:
\begin{eqnarray}\label{5}
k^2T_{ij}+F_T(\mathcal{R},T)T_{ij}+p F_T(\mathcal{R},T)g_{ij}&=&\mathcal{R}_{ij} F_{\mathcal{R}}(\mathcal{R},T)-\frac{1}{2}g_{ij}F(\mathcal{R},T)-
({\nabla}_{i}{\nabla}_{j}-g_{{ij}}{\Box})F_{\mathcal{R}}(\mathcal{R},T),\\\label{6}
\nabla^{i}T_{ij}&=&\frac{-1}{k^2+f_{T}}\bigg[T_{ij} \nabla^{i} f_{T}+g_{ij} \nabla^{i}(\frac{f}{2}+p f_{T})\bigg].
\end{eqnarray}

The flat FRW metric in spherical coordinates with $a(t)$ as expansion radius can be defined as
\begin{equation}\label{7}
ds^2=dt^2-a^{2}(t)\bigg[dr^2+r^{2}(d\theta^{2}+sin^{2}\theta d\phi^{2})\bigg].
\end{equation}
The energy-momentum tensor is considered as perfect fluid and is defined as
\begin{equation}\nonumber
T_{ij}=(\rho+p)u_{i}u_{j}-pg_{ij}.
\end{equation}
For this geometry, the corresponding $00$ component of field equation is given by
\begin{equation}\label{8}
k^2 \rho + (\rho + p) F_{T} + \frac{F}{2} + 3(\dot{H}+H^2)F_{\mathcal{R}}-3H(\dot{\mathcal{R}}f_{\mathcal{RR}}+\dot{T}F_{\mathcal{R}T})=0,
\end{equation}
Here, the dot indicates time derivative while the Hubble parameter, Ricci scalar and trace of energy-momentum tensor can be,
respectively, defined as $H=\frac{\dot{a}}{a},~\mathcal{R}=-6(\dot{H}+2 H^{2})$ and $T=\rho-3p$. The continuity equation can be
re-written as
\begin{equation}\label{9}
\dot{\rho} + 3H(\rho+p)=\frac{-1}{k^2+F_T}\bigg[(\rho+p)\dot{T}F_{TT}+\dot{p}F_{T}+\frac{1}{2}\dot{T}F_{T}\bigg].
\end{equation}
which represents that $\bigtriangledown^{i}T_{ij}\neq 0$ in $F(\mathcal{R},T)$ gravity.
To obtain the continuity equation in standard form, one needs to impose $\bigtriangledown^{i}T_{ij}= 0$ which implies the constraint $\frac{-1}{k^2+F_T}\bigg[(\rho+p)\dot{T}F_{TT}+\dot{p}F_{T}+\frac{1}{2}\dot{T}F_{T}\bigg]=0$ to be satisfied. Applying the EoS $p=\omega\rho$
in the above equation, we obtain:
\begin{eqnarray}\label{10}
&&\dot{\rho}+3 H \rho(1+\omega)=0,\\\label{11}
&&(1+\omega) T F_{TT}+ \frac{1}{2}(1-\omega)F_{T}=0.
\end{eqnarray}

From Eq.(\ref{5}), the two Friedmann equations can be written as
\begin{eqnarray}\label{1*}
3H^2F_\mathcal{R}&=&k^2\rho +\frac{1}{2}(\mathcal{R} F_\mathcal{R}-F)-3H(\dot{\mathcal{R}}F_{\mathcal{RR}}+\dot{T}F_{\mathcal{R} T}),\\\nonumber
(3H^2+2\dot{H})F_\mathcal{R}&=&k^2p-\frac{1}{2}(\mathcal{R} F_\mathcal{R}-F)+2H(\dot{\mathcal{R}}F_{\mathcal{RR}}+\dot{T}F_{\mathcal{R} T})+\ddot{\mathcal{R}}F_{\mathcal{RR}}+(\dot{\mathcal{R}})^2F_{\mathcal{RRR}}+2\dot{\mathcal{R} }\dot{T}F_{\mathcal{RR}T}+\ddot{T}F_{\mathcal{R} T}\\\label{1**} &+&(\dot{T})^2F_{\mathcal{R}TT}.
\end{eqnarray}
The above equations can be re-written in terms of effective energy ($\rho^{eff}$) and effective pressure ($p^{eff}$) by comparing them
with the standard Friedmann equations, i.e., $\rho^{eff}=3(\frac{H}{k})^2$ and $p^{eff}=-\frac{3H^2+2\dot{H}}{k^2}$, where the
effective terms are given by
\begin{eqnarray}\label{2*}
\rho^{eff}&=&\frac{1}{k^2+F_{T}}\bigg((k^2+F_{T})\rho +\frac{1}{2}(\mathcal{R} F_\mathcal{R}-F)-3H(\dot{\mathcal{R}}F_{\mathcal{RR}}+\dot{T}F_{\mathcal{R} T})+3(1-F_\mathcal{R})H^2\bigg),\\\nonumber
p^{eff}&=&\frac{1}{k^2+F_{T}}\bigg((k^2+F_{T})p-\frac{1}{2}(\mathcal{R} F_\mathcal{R}-F)+2H(\dot{\mathcal{R}}F_{\mathcal{RR}}+\dot{T}F_{\mathcal{R} T})+\ddot{\mathcal{R}}F_{\mathcal{RR}}+(\dot{\mathcal{R}})^2F_{\mathcal{RRR}}+2\dot{\mathcal{R}}\dot{T}F_{\mathcal{RR}T}\\\label{2**}
&+&\ddot{T}F_{\mathcal{R} T}+ (\dot{T})^2F_{\mathcal{R}TT}-(1-F_{\mathcal{R}})(3H^2+2\dot{H}))\bigg).
\end{eqnarray}
The corresponding effective EoS parameter can be obtained by dividing $p^{eff}$ with $\rho^{eff}$, i.e., $\omega^{eff}=\frac{p^{eff}}{\rho^{eff}}$.
For the sake of simplicity in further calculations, we interchange the cosmic time with a new variable namely e-folding parameter
given by $N=\ln(\frac{a}{a_0})=-\ln(1+Z)$ and further, the time derivative operator can be converted in terms of new parameter
by the relation $\frac{d}{dt}=H\frac{d}{dN}$. Consequently, the effective terms can be shifted to new variable as follows
\begin{eqnarray}\label{3*}
\rho^{eff}&=&\frac{1}{k^2+F_{T}}\bigg((k^2+F_{T})\rho +\frac{F}{2}+3(H H'+H^2)F_\mathcal{R}+18H(H^2 H^{''} + H (H')^2
+ 4 H^2 H')F_{\mathcal{RR}}+T'F_{\mathcal{R} T}\\\nonumber &&+3(1-F_\mathcal{R})H^2\bigg),\\\nonumber
p^{eff}&=&\frac{1}{k^2+F_{T}}\bigg((k^2+F_{T})p-\frac{F}{2}+3(H H'+H^2) F_\mathcal{R}-12H(H^2 H^{''}
+H (H')^2+4H^2H')F_{\mathcal{R}\mathcal{R}}+T'F_{\mathcal{R} T}-6(H^{'''}H^3 \\\nonumber
&+&3 H^2H'H^{''}+H (H')^3 + 4 H^3 H^{''} +8 H^2(H')^2)F_{\mathcal{R}\mathcal{R}}++36 (H^2H^{''} + H (H')^2 +
4 H^2 H')^2F_{\mathcal{R}\mathcal{R}\mathcal{R}}-12(H^2 H^{''}\\\label{3**}
&+& H(H')^2 + 4 H^2H')T'F_{\mathcal{R}\mathcal{R}T}+T^{''}F_{\mathcal{R} T}+(T')^2F_{\mathcal{R}TT}-(1-F_{\mathcal{R}})(3H^2+2H H'))\bigg),
\end{eqnarray}
where prime indicates derivative with respect to $N$.

\section{Reconstruction of bouncing models}

\begin{figure}
\centering
\includegraphics[width=0.235\textwidth]{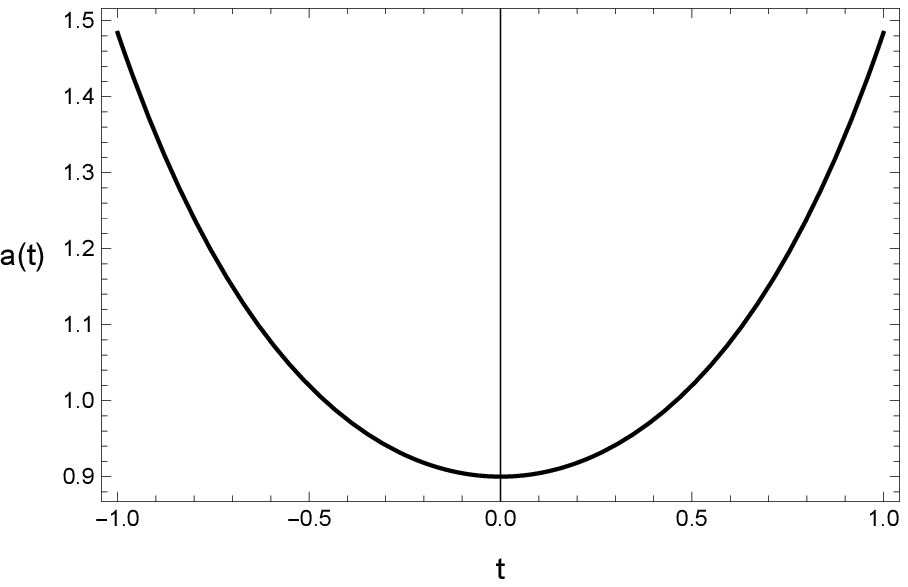}
\includegraphics[width=0.235\textwidth]{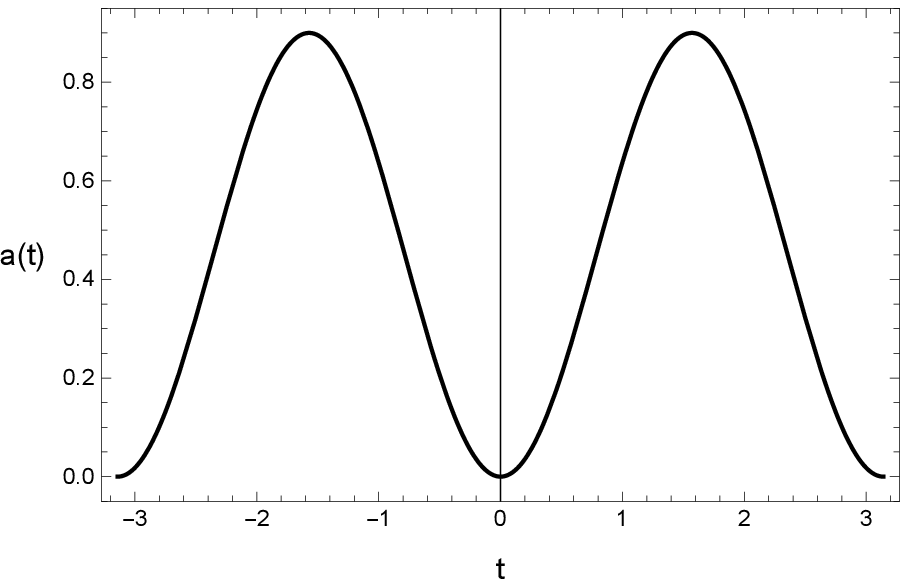}
\includegraphics[width=0.235\textwidth]{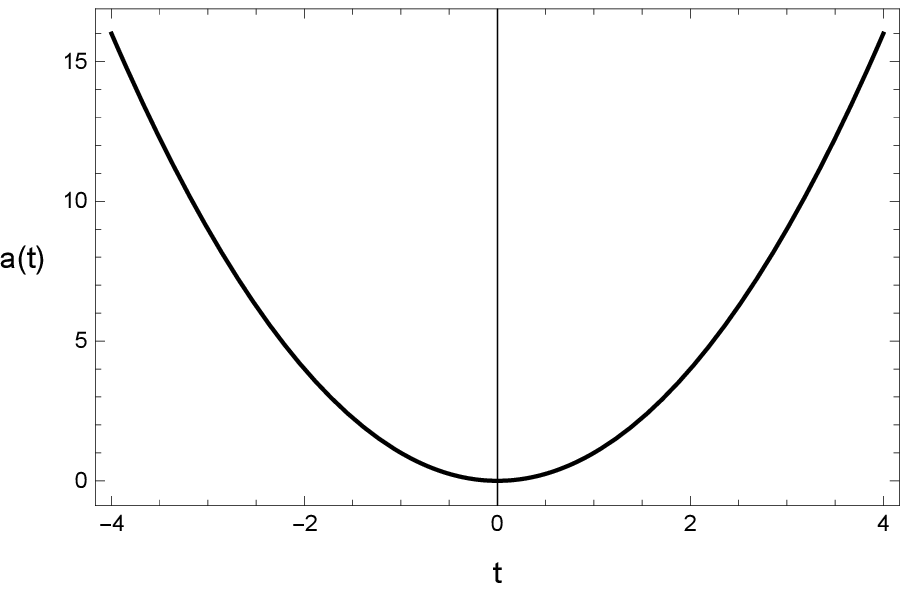}
\includegraphics[width=0.235\textwidth]{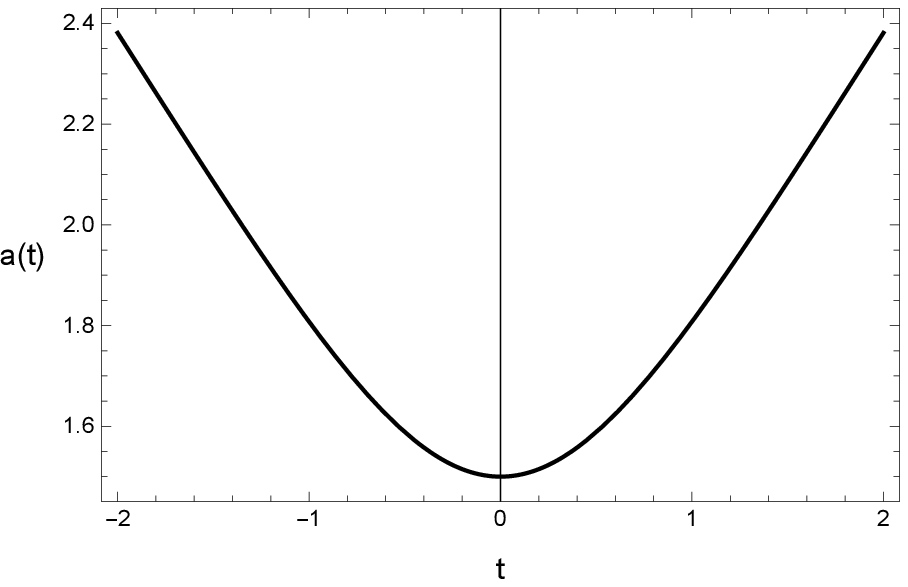}
\includegraphics[width=0.235\textwidth]{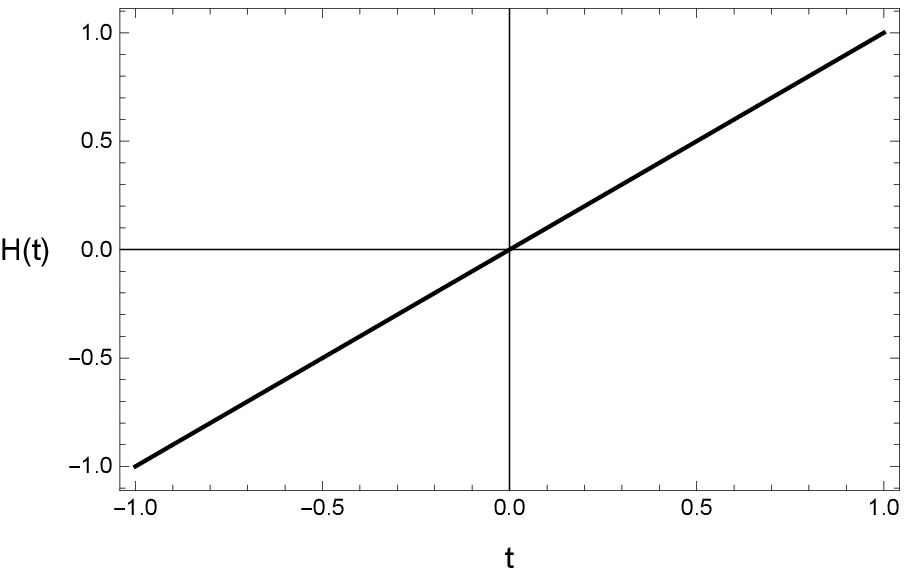}
\includegraphics[width=0.235\textwidth]{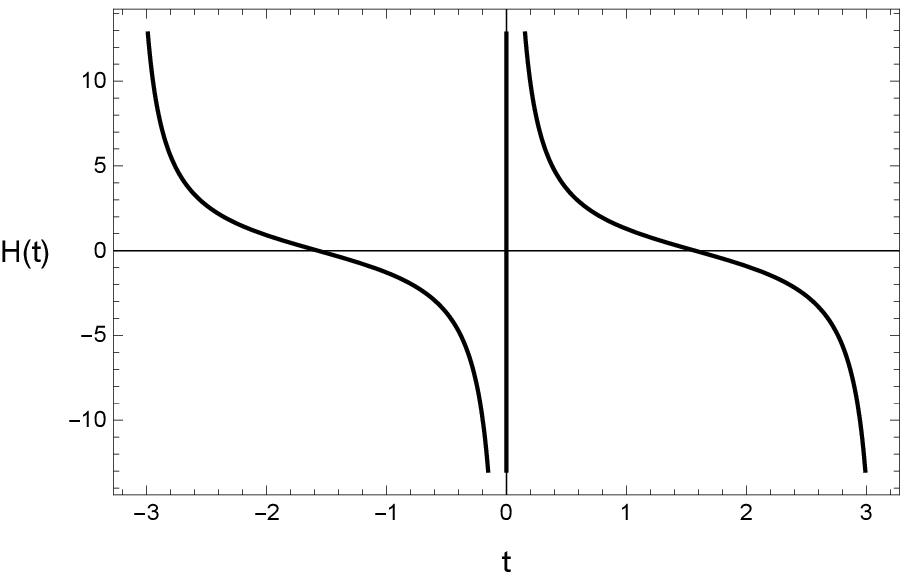}
\includegraphics[width=0.235\textwidth]{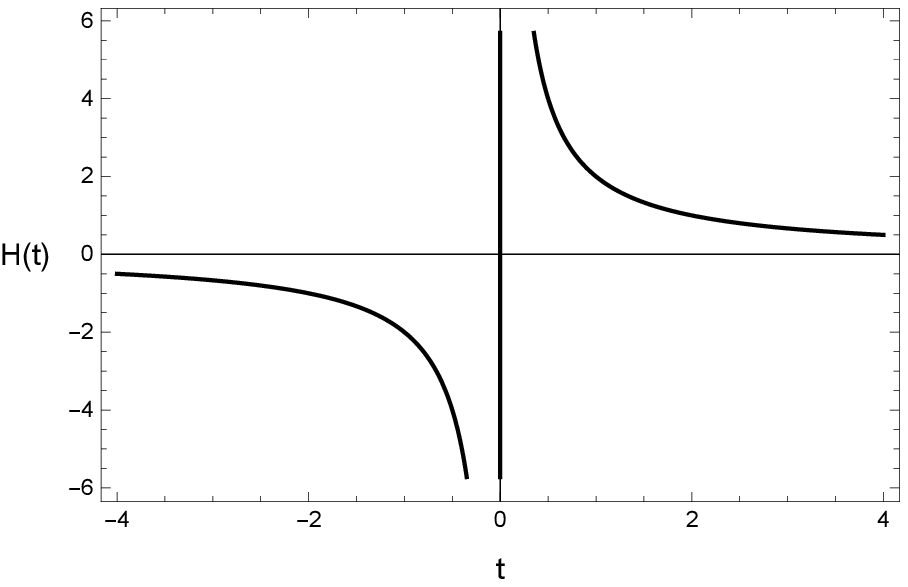}
\includegraphics[width=0.235\textwidth]{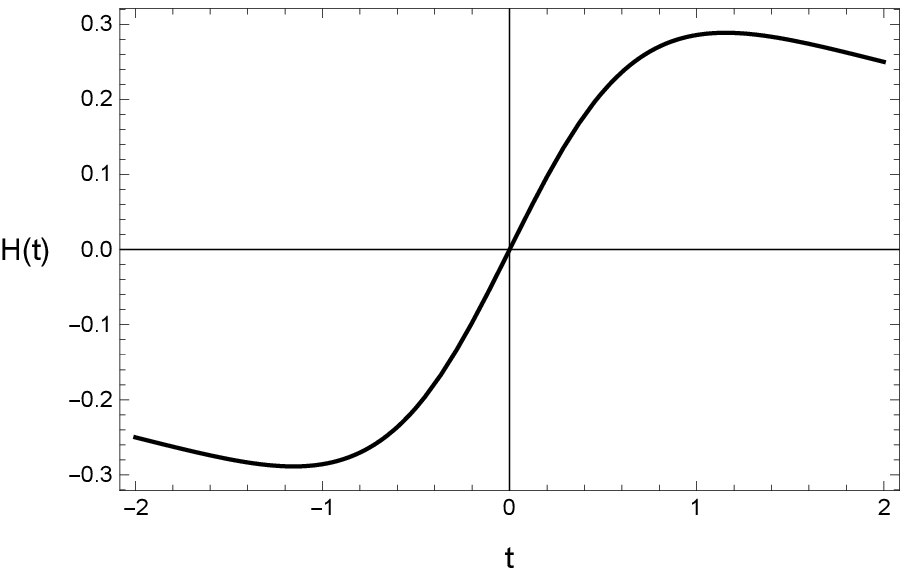}
\caption{\scriptsize{Figure provides the evolution of scale factor and Hubble parameter for four bouncing models.
The plots in the first column refers to exponential evaluation model, second column corresponds to oscillatory model, third column
refers to power law model while fourth column corresponds to matter bounce model. Here, we have chosen free parameters as
$A=.9,~ B=1,~ \sigma=.5$ and $\alpha=2$.}}\label{fig00}
\end{figure}

In this section, we shall adopt the well-known reconstruction scheme to reconstruct the forms of generic function $F(\mathcal{R},T)$ by
taking some interesting bouncing cosmological models defined by
\begin{enumerate}
\item  Exponential evaluation model: $a(t)=A\exp(\sigma\frac{t^2}{t_{\star}^2}),$ where $A$ and $\sigma$ are free constants;
\item  Oscillatory model: $a(t)=A \sin^2(B \frac{t}{t_{\star}}),$ where $A$ and $B$ are arbitrary constants;
\item  Power law model: $a(t)=\bigg(\frac{t_s-t}{t_{0}}\bigg)^{\frac{2}{c^2}}$, where $t_s,~ t_0$ and $c$ are free constants;
\item  Matter bounce model: $a(t)=A\bigg(\frac{3}{2}\rho_{cr}t^2+1\bigg)^{\frac{1}{3}},$ where $A$ and $\rho_{cr}$ are constants.
\end{enumerate}
The graphical behavior of these scale factors along with Hubble parameters are presented in the graphs of Figure \textbf{\ref{fig00}}.
The detail about these models like at what point the bouncing may occur will be provided in the upcoming sections.
We can reconstruct different forms of gravitational Lagrangian with the help of this approach in the desired cosmology
by the above listed form of scale factors along with the corresponding Hubble parameters. It is worthy to mention here that
the resulting partial differential equation (\ref{8}) is very complicated whose analytic solution is not possible, therefore
one require the following ansatz forms of $F(R,T)$ function:\\
(i) $F(\mathcal{R},T)=\beta_{1}g(\mathcal{R})+\beta_{2}h(T)$, ~~~~~~~~~~~~(ii) $F(\mathcal{R},T)=\mathcal{R}+\beta_{1}g(\mathcal{R})+\beta_{2}h(T)$,~~~~~~~~~~
(iii) $F(\mathcal{R},T)=\mathcal{R}g(T)$, \\ (iv) $F(\mathcal{R},T)=F(\mathcal{R})g(T)$,~~~~~~~~~~~~~~~~~~~~~(v) $F(\mathcal{R},T)=\mathcal{R}+Tg(\mathcal{R})$,
~~~~~~~~~~~~~~~~~~~~~~~~(vi) $F(\mathcal{R},T)=\mathcal{R}+\mathcal{R}F(T)$, \\(vii) $F(\mathcal{R},T)=F_1(\mathcal{R})+ F_2(\mathcal{R})F_3(T)$,
~~~~~(viii) $F(\mathcal{R},T)=\mu(\frac{R}{R_0})^{\beta}(\frac{T}{T_0})^{\gamma}$.\\
Here $\beta_1,~\beta_2,~\mu,~\beta,~\gamma,~R_0$ and $T_0$ are free parameters. These models provide minimal or non-minimal
interaction of Ricci scalar and energy-momentum tensor trace and are proved as interesting choices for $F(R,T)$ model in literature.
In the upcoming subsections, we shall reconstruct the form of generic function $F(R,T)$ theory by considering the above given
four cosmological bouncing solutions.

\subsection{Exponential Evaluation Bouncing Model}

Here, we consider the bouncing solution in exponential form which is defined by the following scale factor \cite{13*,25**,35*,9,32},
\begin{equation}\label{12}
a(t) = A \exp (\sigma \frac{t^2}{t_{\star}^2}),
\end{equation}
where $t_{\star}$ being the arbitrary time and $A$, $\sigma$ are positive constants. The corresponding Hubble parameter takes the form
\begin{equation}\label{13}
H = \frac{2\sigma t}{t_{\star}^2}.
\end{equation}
At $t=0$, it can be seen that $a(0)=A$, while the Hubble parameter's value at $t=0$ indicates that it represents bouncing point as $H=0$ there,
and $H$ is negative when $t<0$ whereas it becomes positive for $t>0$ as shown in the first column of Figure \textbf{\ref{fig00}}. Thus the
$00$ component of FRW metric in Eq.(\ref{8}) is given by
\begin{equation}\label{14}
k^2 \rho + (\rho+p)F_{T} + \frac{F}{2} + \frac{1}{4}(a_1-\mathcal{R})F_{\mathcal{R}}-a_1(\mathcal{R}+a_1)F_{\mathcal{RR}}
-\frac{3}{4}(1+\omega)(\mathcal{R}+a_1)TF_{\mathcal{R}T}=0,
\end{equation}
where $a_1=6\dot{H},~ \dot{H}=\frac{2\sigma}{t_{\star}^2}$ and $\dot{T}=-3H(1+\omega)T$. This is a partial differential equation whose
solution is difficult to find, therefore one can consider the above ansatz form of functions. By assuming $(i)$ form of function $F(R,T)$,
it can be easily checked that the above equation becomes separable and the resulting two differential equations (DEs) can be written as
\begin{equation}\label{15}
\frac{\beta_{1} g(\mathcal{R})}{2} + \frac{\beta_1}{4}(a_1-\mathcal{R})g_{\mathcal{R}}-a_1(\mathcal{R}+a_1)\beta_1g_{\mathcal{RR}}=0,
\end{equation}
\begin{equation}\label{16}
k^2 T + (1+\omega) \beta_2 T h_{T} + \frac{\beta_2(1-3\omega)h(T)}{2}=0.
\end{equation}
On solving these DEs, the solution for $g$ is given by
\begin{eqnarray}\label{17}
g(\mathcal{R})=\frac{(a_1^2 + 6 a_\mathcal{R}+\mathcal{R}^2)}{a_1^2}\bigg[c_1-\frac{\exp(-(\mathcal{R}/(4 a_1)))
\sqrt{(a_1 + \mathcal{R})}(3 a_1 + \mathcal{R}) c_2}{16 a_1^2 (a_1^2 + 6 a_1 \mathcal{R} +\mathcal{R}^2)}-\frac{\exp(1/4)\sqrt{\pi}
c_2 Erf[\frac{\sqrt{(a_1 + \mathcal{R})}}{(2 \sqrt{(a_1)})}]}{32 a_1^{5/2}}\bigg],
\end{eqnarray}
and the solution for $h(T)$ can be found by using constraint (\ref{11}) as follows
\begin{eqnarray}\label{18}
h(T)=-\frac{2 k^2 T}{\beta_{2} - 3 \beta_{2} \omega}+T^{1/2-\frac{\sqrt{-1+\omega(3+\omega-3\omega^2)}}{(1+\omega)\sqrt{-2+6\omega}}}c_3+
T^{1/2+\frac{\sqrt{-1+\omega(3+\omega-3\omega^2)}}{(1+\omega)\sqrt{-2+6\omega}}}c_4.
\end{eqnarray}
Thus $F(\mathcal{R},T)$ function can be written as:
\begin{eqnarray}\nonumber
F(\mathcal{R},T)& =&\beta_1 \frac{(a_1^2 + 6 a_1 \mathcal{R} + \mathcal{R}^2)}{a_1^2}\bigg[c_1-\frac{\exp(-(\mathcal{R}/(4 a_1)))\sqrt{(a_1 + \mathcal{R})} (3 a_1 + \mathcal{R}) c_2}{16 a_1^2 (a_1^2 + 6 a_1 \mathcal{R} + \mathcal{R}^2)}-\frac{\exp(1/4) \sqrt{\pi}c_2 Erf[\frac{\sqrt{(a_1 + \mathcal{R})}}{2 \sqrt{(a_1)}}])}{32 a_1^{5/2}}\bigg]\\\label{19} &&
-\frac{2 k^2 T}{1 - 3 \omega}+
T^{1/2 - \frac{\sqrt{-1+\omega(3+\omega-3\omega^2)}}{(1+\omega)\sqrt{-2+6\omega}}}\beta_2c_3+
T^{1/2+\frac{\sqrt{-1+\omega(3+\omega-3\omega^2)}}{(1+\omega)\sqrt{-2+6\omega}}}\beta_2c_4,
\end{eqnarray}
where all $c_i$'s are constants of integration. Now, we will check the possibility of vacuum solutions,
i.e., $F(0,0)=g(0)+h(0)=0$. Here, it is easy to verify that $h(T=0)=0$, thus we must impose $g(R=0)=0$ which further leads to the
relation: $32 (a_1)^{\frac{5}{2}}c_1-(6+e^{1/4}\sqrt{\pi}Erf(\frac{1}{2}))c_2=0$. Thus, the compatibility with the vacuum condition
is only possible when the condition: $c_1=\frac{(6+e^{1/4}\sqrt{\pi}Erf(\frac{1}{2}))c_2}{32 (a_1)^{\frac{5}{2}}}$ holds.

For $(ii)$ form of $F(R,T)$ function, the 00 component results in a similar equation as given in (\ref{14}) with an
additional term $\mathcal{R}$. This equation is a second-order partial differential equation which can also be separated by
introducing this $F(\mathcal{R},T)$ function. Consequently, the solution for $F(\mathcal{R},T)$ can be obtained as
\begin{eqnarray}\nonumber
F(\mathcal{R},T)&=&\frac{27a_1^2+34a_1\mathcal{R}+27\mathcal{R}^2}{128a_1}+ d_1\beta_1\frac{(a_1^2+6a_1\mathcal{R}+\mathcal{R}^2)}{a_1^2}
-\frac{\beta_1}{32a_1^{9/2}} \bigg[\exp(-(\mathcal{R}/(4 a_1))) d_2 (2 \sqrt{a_1(a_1+\mathcal{R})}\\\nonumber
&\times&(3a_1+\mathcal{R})+\exp((a_1+\mathcal{R})/(4a_1))\sqrt{\pi}(a_1^2+6a_1\mathcal{R}+\mathcal{R}^2)
Erf[\sqrt{a_1+\mathcal{R}}/(2\sqrt{ a_1})])\bigg]-\frac{2 k^2 T}{1-3\omega}\\\label{23}
&+&T^{1/2 - \frac{\sqrt{-1+\omega(3+\omega-3\omega^2)}}{(1+\omega)\sqrt{-2+6\omega}}}\beta_2d_3+
T^{1/2+\frac{\sqrt{-1+\omega(3+\omega-3\omega^2)}}{(1+\omega)\sqrt{-2+6\omega}}}\beta_2d_4,
\end{eqnarray}
where $d_i$'s are integration constants. Since $h(0)=0$ and hence to verify the vacuum condition, the condition: $g(R=0)=0$ must be satisfied.
It is easy to check that $g(R=0)=0$ will hold only if $\frac{d_2}{32(a_1)^{\frac{5}{2}}}(6+\sqrt{\pi}e^{1/4}Erf(\frac{1}{2}))-\frac{27a_1}{128\beta_1}=d_1$.

For $(iii)$ form of $F(\mathcal{R},T)$ function (re-scaling type model), Eq.(\ref{11}) leads to
\begin{eqnarray}\label{24}
B_1T^2(A_1\ln(T)+A_2)g_{TT}-B_{2}(\ln(T)-A_3)g(T)=-k^2 T.
\end{eqnarray}
The general solution of this DE is not possible to find, however one can find its complementary solution as follows
\begin{eqnarray}\nonumber
g(T)&=&1/A_1 B_2\exp(-\frac{B_2\sqrt{(1 + 4 B_2/A_1 B_1)/B_2^2}(A_2 + A_1 \ln(T))}{2 A_1})\sqrt{T}\bigg[c_1Hypergeometric1F1[((-A_2
+A_1(-A_3\\\nonumber
&+& A_1B_1\sqrt{(1 + 4 B_2/A_1 B_1)/B_2^2 }))/(A_1^2 B_1\sqrt{(1 + 4 B_2/A_1 B_1)/B_2^2})),2,\frac{B_2\sqrt{(1+4 B_2/A_1 B_1)/B_2^2}(A_2 + A_1 \ln(T))}{ A_1}]
\\\nonumber &+&C_1 HypergeometricU[((-A_2+A_1(-A_3 + A_1 B_1\sqrt{(1 + 4 B_2/A_1 B_1)/B_2^2)}))/( A_1^2B_1 \sqrt{(1 + 4 B_2/A_1 B_1)/B_2^2})),2,\\\label{25}
&-&\frac{B_2\sqrt{(1 + 4 B_2/A_1 B_1)/B_2^2}(A_2 + A_1 \ln(T))}{ A_1}]\bigg](A_2 + A_1 \ln(T)),
\end{eqnarray}
where the involved constants can be defined as $B_1=\frac{8\sigma(1+\omega)}{(\omega-1)t_{*}^2},~ A_1=1+9\omega,~
A_2=2B-4\ln(1-3\omega)+3(1-3\omega)\ln(1-3\omega)\rho_0,~ B=-\frac{3}{2}(1+\omega)-2\ln\rho_0,~ B_2=\frac{-4\sigma(1-3\omega)}{t_{*}^2(1+\omega)},~
A_3 = \ln(1-3\omega)\rho_0$ while $c_1$ and $c_2$ are
integrating constants. Since the general solution does not exist, therefore we can not check the vacuum condition in this case.

Using form $(iv)$ of $F(\mathcal{R},T)$ function (product form), the Friedman equation (\ref{14}) can be re-written as
\begin{eqnarray}\nonumber
\frac{k^2T}{1-3\omega} + \frac{(1+\omega)T}{1-3\omega}F(\mathcal{R})h_{T} + \frac{F(\mathcal{R})h(T)}{2} + \frac{1}{4}(a_1-\mathcal{R})F_{\mathcal{R}}h(T)-a_1(\mathcal{R}+a_1)F_{\mathcal{RR}}h(T)-\frac{3}{4}(1+\omega)(\mathcal{R}+a_1)T F_{\mathcal{R}}h_{T}=0.
\end{eqnarray}
which is a second-order partial differential equation that can not be separated and hence its analytic solution is not possible to find.
For function form $(v)$, the Friedman equation takes the form as
\begin{eqnarray}\label{27}
\hat{a_1} g(\mathcal{R})+\hat{a_3}(b_1-\mathcal{R} b_2)g_{\mathcal{R}}+\hat{a_4}(\mathcal{R}+a_1)g_{\mathcal{RR}}
=-k^2-\frac{1}{4\rho_0}\exp(\hat{a_2}(\mathcal{R}+a_1))(\mathcal{R}+a_1),
\end{eqnarray}
where $\hat{a_1}=\frac{(3-\omega)}{2},~ \hat{a_3}=\frac{1-3\omega}{4},~ b_1=6\dot{H}(9\omega-2), ~b_2=(4-9\omega),~ \hat{a_4}=6\dot{H}(1-3\omega)$
and $\hat{a_2}=\frac{-9t_{*}^2(1+\omega)}{\sigma}$. This equation is second-order DE whose analytical solution (general solution) does not exist
but the complementary solution can be obtained as follows
\begin{eqnarray}\nonumber
g(\mathcal{R})&=&(a_1 + \mathcal{R})^{(\hat{a_4}-\hat{a_3}(b_1 + b_2 a_1))/\hat{a_4}}d_1
HypergeometricU\bigg[-\frac{\hat{a_1}\hat{a_4} - \hat{a_3} \hat{a_4} b_2 + \hat{a_3}^2 b_1 b_2
+\hat{a_3}^2 b_2^2 a_1}{\hat{a_3} \hat{a_4} b_2}, 1+ \\\nonumber && \frac{\hat{a_4} - \hat{a_3} b_1
-\hat{a_3} b_2 a_1}{\hat{a_4}}, \frac{\hat{a_3} b_2 a_1+\hat{a_3} b_2 \mathcal{R}}{\hat{a_4}}\bigg]
+(a_1+\mathcal{R})^{(\hat{a_4} - \hat{a_3} (b_1 + b_2 a_1))/\hat{a_4}}d_2LaguerreL\bigg[\\\label{28}
&&\frac{\hat{a_1}\hat{a_4}-\hat{a_3}\hat{a_4} b_2+\hat{a_3}^2 b_1 b_2+\hat{a_3}^2 b_2^2 a_1}{\hat{a_3}\hat{a_4} b_2},
\frac{\hat{a_4}-\hat{a_3} b_1-\hat{a_3} b_2 a_1}{\hat{a_4}}, \frac{\hat{a_3} b_2 a_1+\hat{a_3} b_2 \mathcal{R}}{\hat{a_4}}\bigg],
\end{eqnarray}
with $d_1$, $d_2$ being integrating constants. It is worthy to mention here that nothing can be inferred about the vacuum condition
due to unavailability of general solution. In case of form $(vi)$, the Friedman equation can be re-defined as
\begin{eqnarray}\label{29}
s_2 T^ 2 (a_1+s_3(\ln(T)-s_1))f_{TT} + \frac{3}{A}(\ln(T) - s_1) f_{T}=\frac{k^2 }{1-3\omega}T - \frac{3}{A}(\ln(T)-s_1),
\end{eqnarray}
where $a_1=6\dot{H},~ s_2=2\frac{(1+\omega)^2}{(1-3\omega)(\omega-1)},~ s_1= \ln(\rho_0(1-3\omega)),~ s_3=\frac{3(1+9\omega)}{A}$ and
$A=\frac{-3t_{*}^2(1+\omega)}{4\alpha}$. Again, general solution to this DE is not possible to determine, therefore one can obtain only
its complementary solution which is given by
\begin{eqnarray}\nonumber
F(T)&=&\frac{1}{s_3} \exp(\frac{\sqrt{1 - 12/(A s_2 s_3)}(-a_1 + s_1 s_3 - s_3 \ln(T))}{2 s_3})\sqrt{T}
\bigg(c_2 Hypergeometric1F1\bigg[1+\frac{3a_1\sqrt{1-\frac{12}{A s_2 s_3}}}{s_3 (-12 + A s_2 s_3)},\\\nonumber
&& 2, \frac{\sqrt{1-\frac{12}{A s_2 s_3}}(a_1 - s_1 s_3 + s_3 \ln(T))}{s_3}\bigg]
+c_1 HypergeometricU\bigg[1+\frac{3a_1\sqrt{1-\frac{12}{A s_2 s_3}}}{s_3 (-12 + A s_2 s_3)}, 2, \\\nonumber
&& \frac{\sqrt{1-\frac{12}{A s_2 s_3}}(a_1 - s_1 s_3 + s_3 \ln(T))}{s_3}\bigg]\bigg)(a_1 - s_1 s_3 + s_3 \ln(T)),
\end{eqnarray}
where $c_1$ and $c_2$ are integrating constants.

Next we utilize $(vii)$ form of $F(\mathcal{R},T)$ function in the Friedman equation which yields the following DE:
\begin{eqnarray}\nonumber
&&\frac{k^2T}{1-3\omega}+\frac{(1+\omega)T}{1-3\omega}F_2(\mathcal{R})F_{3T} + \frac{F_1(\mathcal{R})+F_2(\mathcal{R})F_3(T)}{2} + \frac{1}{4}(a_1-\mathcal{R})(F_{1\mathcal{R}}+F_{2\mathcal{R}}F_3(T))\\\label{31}
&&-a_1(\mathcal{R}+a_1)(F_{1\mathcal{RR}}+F_{2\mathcal{RR}}F_3(T))-\frac{3}{4}(1+\omega)(\mathcal{R}+a_1)T F_{2\mathcal{R}}F_{3T}=0,
\end{eqnarray}
which can be separated into two DEs given by
\begin{eqnarray}\nonumber
&&\frac{k^2T}{1-3\omega}+\frac{(1+\omega)T}{1-3\omega}F_2(\mathcal{R})F_{3T} + \frac{F_2(\mathcal{R})F_3(T)}{2} + \frac{1}{4}(a_1-\mathcal{R})F_{2\mathcal{R}}F_3(T)-a_1(\mathcal{R}+a_1)F_{2\mathcal{RR}}F_3(T)\\\nonumber
&&-\frac{3}{4}(1+\omega)(\mathcal{R}+a_1)TF_{2\mathcal{R}}F_{3T}=0,\\\nonumber
&&\frac{F_1(\mathcal{R})}{2}+\frac{(a_1-\mathcal{R})}{4}F_{1\mathcal{R}}-a_1(a_1+\mathcal{R})F_{1\mathcal{RR}}=0.
\end{eqnarray}
In the first DE, the unknowns depending on $\mathcal{R}$ and $T$ can not be separated, therefore we can not find its solution.
The second ODE can be solved analytically and its solution for $F_1(\mathcal{R})$ is quite similar to the solution of $g(\mathcal{R})$ as
given in Eq.(\ref{17}). For $(viii)$ form of generic function, the Friedman's equation reduces to
\begin{eqnarray}\nonumber
&&\mu\bigg(\frac{R}{R_0}\bigg)^{\beta}\bigg(\frac{T}{T_0}\bigg)^{\gamma}\bigg[(\frac{1+\omega}{1-3\omega})\gamma
+\frac{1}{2}+\frac{\beta (12\sigma-Rt^2_{\star})}{4R t^2_{\star}}-\frac{12\sigma \beta (\beta-1)(12\sigma+R t^2_{\star})}{R^2t^4_{\star}}
-\frac{3(1+\omega)\beta \gamma (12\sigma +R t^2_{\star})}{4R t^2_{\star}}\bigg]\\\label{i}
&&=-T_0 \sum_i\Omega_{\omega_{i},0}a^{-3(1+\omega_i)},
\end{eqnarray}
where $\mu,~\beta$ and $\gamma$ are arbitrary constants. In this case, the vacuum condition is satisfied if $\gamma>0$. At times, when $R=T=0$,
the Friedman's equation yields the condition, i.e., $\sum_i\Omega_{\omega_i,0}A^{-3(1+\omega_i)}=0$, only exists in vacuum.
Which implies that $\mu$ is equal to zero, a contradiction. Thus, there must exist a fluid with $\omega_i=-1$. Then, by
calculating the value of $\mu$ at current times, we obtain:
\begin{eqnarray}\label{ii}
\mu &=& \frac{-T_0\sum_i\Omega_{\omega_i,0}}{(\frac{1+\omega}{1-3\omega})\gamma+\frac{1}{2}+\frac{\beta (12\sigma-R_0t^2_{\star})}{4R_0 t^2_{\star}}-\frac{12\sigma \beta (\beta-1)(12\sigma+R_0 t^2_{\star})}{R_0^2t^4_{\star}}-\frac{3(1+\omega)\beta \gamma (12\sigma +R_0 t^2_{\star})}{4R_0 t^2_{\star}}}\equiv \frac{-T_0\sum_i\Omega_{\omega_i,0}}{\nu},
\end{eqnarray}
where $\nu\neq 0$. By inserting the value of $\mu$ in Eq.(\ref{i}), we get
\begin{eqnarray}\nonumber
\nu &=&\bigg(\frac{R}{R_0}\bigg)^{\beta}\bigg(\frac{T}{T_0}\bigg)^{\gamma}\bigg[(\frac{1+\omega}{1-3\omega})\gamma+\frac{1}{2}+\frac{\beta (12\sigma-Rt^2_{\star})}{4R t^2_{\star}}-\frac{12\sigma \beta (\beta-1)(12\sigma+R t^2_{\star})}{R^2t^4_{\star}}-\frac{3(1+\omega)\beta \gamma (12\sigma +R t^2_{\star})}{4R t^2_{\star}}\bigg].
\end{eqnarray}
Since $\nu$ is constant, therefore all time dependent terms must vanish. It is significant to mention here that there is no possible choice of
parameters which make it constant and hence satisfies the Friedman's equation. Hence, this power law model does not describe the symmetric bouncing
cosmology.

\subsection{Oscillatory Bouncing Model}

Here we shall reconstruct the form of $F(\mathcal{R},T)$ function by using oscillatory bouncing model which
is defined by the following scale factor \cite{9,33}
\begin{equation}\nonumber
a(t)=A \sin^2(B \frac{t}{t_{\star}}),
\end{equation}
where $t_{\star}>0$ represents the reference time while $A$ and $B$ are positive constants. For such model, the Hubble parameter is defined as
\begin{eqnarray}\nonumber
H=\frac{2 B}{t_{\star}}\cot(B\frac{t}{t_{\star}}),
\end{eqnarray}
This model results into two types of bouncing behavior: firstly $t=\frac{n\pi t_{\star}}{B}, ~n\in Z$ implies that $a(t)=0,~H=-\infty ~ or ~\infty$
which corresponds to super bounce and secondly, for $t=\frac{2(n+1)\pi t_{\star}}{2B},~ n\in Z$, we obtain: $a(t)=A, ~H=0$ which indicates the
simple bounce where universe comes to its maximum size. Using this form of scale factor along with Hubble parameter in Eq.(\ref{8}), we can
re-write the Friedman equation as follows
\begin{eqnarray}\label{32}
k^2 \rho + (1+\omega) \rho F_{T}+\frac{F}{2}-(\frac{\mathcal{R}}{6}+a_2)F_{\mathcal{R}}
+6(\frac{\mathcal{R}}{6}-2a_2)(-\frac{\mathcal{R}}{3}+a_2)F_{\mathcal{RR}}+9T(1+\omega)(-\frac{\mathcal{R}}{9}+a_2)F_{\mathcal{R}T}=0,
\end{eqnarray}
where $a_2=4(\frac{B}{t_{*}})^2$. Again it is checked that the solution of above equation is not possible to find, so for simplicity reasons,
we use specific 7 forms of $F(\mathcal{R},T)$ function as listed before.

Applying $(i)$ form of Lagrangian function, the above equation can be easily separated into two ODEs for unknowns $g$ and $h$ whose
solutions are given by
\begin{eqnarray}\nonumber
g(\mathcal{R})&=&\frac{1}{243 a_2^{5/2}\sqrt{-3a_2+\mathcal{R}}}\bigg[-2(-3a_2+\mathcal{R})^2ArcTanh(\frac{\sqrt{-3a_2
+\mathcal{R}}}{3\sqrt{a_2}})c_2+3\sqrt{a_2}(729 a_2^4 c_1\\\label{34}
&-&486a_2^3\mathcal{R}c_1+81a_2^2 \mathcal{R}^2c_1+2\mathcal{R}\sqrt{-3a_2+\mathcal{R}}c_2)\bigg],
\end{eqnarray}
while the solution for $h(T)$ is same as given in Eq.(\ref{18}). Thus, the form of $F(\mathcal{R},T)$ function is given by
\begin{eqnarray}\nonumber
F(\mathcal{R},T)&=& \frac{1}{243 a_2^{5/2}\sqrt{-3 a_2 + \mathcal{R}}}\bigg[-2 (-3 a_2
+ \mathcal{R})^2ArcTanh(\frac{\sqrt{-3 a_2 + \mathcal{R}}}{3\sqrt{a_2}})c_2\\\nonumber
&+&3\sqrt{a_2}(729 a_2^4 c_1-486a_2^3 \mathcal{R} c_1 + 81 a_2^2 \mathcal{R}^2c_1+2\mathcal{R}\sqrt{-3 a_2+\mathcal{R}}c_2)\bigg]
-\frac{2 k^2 T}{1 - 3 \omega}\\\label{35}
&+& T^{1/2 - \frac{\sqrt{-1+\omega(3+\omega-3\omega^2)}}{(1+\omega)\sqrt{-2+6\omega}}}\beta_2c_3+ T^{1/2+\frac{\sqrt{-1+\omega(3+\omega-3\omega^2)}}{(1+\omega)\sqrt{-2+6\omega}}}\beta_2c_4.
\end{eqnarray}
The vacuum limit $T\rightarrow 0$ implies that $h(0)=0$, whereas in the limit $R\rightarrow 0$, the function
$g(0)=0$ is only possible when $ c_1=\frac{18 c_2\tanh^{-1}(\frac{\iota}{\sqrt{3}})}{2187 a_2^{5/2}}$. Hence vacuum condition can be
satisfied by imposing this constraint on constants $c_1$ and $c_2$.

Next we use $(ii)$ of Lagrangian function in the Friedman equation and obtain second-order separable partial differential
equation whose solution is given by
\begin{eqnarray}\nonumber
F(\mathcal{R},T)&=&\mathcal{R}+\frac{1}{486 a_2^{5/2}\sqrt{-3 a_2 + \mathcal{R}}}
\bigg[2 (-3 a_2 + \mathcal{R})^2 ArcTanh(\frac{\sqrt{-3 a_2 + \mathcal{R}}}{3\sqrt{a_2}})(9 a_2^2 - 2 \beta_1 c_2)+81 a_2^2
\\\nonumber &\times& (-3a_2+ \mathcal{R})^2 \ln(-3\sqrt{a_2}+\sqrt{-3a_2+\mathcal{R}})+3\sqrt{a_2}(2 (729 a_2^4 \beta_1 c_1
-486 a_2^3 \mathcal{R}\beta_1c_1-9 (a_2^2)\mathcal{R}\\\nonumber
&\times&(\sqrt{-3a_2+\mathcal{R}}-9 \mathcal{R} \beta_1 c_1)+2\mathcal{R}\sqrt{-3a_2+\mathcal{R}}\beta_1 c_2)
-27 a_2^{3/2} (-3 a_2 + \mathcal{R})^2 \ln(3 \sqrt{a_2} + \sqrt{-3 a_2 + \mathcal{R}}))\\\label{37}
&-&\frac{2 k^2 T}{1 - 3 \omega}+  T^{1/2 - \frac{\sqrt{-1+\omega(3+\omega-3\omega^2)}}{(1+\omega)\sqrt{-2+6\omega}}}\beta_2c_3+
T^{1/2+\frac{\sqrt{-1+\omega(3+\omega-3\omega^2)}}{(1+\omega)\sqrt{-2+6\omega}}}\beta_2c_4\bigg].
\end{eqnarray}
In the vacuum limit: $R\rightarrow 0$, we get $g(0)=\frac{1}{486 \sqrt{3}\iota a_2^3}[18a_2^2\tanh^{-1}(\iota/\sqrt{3})(9a_2^2-2\beta_1c_2)+729 a_2^4(\ln(\frac{-3\sqrt{a_2}+\sqrt{3a_2}\iota}{3\sqrt{a_2}+\sqrt{3a_2}\iota}))+4374a_2^{9/2}c_1\beta_1]\neq 0$. Here, the vacuum
condition is only satisfied when $c_1$=$\frac{-1}{4374 a_2^{9/2}\beta_1}[18a_2^2\tanh^{-1}(\iota/\sqrt{3})(9a_2^2-2\beta_1c_2)+\\729 a_2^4(\ln(\frac{-3\sqrt{a_2}+\sqrt{3a_2}\iota}{3\sqrt{a_2}+\sqrt{3a_2}\iota}))]$ is imposed. If we consider $(iii)$ form of $F(\mathcal{R},T)$
function, the Friedman equation along with constraint (\ref{11}) yields
\begin{eqnarray}\label{38}
s_2 T^ 2 (s_3 T ^c + s_4)F_{TT}+(s_5 T^c +3 a_2)F(T)= -\hat{a} T,
\end{eqnarray}
where $s_2=\frac{2(1+\omega)^2}{\omega-1},~ s_3=\frac{-108\omega B^2}{(1-3\omega)^{c+1}t_{*}^2\rho_0^c},~ s_4=\frac{144\omega B^2}{(1-3\omega)t_{*}^2}+a_2, ~
s_5=\frac{-3a_2}{((1-3\omega)\rho_0)^c},~ \hat{a}=\frac{k^2}{1-3\omega}$ and $c=\frac{1}{3+3\omega}$. Again, it is non-homogeneous ODE of
second-order whose analytical solution is not possible, however its complementary solution is given by
\begin{eqnarray}\nonumber
F(T)&=&\iota^{\frac{cs_2s_4-c\sqrt{s_2s_4(-12 a_2 + s_2 s_4)}}{c^2 s_2 s_4}}s_3^{\frac{cs_2s_4
-c\sqrt{s_2s_4(-12 a_2 + s_2 s_4)}}{2c^2 s_2 s_4}}s_4^{-\frac{cs_2s_4-c\sqrt{s_2s_4(-12 a_2 + s_2 s_4)}}{2c^2 s_2 s_4}}(T^{c})^{\frac{cs_2s_4
-c\sqrt{s_2s_4(-12 a_2 + s_2 s_4)}}{2c^2 s_2 s_4}}\\\nonumber
&&C_1 Hypergeometric2F1\bigg[-\frac{\sqrt{-12 a_2 c^2 s_2 s_4 + c^2 s_2^2 s_4^2}}{2 c^2 s_2 s_4}-\frac{\sqrt{s_2 s_3 + 4 s_5}}{2 c \sqrt{s_2s_3}},
-\frac{\sqrt{-12 a_2 c^2 s_2 s_4 + c^2 s_2^2 s_4^2}}{2 c^2 s_2 s_4}+\frac{\sqrt{s_2 s_3 + 4 s_5}}{2 c \sqrt{s_2s_3}},\\\nonumber
&&1-\frac{\sqrt{-12 a_2 c^2 s_2 s_4 + c^2 s_2^2 s_4^2}}{2 c^2 s_2 s_4}, \frac{s_3T^c}{s_4}\bigg]+\iota^{\frac{cs_2s_4+c\sqrt{s_2s_4(-12 a_2 + s_2 s_4)}}{c^2 s_2 s_4}}s_3^{\frac{cs_2s_4+c\sqrt{s_2s_4(-12 a_2 + s_2 s_4)}}{2c^2 s_2 s_4}}s_4^{-\frac{cs_2s_4+c\sqrt{s_2s_4(-12 a_2 + s_2 s_4)}}{2c^2 s_2 s_4}}\\\nonumber &\times&(T^{c})^{\frac{cs_2s_4+c\sqrt{s_2s_4(-12 a_2 + s_2 s_4)}}{2c^2 s_2 s_4}}C_1
Hypergeometric2F1\bigg[-\frac{\sqrt{-12 a_2 c^2 s_2 s_4 + c^2 s_2^2 s_4^2}}{2 c^2 s_2 s_4}-\frac{\sqrt{s_2 s_3 + 4 s_5}}{2 c \sqrt{s_2s_3}},\\\label{39}
&&-\frac{\sqrt{-12 a_2 c^2 s_2 s_4 + c^2 s_2^2 s_4^2}}{2 c^2 s_2 s_4}+\frac{\sqrt{s_2 s_3 + 4 s_5}}{2 c \sqrt{s_2s_3}}, 1
+\frac{\sqrt{-12 a_2 c^2 s_2 s_4 + c^2 s_2^2 s_4^2}}{2 c^2 s_2 s_4}, \frac{s_3T^c}{s_4}\bigg].
\end{eqnarray}
It is easy to check that in the limit $T\rightarrow 0$, the above solution results in $F(0)=0$ but nothing can
be inferred about general solution of $F(T)$. Therefore, vacuum condition is not valid in this type of Lagrangian. Similarly,
imposing $(iv)$ form of $F(\mathcal{R},T)$ in Friedman equation, we obtain a new partial differential equation whose analytical solution is
not possible.

In case of $(v)$ form of function $F(\mathcal{R},T)$, the Friedman equation becomes
\begin{eqnarray}\label{41}
\frac{k^2 T}{1-3\omega}+T(1+\omega)(\frac{3\omega \mathcal{R}}{1-3\omega}+\frac{12B^2}{t_{\star}^2})F_{T}+(\frac{\mathcal{R}}{3}-\frac{4B^2}{t_{\star}^2})+(\frac{\mathcal{R}}{3}
-\frac{4B^2}{t_{\star}^2})F(T)=0.
\end{eqnarray}
Using equation of continuity, we obtain a relation between $\mathcal{R}$ and $T$ given by
$\mathcal{R}=\frac{-36B^2}{t_{\star}^2}(\frac{T}{(1-3\omega)\rho_0})^{\frac{1}{3+3\omega}}$ + $\frac{48B^2}{t_{\star}^2}$.
Inserting value of $\mathcal{R}$ and applying the constraint (\ref{11}), the above DE can also be expressed in terms of $T$ and can be
written as
\begin{eqnarray}\label{42}
r_1 T^ 2 (r_2 T ^c + r_3)F_{TT}-r_4(r_5 T^c - 1)F(T)=r_4(r_5 T^c - 1)-\frac{k^2T}{1-3\omega},
\end{eqnarray}
which is non-homogeneous DE and its general solution does not exist, but one can obtain the following complementary solution:
\begin{eqnarray}\nonumber
F(T)&=&r_2^{\frac{\sqrt{r_1r_3}-\sqrt{r_1r_3-4r_4}}{2c\sqrt{r_1r_3}}}r_3^{-\frac{\sqrt{r_1r_3}-\sqrt{r_1r_3-4r_4}}{2c\sqrt{r_1r_3}}}(T^c)^{\frac{\sqrt{r_1r_3}-
\sqrt{r_1r_3-4r_4}}{2c\sqrt{r_1r_3}}} C_1 Hypergeometric2F1\bigg[-\frac{\sqrt{r_1 r_3 - 4 r_4}}{2c\sqrt{r_1r_3}}\\\nonumber
&&-\frac{\sqrt{r_1 r_2 + 4 r_4r_5}}{2c\sqrt{r_1r_2}}, -\frac{\sqrt{r_1 r_3 - 4 r_4}}{2c\sqrt{r_1r_3}}+\frac{\sqrt{r_1 r_2 + 4 r_4r_5}}{2c\sqrt{r_1r_2}},
1-\frac{\sqrt{r_1 r_3 - 4 r_4}}{c\sqrt{r_1r_3}}, -\frac{r_2}{r_3}T^c\bigg]\\\nonumber &+& r_2^{\frac{\sqrt{r_1r_3}+\sqrt{r_1r_3-4r_4}}{2c\sqrt{r_1r_3}}}r_3^{-\frac{\sqrt{r_1r_3}+\sqrt{r_1r_3-4r_4}}{2c\sqrt{r_1r_3}}}(T^c)^{\frac{\sqrt{r_1r_3} +
\sqrt{r_1r_3-4r_4}}{2c\sqrt{r_1r_3}}} C_2 Hypergeometric2F1\bigg[-\frac{\sqrt{r_1 r_3 - 4 r_4}}{2c\sqrt{r_1r_3}}\\\label{43}
&&-\frac{\sqrt{r_1 r_2 + 4 r_4r_5}}{2c\sqrt{r_1r_2}}, -\frac{\sqrt{r_1 r_3 - 4 r_4}}{2c\sqrt{r_1r_3}}+\frac{\sqrt{r_1 r_2 + 4 r_4r_5}}{2c\sqrt{r_1r_2}},
1-\frac{\sqrt{r_1 r_3 - 4 r_4}}{c\sqrt{r_1r_3}}, -\frac{r_2}{r_3}T^c\bigg],
\end{eqnarray}
where $r_1=\frac{2(1+\omega)^2}{\omega-1}, ~r_2=\frac{-108\omega B^2}{(1-3\omega)^{c+1}t_{*}^2\rho_0^c},~ r_3=\frac{144\omega B^2}{(1-3\omega)t_{*}^2}+r_4,
~r_4=\frac{12B^2}{t_{*}^2},~ r_5=(\frac{1}{(1-3\omega)\rho_0})^c,~ c=\frac{1}{3+3\omega}$ while $c_1$ and $c_2$ are integrating constants.
Since the analytical solution for $F_{part}(T)$ is not obtained so vacuum condition is not satisfied. Similarly, inserting the form $(vii)$
of generic function in Friedman equation, we obtain a new partial differential equation whose analytic solution is impossible to find.

Taking $(viii)$ power law type of Lagrangian function, the Friedman's equation becomes
\begin{eqnarray}\nonumber
&&\mu \bigg(\frac{R}{R_0}\bigg)^{\beta}\bigg(\frac{T}{T_0}\bigg)^{\gamma}\bigg[(\frac{1+\omega}{1-3\omega})\gamma+\frac{1}{2}
+\frac{\beta (24B^2+Rt^2_{\star})}{6R t^2_{\star}}+\frac{6 \beta (\beta-1)(12B^2-R t^2_{\star})(R t^2_{\star}-48B^2)}{18R^2t^4_{\star}}
+\frac{9(1+\omega)\beta \gamma}{6R t^2_{\star}}\\\nonumber &&\times (24B^2 -R t^2_{\star})\bigg]=-T_0 \sum_i\Omega_{\omega_{i},0}a^{-3(1+\omega_i)},
\end{eqnarray}
In this case, the vacuum condition is satisfied if $\gamma>0$. At times when $R=T=0$, the Friedman's equation yields the condition:
$\sum_i\Omega_{\omega_i,0}A^{-3(1+\omega_i)}=0$, which is possible in vacuum. Then, the value of $\mu$ at the current time is given by
\begin{eqnarray}\nonumber
\mu &=& \frac{-T_0\sum_i\Omega_{\omega_i,0}}{(\frac{1+\omega}{1-3\omega})\gamma+\frac{1}{2}
+\frac{\beta (24B^2+R_0t^2_{\star})}{6R_0 t^2_{\star}}+\frac{6 \beta (\beta-1)(12B^2
-R_0 t^2_{\star})(R_0 t^2_{\star}-48B^2)}{18R_0^2t^4_{\star}}+\frac{9(1+\omega)\beta \gamma (24B^2
-R_0 t^2_{\star})}{6R_0 t^2_{\star}}}\equiv \frac{-T_0\sum_i\Omega_{\omega_i,0}}{\nu},
\end{eqnarray}
where $\nu\neq 0$ and $\omega_i=-1,~ \forall i$. By inserting the value of $\mu$ in the above Friedman's equation, we obtain:
\begin{eqnarray}\nonumber
\nu &=&\bigg(\frac{R}{R_0}\bigg)^{\beta}\bigg(\frac{T}{T_0}\bigg)^{\gamma}\bigg[(\frac{1+\omega}{1-3\omega})\gamma
+\frac{1}{2}+\frac{\beta (24B^2+Rt^2_{\star})}{6R t^2_{\star}}+\frac{6 \beta (\beta-1)(12B^2
-R t^2_{\star})(R t^2_{\star}-48B^2)}{18R^2t^4_{\star}}+\\\nonumber &&\frac{9(1+\omega)\beta \gamma (24B^2 -R t^2_{\star})}{6R t^2_{\star}}\bigg].
\end{eqnarray}
Since $\nu$ is a constant, therefore all time dependent terms must vanish. Clearly there is no possible choice of parameters which can make it
constant and satisfy the Friedman's equation. Hence, it can be concluded that this power law model does not describe the oscillatory bouncing
cosmology.

\subsection{Power Law Model}

Here we shall reconstruct the form of generic function by taking power law bouncing model into account. It is defined by the following
scale factor: \cite{13*,25**,35*,9,29e}
\begin{eqnarray}\nonumber
a(t)=\bigg(\frac{t_s-t}{t_{0}}\bigg)^{\frac{2}{c^2}},
\end{eqnarray}
where $t_s$ represents the bouncing point, $t_0$ is the arbitrary time and $c$ is an arbitrary constant. Here, by re-scaling
$t_1=t_s-t$ and $\alpha=\frac{2}{c^2}$, we can re-write above scale factor as follows
\begin{eqnarray}\nonumber
a(t_1)=\bigg(\frac{t_1}{t_0}\bigg)^{\alpha} \rightarrow H=-\frac{\alpha}{t_1}\rightarrow \mathcal{R}=6\alpha(1-2\alpha)t_1^{-2}.
\end{eqnarray}
Here $\alpha>0$. Using this scale factor along with Hubble parameter in 00 component of field equation, we obtain
\begin{eqnarray}\nonumber
k^2 \rho + (1+\omega) \rho F_{T} + \frac{F}{2} + \frac{1-\alpha}{2(2\alpha -1)} \mathcal{R} F_{\mathcal{R}}
+\frac{1}{1-2\alpha}\mathcal{R}^2F_{\mathcal{RR}} + \frac{3\alpha(1+\omega)}{2(1-2\alpha)}\mathcal{R} T F_{\mathcal{R}T}=0.
\end{eqnarray}
Since this equation is difficult to solve for unknown, therefore we impose $F(\mathcal{R},T)$ model $(i)$ to find its solution. The
corresponding DEs can be written as
\begin{eqnarray}\label{46}
\beta_{1}\mathcal{R}^2F_{\mathcal{RR}}+\beta_{1}(\frac{\alpha-1}{2})\mathcal{R} F_{\mathcal{R}}+\beta_{1}\frac{1-2\alpha}{2}F(\mathcal{R})=0,
\end{eqnarray}
which is Euler's equation whose solution is given by
\begin{eqnarray}\nonumber
F(\mathcal{R})=w_1 \mathcal{R}^{\mu+}+w_2 \mathcal{R}^{\mu-},
\end{eqnarray}
where $\mu_{\pm}=\frac{3-7\alpha+2\alpha^2}{4(1-2\alpha)}\pm\sqrt{1+\alpha(10+\alpha)}$ and further, by using perfect fluid with
$\omega=\frac{p}{\rho}$ in the DE for unknown $h(T)$ which implies
\begin{eqnarray}\label{47}
F(\mathcal{R},T)=\beta_1w_1 \mathcal{R}^{\mu+}+\beta_1w_2 \mathcal{R}^{\mu-}+\frac{2k^2}{-1+3\omega}T+
T^{1/2 - \frac{\sqrt{-1+\omega(3+\omega-3\omega^2)}}{(1+\omega)\sqrt{-2+6\omega}}}\beta_2w_3+
T^{1/2+\frac{\sqrt{-1+\omega(3+\omega-3\omega^2)}}{(1+\omega)\sqrt{-2+6\omega}}}\beta_2w_4.
\end{eqnarray}
In the vacuum limit: $R\rightarrow 0, ~T\rightarrow 0$, the condition $F(0,0)=0$ trivially holds.

Next let us assume the Lagrangian of type $(ii)$, and consequently, the Friedman equation becomes separable for
unknowns $g(\mathcal{R})$ and $h(T)$ which yield following solution:
\begin{eqnarray}\nonumber
F(\mathcal{R},T)&=&\mathcal{R}^{1/4 (3 - \alpha- \sqrt{1 - 2 \alpha} \sqrt{(1 + 10 \alpha + \alpha^2)/(1 - 2 \alpha)}}\beta_1 c_1+
\mathcal{R}^{1/4 (3 - \alpha+ \sqrt{1 - 2 \alpha} \sqrt{(1 + 10 \alpha + \alpha^2)/(1 - 2 \alpha)}}\beta_1 c_2\\\label{51}
&+&\beta_2\bigg[-\frac{2 k^2 T}{\beta_{2} - 3 \beta_{2} \omega}+T^{1/2 - \frac{\sqrt{-1+\omega(3+\omega-3\omega^2)}}{(1+\omega)\sqrt{-2+6\omega}}}c_3+
T^{1/2+\frac{\sqrt{-1+\omega(3+\omega-3\omega^2)}}{(1+\omega)\sqrt{-2+6\omega}}}c_4\bigg].
\end{eqnarray}
It is interesting to mention here that the vacuum condition also holds trivially in this case. Applying $(iii)$ form of generic
function in first field equation and using constraint (\ref{11}), we obtain the following DE:
\begin{eqnarray}\label{55}
A_1T^2 h_{TT} + A_2 h(T) = -A_3 T^D,
\end{eqnarray}
where $A_1=2(\frac{1}{1-3\omega}+\frac{3\alpha}{2(1-2\alpha)})\frac{(1+\omega)^2}{\omega-1}, ~ A_2=\frac{\alpha}{2(2\alpha-1)}, ~
A_3=\frac{k^2\rho_0^d(1-3\omega)^{d-1}}{6\alpha(1-2\alpha)}$ and $D=1-d$, where $d=\frac{2}{3\alpha(1+\omega)}$. This is non-
homogeneous ODE whose solution can be written as
\begin{eqnarray}\nonumber
h(T)&=&A_1^{\frac{1}{2 (2 + D) (-1 + 2/(2 + D))}}A_2^{-\frac{1}{D}-\frac{1}{2 (2 + D) (-1 + 2/(2 + D))}}A_3^{\frac{1}{2 (2 + D) (-1 + 2/(2 + D))}}
(2+D)^{\frac{1}{(2 + D) (-1 + \frac{2}{2 + D})}}\\\nonumber
&\times&(4 A_1 A_3 + 4 A_1 A_3 D + A_1 A_3 D^2)^{\frac{1}{D}}(-1 + \frac{2}{2 + D})^{\frac{2}{D} +\frac {1}{2 + D} (-1 + \frac{2}{2 + D})}
(T^{2 + D})^{-\frac{-\frac{1}{2} + \frac{1}{2 + D}}{(2 + D) (-1 + \frac{2}{2 + D})}-\frac{-1+\frac{2}{2+D}}{D}}\\\nonumber
&\times&BesselJ\bigg[-\frac{1}{D}, \frac{2\sqrt{A_2(T^(2 + D))^{\frac{1}{2}(-1+\frac{2}{2+D})}}}{\sqrt{4 A_1 A_3 + 4 A_1 A_3 D + A_1 A_3 D^2}(-1
+ \frac{2}{2 +D})}\bigg]c_2Gamma[1-\frac{1}{D}]+A_1^{-\frac{1}{2 (2 + D) (-1 + 2/(2 + D))}}\\\nonumber
&\times& A_2^{\frac{1}{2 (2 + D) (-1 + 2/(2 + D))}}A_3^{-\frac{1}{2 (2 + D) (-1 + 2/(2 + D))}}
(2+D)^{-\frac{1}{(2 + D) (-1 + \frac{2}{2 + D})}}(-1 + \frac{2}{2 + D})^{-\frac {1}{2 + D} (-1 + \frac{2}{2 + D})}\\\nonumber
&\times&(T^{2 + D})^{\frac{\frac{1}{2} + \frac{1}{2 + D}}{(2 + D) (-1 + \frac{2}{2 + D})}-\frac{-1+\frac{2}{2+D}}{D}}
BesselJ\bigg[-\frac{1}{D}, \frac{2\sqrt{A_2(T^(2 + D))^{\frac{1}{2}(-1+\frac{2}{2+D})}}}{\sqrt{4 A_1 A_3 + 4 A_1 A_3 D + A_1 A_3 D^2}(-1 + \frac{2}{2 + D})}\bigg]
\\\label{56}&\times&c_1Gamma[1+\frac{1}{D}],
\end{eqnarray}
with $c_1$ and $c_2$ as integrating constants. Applying vacuum limit in $h(T)$ yields zero which shows that vacuum condition is satisfied
in this case. Applying $(v)$ form of generic function, the first Friedman equation yields
\begin{eqnarray}\label{52}
\frac{k^2 T}{1-3\omega}+\frac{3-\omega}{2(1-3\omega)}T F(\mathcal{R})+\frac{\mathcal{R} \alpha}{2(-1+2\alpha)}
+\frac{(-1+4\alpha+3\alpha\omega)}{2(-1+2\alpha)}\mathcal{R}T F_{\mathcal{R}}+\frac{\mathcal{R}^2 T}{(1-2\alpha)}F_{\mathcal{RR}}=0.
\end{eqnarray}
From the equation of continuity, we get $T$ = $(\rho_0(1-3\omega))\bigg[\frac{\mathcal{R}}{6\alpha(1-2\alpha)}\bigg]^{\frac{3\alpha(1+\omega)}{2}}$,
and further by using this relation, we can re-write the first Friedman equation as follows
\begin{eqnarray}\label{53}
k_1 + k_2f(\mathcal{R})-k_3 \mathcal{R}^{k_4}+ k_5 \mathcal{R} F_\mathcal{R} + k_6 \mathcal{R}^{2}F_{\mathcal{RR}}=0,
\end{eqnarray}
whose solution is given by
\begin{eqnarray}\nonumber
F(\mathcal{R})&=&-\frac{k_1}{k_2}+\frac{k_3 \mathcal{R}^{k_4}}{k_2 + k_4 (k_5 + (-1 + k_4) k_6)}
+\mathcal{R}^{\frac{k_5-k_6+\sqrt{k_2}\sqrt{(k_5^2 - 4 k_2 k_6 - 2 k_5 k_6 + k_6^2)/(k_2 )}}{2 k_6}}\\\nonumber
&\times&\bigg[c_1+ \mathcal{R}^{\frac{\sqrt{(k_5^2 - 4 k_2 k_6 - 2 k_5 k_6 + k_6^2)/( k_6)}}{\sqrt{k_6}}}c_2\bigg],
\end{eqnarray}
where $k_1=\frac{k^2}{1+\omega},~ k_2=\frac{3 - \omega}{2 (1 - 3 \omega)},~
k_3=\frac{\sqrt{216}\alpha^{1+3/2\alpha(1+\omega)}}{2\rho_0(1-3\omega)(1-2\alpha)^{k_4}}, ~k_4=1-3/2\alpha(1+\omega), ~
k_5=\frac{3\alpha\omega+4\alpha-1}{2(-1+2\alpha)}, ~k_6=1/(1 - 2 \alpha)$ while $c_1$ and $c_2$ are integrating constants.
Thus we get
\begin{eqnarray}\nonumber
F(\mathcal{R},T)&=&\mathcal{R}+T\bigg[-\frac{k_1}{k_2}+\frac{k_3 \mathcal{R}^{k_4}}{k_2 + k_4 (k_5 + (-1 + k_4) k_6)}
+\mathcal{R}^{\frac{k_5-k_6+\sqrt{k_2}\sqrt{(k_5^2 - 4 k_2 k_6 - 2 k_5 k_6 + k_6^2)/(k_2 )}}{2 k_6}}\\\label{54}
&\times&\bigg(c_1+ \mathcal{R}^{\frac{\sqrt{(k_5^2 - 4 k_2 k_6 - 2 k_5 k_6 + k_6^2)/( k_6)}}{\sqrt{k_6}}}c_2\bigg)\bigg].
\end{eqnarray}
The vacuum limit $R\rightarrow 0$ implies that $g(0)=\frac{2k^2(3\omega-1)}{(1+\omega)(3-\omega)}$. Consequently, it is concluded that
the vacuum condition is only possible when $\omega=1/3$. Next we impose $(vi)$ form of Lagrangian function in the Friedman equation which turns
out as
\begin{eqnarray}\label{57}
\frac{k^2\rho}{\mathcal{R}} +  T const_1 h_{T} + const_{2} (1+h(T)) = 0.
\end{eqnarray}
Here by using equation of continuity, the relation: $\mathcal{R}$ = $6\alpha(1-2\alpha) (\frac{T}{\rho_0(1-3\omega)})^{2/3\alpha(1+\omega)}$
and constraint (\ref{11}), the above equation results into a DE whose solution is given by
\begin{eqnarray}\nonumber
F(\mathcal{R},T)&=& \mathcal{R}+\mathcal{R}\bigg[\frac{1}{-A_2 A_3 + A_2^2 A_3 + A_4}
\bigg[T^{\frac{\sqrt{A_3 - 4 A_4}}{2 \sqrt{A_3}}}\bigg(-A_1 T^{A_2 + \sqrt{A_3 - 4 A_4}/(2 \sqrt{A_3})}+(-A_2 A_3 + A_2^2 A_3 \\\label{58}
&+& A_4) \bigg(-T^{\frac{\sqrt{A_3 - 4 A_4}}{2 \sqrt{A_3}}}+\sqrt{T} C_1+T^{1/2 + \sqrt{A_3 - 4 A_4}/\sqrt{A_3}} C_2\bigg)\bigg)\bigg]\bigg].
\end{eqnarray}
Here $A_1=\frac{K^2\rho_0^{2/3\alpha(1+\omega)}}{(1-3\omega)^{1-2/3\alpha(1+\omega)}6\alpha(1-2\alpha)},~ A_2=1-2/3\alpha(1+\omega),
~A_3=\frac{2(1+\omega)}{(\omega-1)}const_1$ and $A_4=const_2=\frac{-\alpha}{2(1-2\alpha)}$. Also,
$const_1=\frac{1+\omega}{1-3\omega}+\frac{3\alpha(1+\omega)}{2(1-2\alpha)}$. Here, the vacuum condition is trivially satisfied in
the limit of $R\rightarrow 0, ~T\rightarrow 0$.

Now using $(vii)$ form of function in the Friedman equation, we obtain the following DE:
\begin{eqnarray}\nonumber
&&k^2 \rho + (1+\omega) \rho F_2(\mathcal{R})F_{3T} + \frac{F_1(\mathcal{R})+F_{2}(\mathcal{R})F_3(T)}{2} + \frac{1-\alpha}{2(2\alpha -1)} \mathcal{R}(F_{1\mathcal{R}}+F_{2\mathcal{R}}F_3(T))\\\label{59}&&+
\frac{1}{1-2\alpha}\mathcal{R}^2(F_{1\mathcal{RR}}+F_{2\mathcal{RR}}F_3(T))+ \frac{3\alpha(1+\omega)}{2(1-2\alpha)}\mathcal{R} T F_{2\mathcal{R}}F_{3T}=0.
\end{eqnarray}
It can be transformed into set of two DEs whose solution is not possible to find. Applying $(viii)$ form of Lagrangian function
in the Friedman's equation, we obtain
\begin{eqnarray}\nonumber
\mu \bigg(\frac{R}{R_0}\bigg)^{\beta}\bigg(\frac{T}{T_0}\bigg)^{\gamma}\bigg[(\frac{1+\omega}{1-3\omega})\gamma+\frac{1}{2}
+\frac{\beta(1-\alpha)}{2(2\alpha-1)}+\frac{\beta(\beta-1)}{1-2\alpha}+\frac{3\alpha(1+\omega)\beta\gamma}{2(1-2\alpha)}\bigg]
=-T_0\sum_i\Omega_{\omega_i,0}a^{-3(1+\omega_i)},
\end{eqnarray}
where the value of $\mu$ can be calculated by assuming present time and is given by
\begin{eqnarray}\nonumber
\mu &=& \frac{-T_0\sum_i\Omega_{\omega_i,0}}{(\frac{1+\omega}{1-3\omega})\gamma+\frac{1}{2}+\frac{\beta(1-\alpha)}{2(2\alpha-1)}
+\frac{\beta(\beta-1)}{1-2\alpha}+\frac{3\alpha(1+\omega)\beta\gamma}{2(1-2\alpha)}}.
\end{eqnarray}
It is worthy to mention here that the expression in denominator must not be zero. By inserting value of $\mu$ in the first
Friedman's equation, we obtain:
\begin{eqnarray}\nonumber
\bigg(\frac{R}{R_0}\bigg)^{\beta}\bigg(\frac{T}{T_0}\bigg)^{\gamma}\sum_i\Omega_{\omega_i,0}=\sum_i\Omega_{\omega_i,0}
\bigg(\frac{R}{R_0}\bigg)^{\frac{3}{2}\alpha(1+\omega_i)}.
\end{eqnarray}
The above expression is satisfied when the time dependent terms and their powers will be canceled. It further leads to the conditions, i.e.,
$\gamma=0$ and $3\alpha(1+\omega_i)=2\beta, \forall i$, where the second condition will be satisfied when a fluid is present.
Also, since $\alpha>o$, the EoS parameter must be greater than $-1$, i.e., $\omega>-1$. Hence this Lagrangian can describe the
super-bounce cosmology.

\subsection{Matter Bounce Model}

The matter bounce model is defined by the following scale factor \cite{30*,33*,33,9,29e}
\begin{eqnarray}\nonumber
a(t)=A \bigg(\frac{3}{2}\rho_{cr}t^2+1\bigg)^{\frac{1}{3}},
\end{eqnarray}
where $\rho_{cr}$ stands for critical density and A is an arbitrary positive constant. Also, the corresponding Hubble parameter
and Ricci scalar are given by
\begin{eqnarray}
H&=&\frac{2t \rho_{cr}}{2+3\rho_{cr}t^2},~~~~~~~~~~~~R= \frac{-12\rho_{cr}(2+\rho_{cr}t^2)}{(2+3\rho_{cr}t^2)^2}.
\end{eqnarray}
Here, bouncing point is $t=0$ $\Rightarrow$ $H=0$. Since Eq.(\ref{8}) is partial differential equation whose solution is
difficult to find, therefore, one can consider the forms of generic function as considered in the last sections.
By assuming $(i)$ form of Lagrangian function, we obtain two DEs which are given by
\begin{eqnarray}\nonumber
k^2T+(1+\omega)\beta_2T h_T+\frac{\beta_2(1-3\omega)g(T)}{2}=0,\\\nonumber
\frac{g(R)}{2}+3(\dot{H}+H^2)g_R-3H(\dot{R}g_{RR})=0.
\end{eqnarray}
Here the first DE is similar to Eq.(\ref{16}) whose solution is given in Eq.(\ref{18}). By inserting values of
$H, ~\dot{H}$ and $R$ in the second DE, we obtain
\begin{eqnarray}\label{m1}
g(t)-\frac{(2+3t^2\rho_{cr})(-20+68t^2\rho_{cr}+15t^4(\rho_{cr})^2)}{t\rho_{cr}(10+3t^2\rho_{cr})^2}\dot{g}(t)-
\frac{(2+3t^2\rho_{cr})^2}{2\rho_{cr}(10+3t^2\rho_{cr})}\ddot{g}(t)=0,
\end{eqnarray}
which can be solved by Frobenius method as follows
\begin{eqnarray}\nonumber
g(t)=\sum_{n=0}^{\infty}a_nt^{n+r}.
\end{eqnarray}
For $r=3$ and $r=0$, the following recurrence relation can be obtained
\begin{eqnarray}\nonumber
40(n+r+1)(n+r-2)&=&-4\rho_{cr}a_{n-1}(33(n+r-1)^2-50)-6\rho_{cr}^2a_{n-3}((n+r-3)(21(n+r-3)+\\\nonumber
&&15\rho_{cr}+57)+20)-9a_{n-5}\rho_{cr}^3(3(n+r-5)(n+r-6)+2).
\end{eqnarray}
with $a_{-1}=a_{-3}=a_{-5}=0$ and consequently, we get the following solution:
\begin{eqnarray}\nonumber
g(t)&=&c_1\bigg[t^3-\frac{131}{50}\rho_{cr}t^5+\frac{\rho_{cr}^2(2188+135\rho_{cr})}{560}t^7+...+\infty\bigg]
+c_2\bigg[160 \ln(t)-80\rho_{cr}(2+5\ln(t))t^2\\\label{mb1}
&+&((339+350\ln(t))\rho_{cr}^2+90\rho_{cr}^3)t^4+...+\infty\bigg],
\end{eqnarray}
where $t=\pm \sqrt{\frac{2}{3}}\sqrt{-\frac{1}{R}-\frac{1}{\rho_{cr}}\pm \frac{\sqrt{-(4R-\rho_{cr})\rho_{cr}^3}}{R\rho_{cr}^2}}$,
which yields the $g(R)$. Now to check vacuum condition, we put vacuum limit $T\rightarrow 0, R\rightarrow 0$ which
implies that $h(0)=0$ while $g(0)$ diverges. Therefore, vacuum condition is not satisfied for this solution.

Similarly, by applying $(ii)$ form of Lagrangian function in the first Friedman's equation, we obtain same DEs with an
extra term, i.e., $g(R)=-\frac{6H^2}{\beta_2}$, whose complementary solution is identical to the previous one but
general solutions is not possible. Therefore, nothing can be concluded about $g_{part}(0)=0$ and thus vacuum condition
does not hold and also $g_{hom}(0)$ diverges.

Further we insert $(iii)$ form of generic function in the first Friedman's equation and by using values of $a(t),~H$ and $R$,
we obtain
\begin{eqnarray}\nonumber
\frac{24\rho_{cr}^2t^2}{(2+3t^2\rho_{cr})^2}g(t)+\frac{2(2+\rho_{cr}t^2(-2+9\omega))}{(3\omega-1)(2+3t^2\rho_{cr})}\dot{g}(t)=k^2 \rho_0\bigg(A\bigg(1+\frac{3t^2\rho_{cr}}{2}\bigg)^{1/3}\bigg)^{-3(1+\omega)},
\end{eqnarray}
whose solution is given by
\begin{eqnarray}\nonumber
g(t)&=&C_1\exp(\frac{2(1-3\omega)((9\omega-2)\ln(2+3t^2\rho_{cr})-3\ln(2+(9\omega-2)t^2\rho_{cr}))}{(9\omega-5)(9\omega-2)})
+\frac{1}{A^3(7 - 15\omega + \omega (-5 + 9 \omega))\rho_{cr}}\\\nonumber && 2^{\omega-\frac{2 (7 - 54 \omega + 81 \omega^2)}{(-5 + 9 \omega) (-2 + 9 \omega)}}\exp(\frac{2(1-3\omega)((9\omega-2)\ln(2+3t^2\rho_{cr})-3\ln(2+(9\omega-2)t^2\rho_{cr}))}{(9\omega-5)(9\omega-2)})k^2(3\omega-1)\\\nonumber
&&Hypergeometric2F1\bigg[\frac{7 - 20 \omega + 9 w^2}{5-9\omega},1+\frac{6 (-1 + 3 \omega)}{(-5 + 9 \omega) (-2 + 9 \omega)},
\frac{12 - 29 \omega + 9 \omega^2}{5 - 9 \omega}, \frac{(-2 + 9 \omega) (2 + 3 t^2 \rho_{cr})}{2(9\omega-5)}\bigg]\rho_0\\\nonumber
&& (A^{3}(2+3t^2\rho_{cr}))^{-\omega}(2+3t^2\rho_{cr})^{\frac{7 - 15 \omega}{5-9\omega}}
\bigg(\frac{-6 + 3 t^2 (2 - 9 \omega)\rho_{cr}}{9\omega-5}\bigg)^{\frac{6 (-1 + 3 \omega)}{10 - 63\omega + 81 \omega^2}}
\bigg(2 +  t^2 (-2 + 9 \omega)\rho_{cr}\bigg)^{\frac{6 - 18 \omega)}{10 - 63\omega + 81 \omega^2}},
\end{eqnarray}
where $2+3t^2\rho_{cr}=\frac{2}{A^3}\bigg(\frac{T}{\rho_0(1-3\omega)}\bigg)^{\frac{-1}{(1+\omega)}}$, which yields the following solution
for $g(T)$
\begin{eqnarray}\nonumber
g(T)&=&\exp \bigg(-\frac{2(3\omega-1)(\ln(27)+(9\omega-2)\ln(\frac{2(\frac{T}{\rho_0(1-3\omega)})^{\frac{-1}{\omega+1}}}{A^3})-
3\ln(6+(9 \omega-2)(\frac{2(\frac{T}{\rho _0-3 \rho _0 \omega}){}^{-\frac{1}{\omega+1}}}{A^3}-2)}{(-5 + 9 \omega) (-2 + 9 \omega)}\bigg)\\\nonumber
&\times&\bigg[C_1+\frac{1}{A^3 (\omega (9 \omega-20)+7) \rho _{\text{cr}}}8^{\frac{\omega}{2-9 \omega}}k^2 (3 \omega-1)
\text{Hypergeometric2F1}[-\omega+\frac{4}{3 (9 \omega-5)}+\frac{5}{3},\frac{6 (3 \omega-1)}{81 \omega^2-63 \omega+10}\\\nonumber
&&+1,\frac{4}{3}(\frac{1}{9 \omega-5}+2)-\omega,\frac{(9 \omega-2)(\frac{T}{\rho _0-3 \rho _0 \omega}){}^{\frac{-1}{\omega+1}}}{A^3 (9 \omega-5)}]
\rho _0 (A \sqrt[3]{\frac{(\frac{T}{\rho _0-3 \rho _0 \omega}){}^{\frac{-1}{\omega+1}}}{A^3}}){}^{-3 \omega} (\frac{(\frac{T}{\rho _0-3 \rho _0 \omega}){}^{\frac{-1}{\omega+1}}}{A^3}){}^{\frac{7-15 \omega}{5-9 \omega}}\\\label{mb2}
&\times& (6-\frac{6 (9 \omega-2) (\frac{T}{\rho _0-3 \rho _0 \omega}){}^{\frac{-1}{\omega+1}}}{A^3 (9 \omega-5)})
{}^{\frac{6 (3 \omega-1)}{81 \omega^2-63 \omega+10}}((9\omega-2)(\frac{2 (\frac{T}{\rho _0-3 \rho _0 \omega}){}^{\frac{-1}{\omega+1}}}{A^3}-2)+6)
{}^{\frac{6-18 \omega}{81 \omega^2-63 \omega+10}}\bigg].
\end{eqnarray}
The vacuum limit $T\rightarrow 0$ implies that $g(o)$ diverges, so the vacuum condition does not hold in this case.

By applying $(vi)$ of Lagrangian function in the first Friedman's equation, we obtain
\begin{eqnarray}\nonumber
k \rho _0 \left(A \sqrt[3]{\frac{3 t^2 \rho _{\text{cr}}}{2}+1}\right){}^{-3 (\omega+1)}
-\dot{g}(t) \left(\frac{6 t \rho _{\text{cr}}}{3 t^2 \rho _{\text{cr}}+2}-\frac{2 \left(t^2 \rho _{\text{cr}}+2\right)}{t (1-3\omega)
\left(3 t^2 \rho_{\text{cr}}+2\right)}\right)-\frac{12 t^2 \rho _{\text{cr}}^2 g(t)}{\left(3 t^2 \rho _{\text{cr}}+2\right){}^2}
-\frac{12 t^2 \rho _{\text{cr}}^2}{\left(3 t^2 \rho _{\text{cr}}+2\right){}^2}=0,
\end{eqnarray}
whose analytical solution is given by
\begin{eqnarray}\nonumber
g(t)&=&u\bigg[c_1+\frac{(3\omega-1)(A\sqrt[3]{3t^2\rho_{\text{cr}}+2}){}^{-3\omega}}{4A^3(5-9\omega)^2}\bigg[
\frac{1}{(9\omega^2-17\omega+6)\rho_{\text{cr}}}2^{\frac{81\omega^3-63\omega^2+\omega+3}{81\omega^2-63\omega+10}}
3^{\frac{162\omega^2-117\omega+17}{81\omega^2-63\omega+10}}k\omega(9\omega-5)u_1\rho_0\\\nonumber
&\times&(3t^2\rho_{\text{cr}}+2){}^{\frac{6-12\omega}{5-9\omega}}(\frac{t^2(2-9\omega)
\rho_{\text{cr}}-2}{9\omega-5}){}^{\frac{9\omega-3}{81\omega^2-63\omega+10}}
(t^2(9\omega-2)\rho_{\text{cr}}+2){}^{\frac{3-9\omega}{81\omega^2-63\omega+10}}
-\frac{A^3(\frac{3}{2})^{\frac{9\omega-3}{81\omega^2-63\omega+10}}}{2\omega-1}\\\nonumber
&\times&(9\omega-5)u_2(A\sqrt[3]{3t^2\rho_{\text{cr}}+2}){}^{3\omega}(3t^2\rho_{\text{cr}}+2){}^{\frac{6-12\omega}{5-9\omega}}
(\frac{t^2(2-9\omega)\rho_{\text{cr}}-2}{9\omega-5}){}^{\frac{9\omega-3}{81\omega^2-63\omega+10}}(t^2(9\omega-2)\rho _{\text{cr}}+2)
{}^{\frac{3-9\omega}{81\omega^2-63\omega+10}}\\\nonumber
&-&\frac{1}{(9\omega^2-17\omega+6)\rho_{\text{cr}}}5 k 2^{\frac{81\omega^3-63\omega^2+\omega+3}{81\omega^2-63\omega+10}}
(9\omega-5)u_1\rho_0(3t^2\rho _{\text{cr}}+2){}^{\frac{6-12\omega}{5-9\omega}}(\frac{3t^2(2-9\omega)\rho_{\text{cr}}-6}{9\omega-5})
{}^{\frac{9\omega-3}{81\omega^2-63\omega+10}}\\\nonumber
&\times&(t^2(9\omega-2)\rho_{\text{cr}}+2){}^{\frac{3-9\omega}{81\omega^2-63\omega+10}}-\frac{1}{81\omega^2-72\omega+13}
3A^32^{\frac{6-12\omega}{5-9\omega}}(9\omega-5)(9\omega-2)u_3(3t^2\rho_{\text{cr}}+2){}^{\frac{1-3\omega}{5-9\omega}}\\\label{mb3}
&\times&(\frac{(9\omega-2)(3t^2\rho_{\text{cr}}+2)}{9\omega-5}){}^{\frac{1-3\omega}{9\omega-5}}(t^2(9\omega-2)\rho_{\text{cr}}+2)
{}^{\frac{3-9\omega}{81\omega^2-63\omega+10}+1}(A\sqrt[3]{3t^2\rho_{\text{cr}}+2}){}^{3\omega}\bigg]\bigg],
\end{eqnarray}
where
\begin{eqnarray}\nonumber
u&=&\exp\left(\frac{(1-3\omega)((9\omega-2)\log(3t^2\rho_{\text{cr}}+2)-3\log(t^2(9\omega-2)\rho_{\text{cr}}+2))}{(9\omega-5)(9\omega-2)}\right),\\\nonumber
u_1&=&\,_2F_1\left(\frac{9\omega^2-17\omega+6}{5-9\omega}; \frac{9\omega-3}{(9\omega-5)(9\omega-2)}+1; \frac{9\omega^2-26\omega+11}{5-9\omega};
\frac{(9\omega-2)(3\rho_{\text{cr}} t^2+2)}{2 (9 \omega -5)}\right),\\\nonumber
u_2&=&\,_2F_1\left(\frac{6-12\omega}{5-9\omega},\frac{9\omega-3}{(9\omega-5)(9\omega-2)}+1; \frac{11-21\omega}{5-9\omega};
\frac{(9\omega-2)\left(3\rho_{\text{cr}}t^2+2\right)}{2(9\omega-5)}\right),\\\nonumber
u_3&=&\,_2F_1\left(\frac{4-6\omega}{5-9\omega}; \frac{81\omega^2-72\omega+13}{81\omega^2-63\omega+10}; \frac{162\omega^2-135\omega+23}{81\omega^2-63\omega+10};
\frac{3t^2(2-9\omega)\rho_{\text{cr}}-6}{2(9\omega-5)}\right).
\end{eqnarray}
By inserting $2+3t^2\rho_{cr}$ in terms of $T$ yields $g(T)$. In the vacuum limit $T\rightarrow 0$, $g(0)\rightarrow \infty$. Thus, vacuum condition is not satisfied. Similarly, by inserting $(iv), ~(v), ~(vii)$ and
$(viii)$ forms of Lagrangian function in the first Friedman's equation, we obtain very complicated DEs whose
analytical solutions are not possible to find therefore, nothing can be inferred about vacuum condition in
these cases.

\section{Energy Conditions}

The energy conditions are some well-known constraints which are based on energy-momentum tensor and
possess some physical features used to examine the physical consistency of cosmic models. These can be
categorized into four types namely: the null, weak, strong and dominant energy conditions. In $F(R,T)$ gravity,
the energy conditions for effective fluid can be defined as
\begin{itemize}
\item NEC: $\rho^{eff}+p^{eff}\geq 0$,
\item WEC: $\rho^{eff}\geq 0$, $\rho^{eff}+p^{eff}\geq 0$,
\item SEC: $\rho^{eff}\geq 0$, $\rho^{eff}+3p^{eff}\geq 0$,
\item DEC: $\rho^{eff}\geq 0$, $\rho^{eff}-p^{eff}\geq 0$.
\end{itemize}
\begin{figure}
\centering
\includegraphics[width=0.235\textwidth]{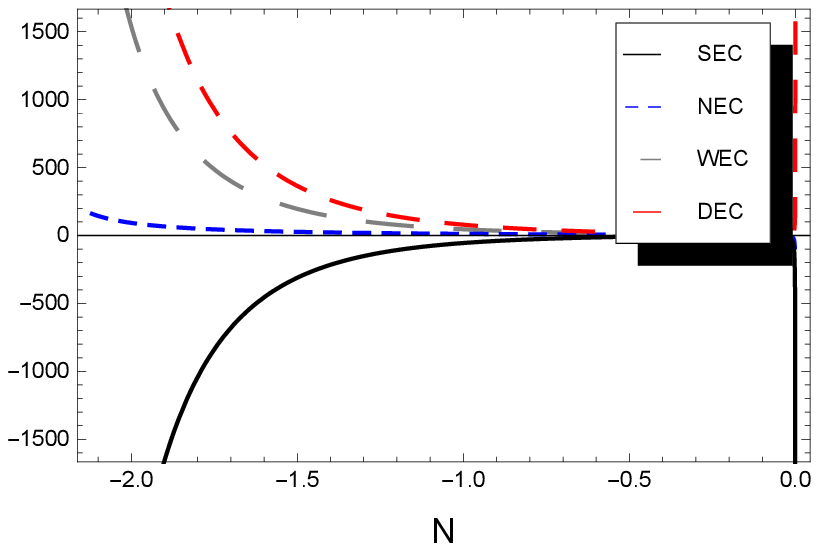}
\includegraphics[width=0.235\textwidth]{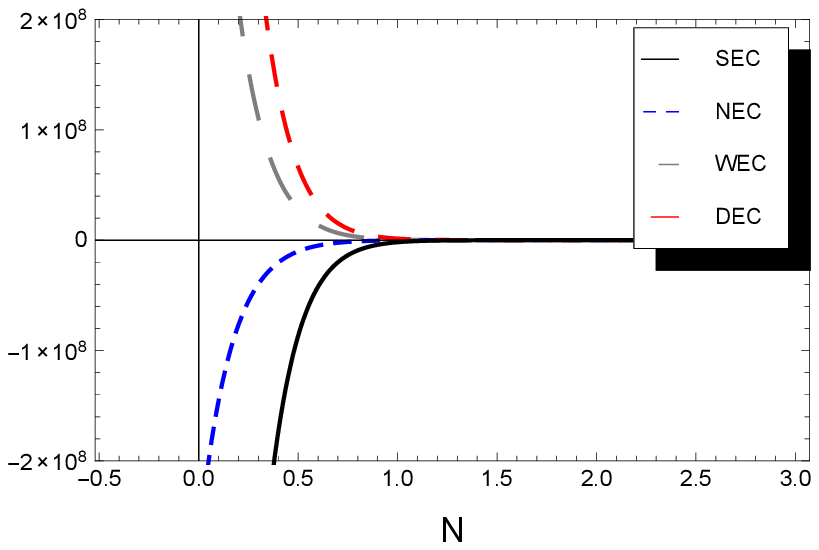}
\includegraphics[width=0.235\textwidth]{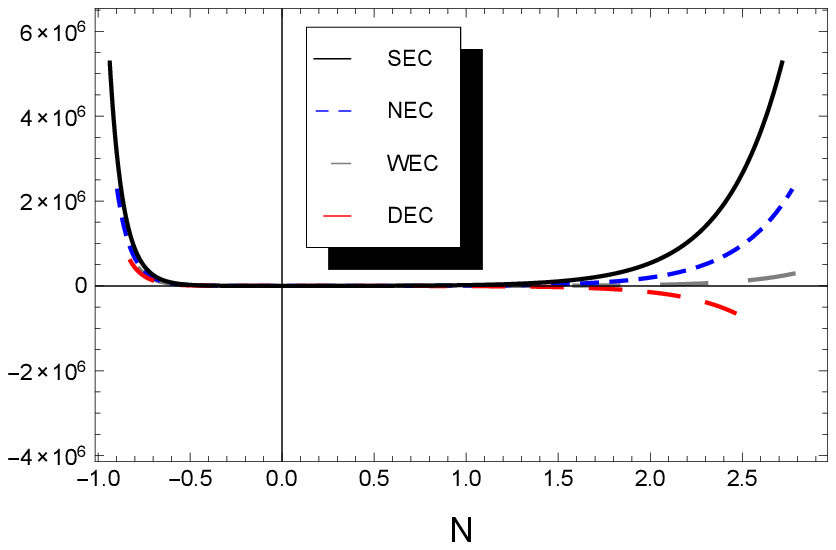}
\includegraphics[width=0.235\textwidth]{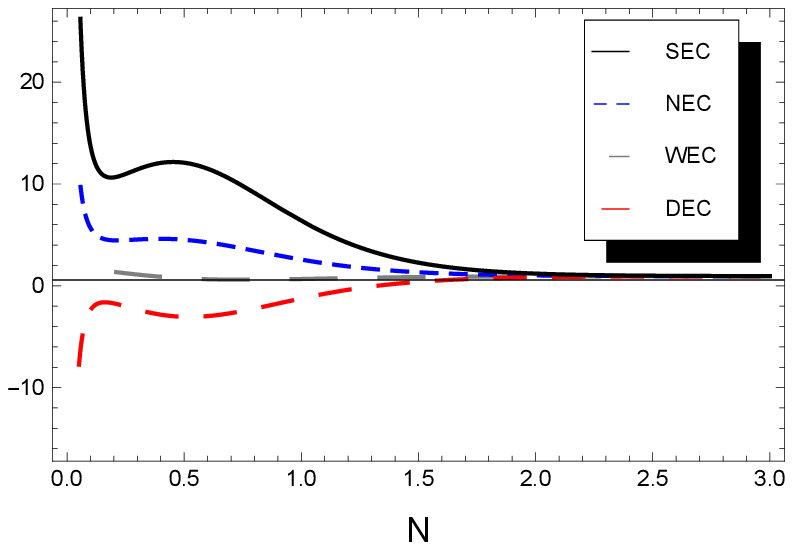}
\caption{\scriptsize{Evolution of energy constraints for three bouncing models with $(i)$ form of function $F(\mathcal{R},T)$.
The first (on left) plot refers to exponential evaluation model, second plot corresponds to power law model with $\alpha=2$,
third plot represents power law model with $\alpha=\frac{2}{3}$ and last plot (on right) refers to oscillatory model.
Here we have fixed free parameters as $A=.9,~ B=.5,~ \sigma=1,~c_1=c_2=c_3=c_4=w_1=w_2=w_3=w_4=.2$ and $\beta_1=\beta_2=2$.}}\label{figECS}
\end{figure}
\begin{figure}
\centering
\includegraphics[width=0.23\textwidth]{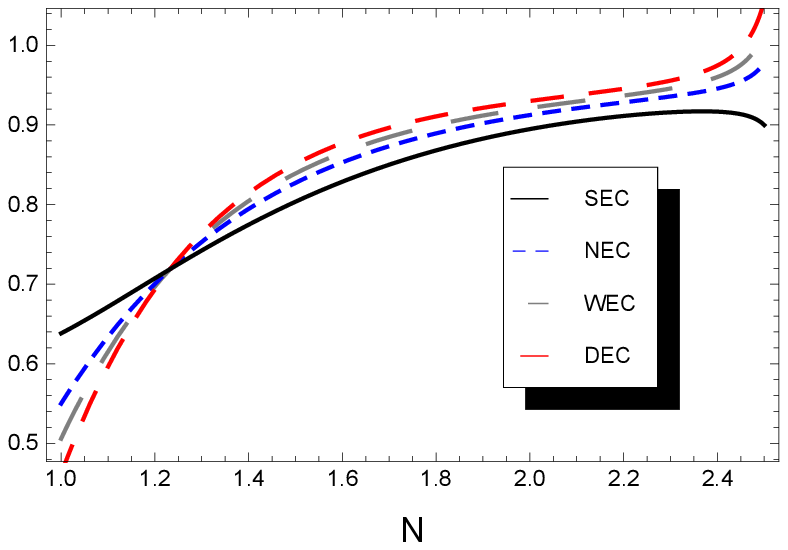}
\includegraphics[width=0.23\textwidth]{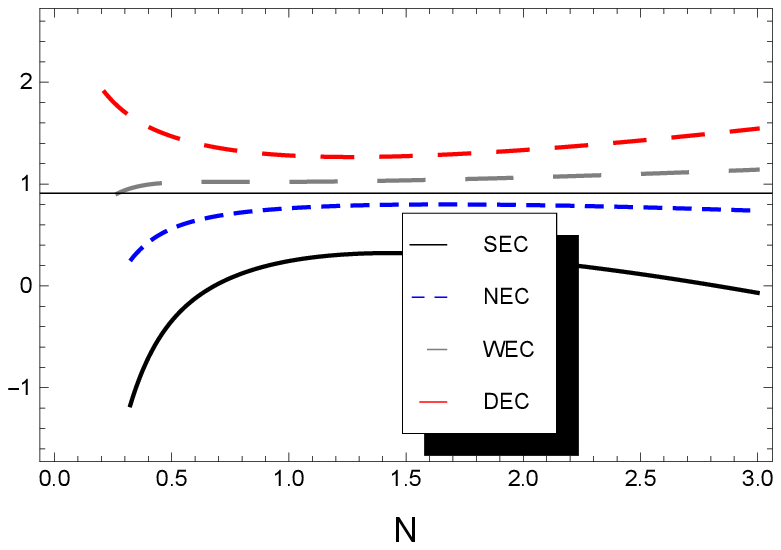}
\includegraphics[width=0.23\textwidth]{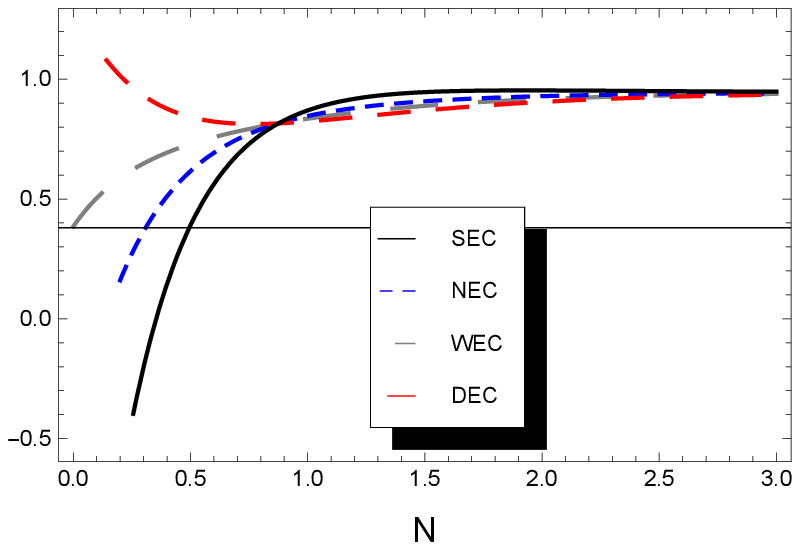}
\includegraphics[width=0.23\textwidth]{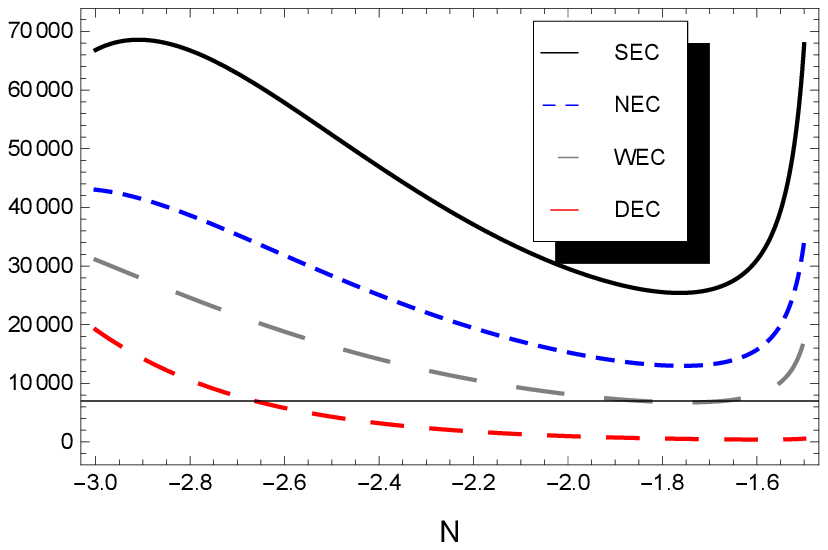}
\caption{\scriptsize{Evolution of energy constraints for power law bouncing model with $(v)$ and $(vi)$ forms of
generic function and matter bounce model. The first plot (on left) indicates the behavior of energy conditions for $v$ form of
Lagrangian function with $\alpha=2$, second plot refers to $v$ form of generic function with $\alpha=2/3$,
third plot corresponds to $vi$ form of function $F(R,T)$ and last plot (on right) represents matter bounce solution with
$(iii)$ form of Lagrangian function. Here the chosen values of free parameters are $c_1=c_2=c_3=c_4=.2$.}}\label{figECS3}
\end{figure}
Here, we will discuss the graphical behavior of these energy conditions for all reconstructed bouncing models of previous section.
The plots of energy constraints for all bouncing models with $(i)$ form of Lagrangian are shown in Figure \ref{figECS}.
From the graphs, it is clear that NEC and SEC violate for exponential and power law models when $\alpha=2$ at the neighborhood
of bouncing point while WEC and DEC remain valid there. In case of oscillatory and power law models, NEC, WEC and SEC are
valid for $\alpha=2/3$ while DEC violates. Similarly, we analyze the energy conditions for these bouncing models for $(ii)$
form of generic function and similar behavior can be obtained. In case of exponential and oscillatory models with other
ansatz forms of function $F(\mathcal{R},T)$, we cannot analyze their energy constraints as analytical solution was not
possible in these cases. Since we have obtained analytical solutions for other forms of function in case of power law,
therefore, we have analyzed all energy constraints for this model with $(v)$ and $(vi)$ forms which are given in by
the graphs of Figure \ref{figECS3}. It is observed that for accelerating universe ($\alpha=2$) all energy conditions are valid.
In case of dust dominated era ($\alpha=2/3$) for $(v)$ form of function, it is seen that NEC and SEC are violated while DEC and
WEC are valid. For the reconstructed function using $(vi)$ form, graphs indicate that WEC, NEC and DEC are valid while SEC may violate
near bouncing point. It is worthy to mention here that for $(vi)$ and $(iii)$ forms of reconstructed function $F(\mathcal{R},T)$,
imaginary solutions are obtained and hence graphical analysis of energy constraints is not possible. For matter bounce model,
we have obtained analytical solutions for $(i),~ (iii)$ and $(vi)$ ansatz forms but it is found that for $(i)$ and $(vi)$ solutions,
energy constraints involve imaginary terms and hence cannot be discussed graphically. While for $(iii)$ Lagrangian,
energy constraints are shown in Figure \ref{figECS3} where it is seen that all energy constraints are valid in this case.

\section{Perturbations and Stability}

In this section, we shall examine the stability of reconstructed bouncing solutions by applying linear and homogeneous
perturbations in $F(\mathcal{R},T)$ gravitational framework. In this respect, we find the corresponding perturbed field
equations and continuity equation using FRW universe model for exponential, oscillatory, power law and matter bouncing solutions.
For this purpose, let us suppose a solution
\begin{eqnarray}\label{60}
H(t)=H_{\star}(t)
\end{eqnarray}
compatible with the background field equations which further leads to form of Ricci scalar given by
$\mathcal{R}_{\star}=-6H_{\star}(H_{\star}^{'}+2H_{\star})$, where prime indicates derivative with respect to
e-folding parameter. For a particular $F(\mathcal{R},T)$ model which is consistent with solution (\ref{60}), then
following equation of motion and $\bigtriangledown^{i}T_{ij}\neq0$ must hold:
\begin{eqnarray}\label{61}
&&k^2\rho_{\star}+(1+\omega)\rho_{\star}F_{T}^{\star}+\frac{F^{\star}}{2}+3(H_{\star}H_{\star}^{'}+H_{\star}^{2})F_{\mathcal{R}}^{\star}
+3H_{\star}[(6H_{\star}^2H_{\star}^{''}+H_{\star}H_{\star}^{'^2}+4H_{\star}^2H_{\star}^{'})F_{\mathcal{RR}}^{\star}-H_{\star}T_{\star}^{'}
F_{\mathcal{R}T}^{\star}]=0,\\\label{62}
&&\rho_{\star}^{'}+3(1+\omega)\rho_{\star}=\frac{-1}{k^2+F_{T}^{\star}}\bigg[(1+\omega)\rho_{\star}T_{\star}^{'}F_{TT}^{\star}
+\omega\rho^{'}F_{T}^{\star}+\frac{1}{2}T_{\star}^{'}F_{T}^{\star}\bigg],
\end{eqnarray}
where $F^{\star}$ indicates the function corresponding to solution (\ref{60}). In case the usual conservation law holds
(generally it does not remain valid $F(\mathcal{R},T)$ gravity), we find $\rho$ in terms of $H_{\star}(t)$ as follows
\begin{eqnarray}\nonumber
\rho_{\star}(t)=\rho_{0}e^{(-3(1+\omega)\int H_{\star}(t)dt)}=\rho_{0}e^{(-3(1+\omega)N)}.
\end{eqnarray}
Further, the perturbed $H(t)$ and $\rho(t)$ can be defined as
\begin{eqnarray}\label{63}
H(t)=H_{\star}(t)(1+\delta(t)), ~~~~~~~~~~~ \rho(t)=\rho_{\star}(t)(1+\delta_m(t)),
\end{eqnarray}
where $\delta_m(t)$ and $\delta(t)$ stand for perturbation functions. In order to analyze introduced perturbations about
the solution (\ref{60}), let us expand $F(\mathcal{R},T)$ function as a series in terms of Ricci scalar and
energy-momentum tensor trace given by
\begin{eqnarray}\label{64}
F(\mathcal{R},T)=F^{\star}+F_{\mathcal{R}}^{\star}(\mathcal{R}-\mathcal{R}_{\star})+F_{T}^{\star}(T-T_{\star})+o^2.
\end{eqnarray}
Here only the linear terms are to be considered for further calculations while $o^2$, denoting the quadratic or higher power terms
of $\mathcal{R}$ and $T$, will be ignored. Using Eqs.(\ref{63}) and (\ref{64}) in FRW equation, we obtain the following perturbed
field equation:
\begin{eqnarray}\label{65}
b_2\delta^{''}+b_1\delta^{'}+b_0\delta=c_{m1}\delta_m+c_{m2}\delta_m^{'},
\end{eqnarray}
where $b_i$'s and $c_{mi}$'s are listed in Appendix \textbf{A}. Applying above perturbations in Eq.(\ref{9}), we obtain the
perturbed form of continuity equation as
\begin{eqnarray}\label{66}
d_1\delta_m^{'}+d_2\delta_m+d_3\delta^{'}+d_4\delta=0,
\end{eqnarray}
where $d_i$'s are also defined in Appendix \textbf{A}. If the usual conservation law holds, then Eq.(\ref{65}) reduces to
\begin{eqnarray}\label{67}
\delta_m^{'}+3\delta=0.
\end{eqnarray}
These perturbed Eqs.(\ref{65}) and (\ref{66}) will play the role of key equations in analyzing the stability of FRW model.
For $F(\mathcal{R},T)=F_1(\mathcal{R})+F_2(T)$ type models, these perturbed equations reduce to
\begin{eqnarray}\label{68}
\hat{b_2}\delta^{''}+\hat{b_1}\delta^{'}+\hat{b_0}\delta=\hat{c_{m1}}\delta_m,~~~~~~~~~~
\hat{d_1}\delta_m^{'}+\hat{d_2}\delta_m+\hat{d_3}\delta=0,
\end{eqnarray}
where $b_i's$ and $d_i's$, ${i=0,1,2,3}$ are given in Appendix \textbf{A}. In following sections, we present the stability
of exponential, oscillatory, matter bounce and power law models for this form of generic function.

\subsection{Stability of Exponential Solutions}

\begin{figure}
\centering
\includegraphics[width=0.40\textwidth]{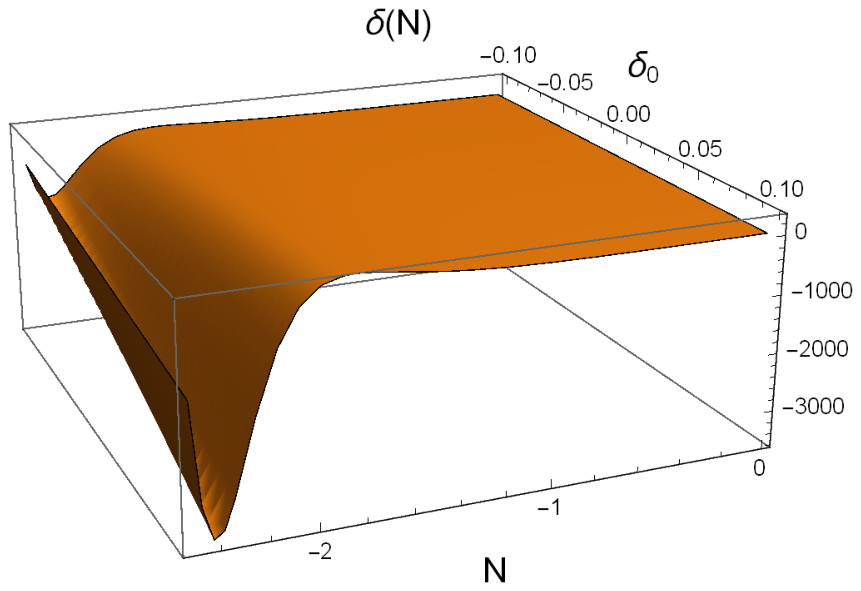}\textbf{(a)} \includegraphics[width=0.40\textwidth]{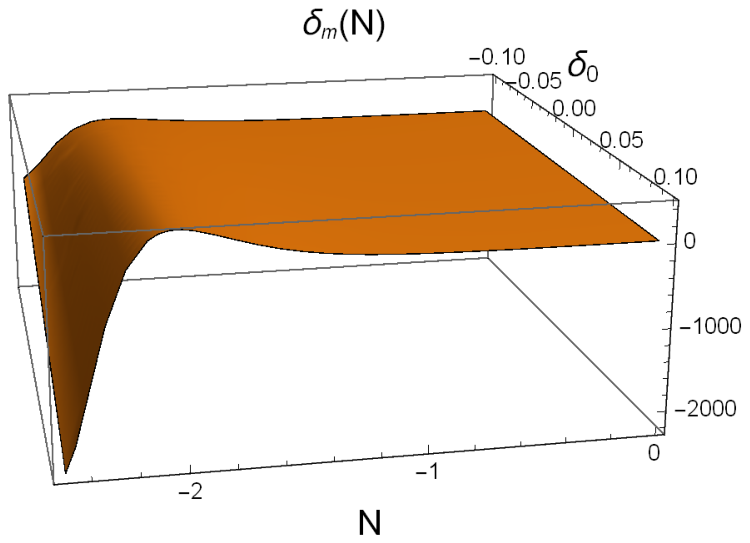}
\includegraphics[width=0.40\textwidth]{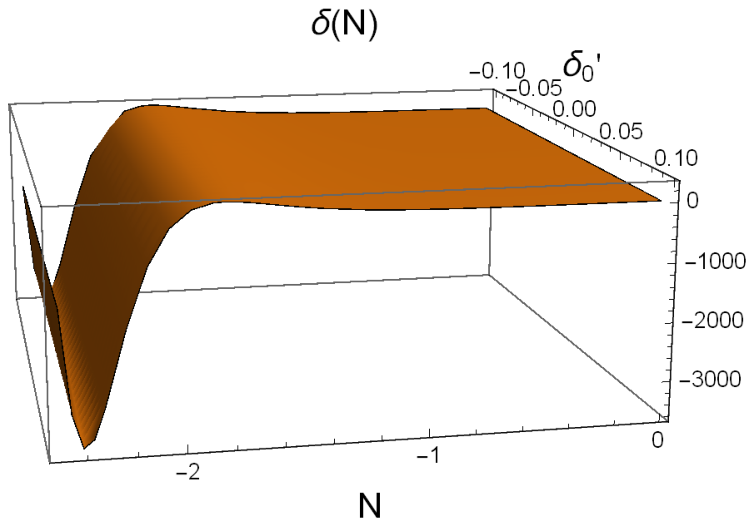} \textbf{(b)} \includegraphics[width=0.40\textwidth]{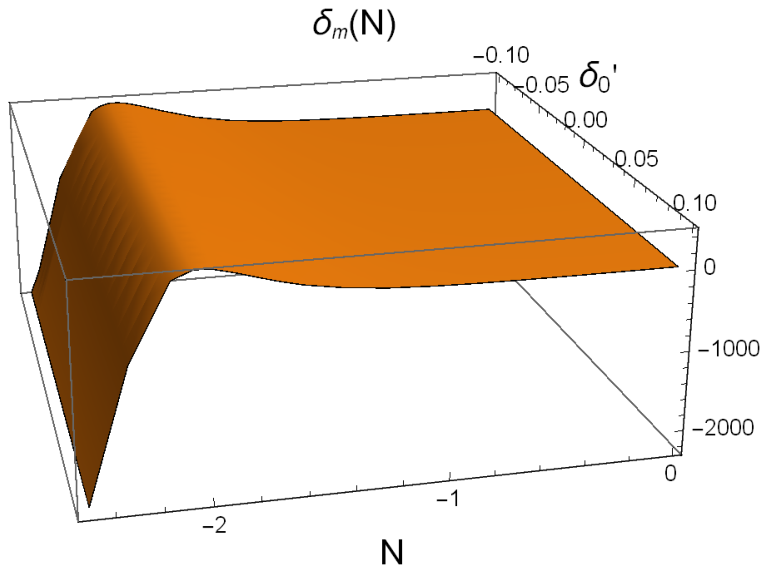}
\includegraphics[width=0.45\textwidth]{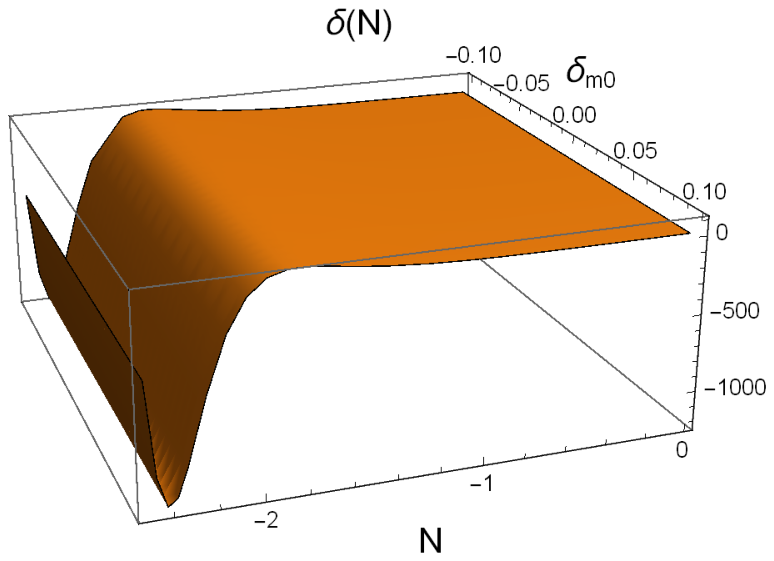} \textbf{(c)} \includegraphics[width=0.40\textwidth]{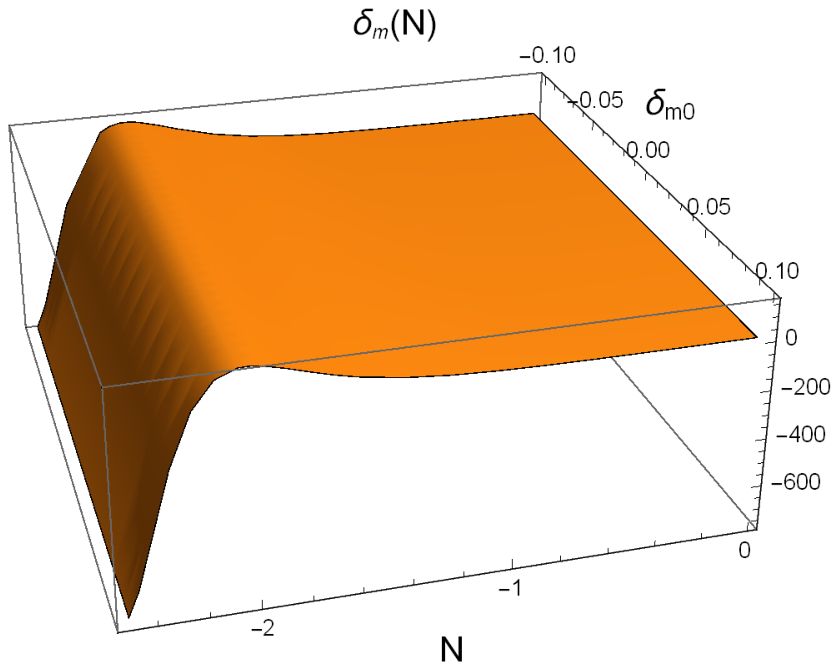}
\caption{\label{fig0} Stability of $F(\mathcal{R},T)=\beta_1g(\mathcal{R})+\beta_2h(T)$ model for exponential case with $c_i=.2$.
Distribution of $\delta(N)$ and $\delta_m(N)$ verses N with $\omega=0$ is presented. We set \textbf{(a)} $\delta_{m0}=\delta_0^{'}=.1$
and vary $\delta_0$, \textbf{(b)} $\delta_0=.1=\delta_{m0}$ and vary $\delta_0^{'}$, \textbf{(c)} $\delta_{m0}$ is varied
with same initial values for $\delta_0$ and $\delta_0^{'}$. These curves show smooth behavior having small value near $z=0$.}
\end{figure}
\begin{figure}
\centering \epsfig{file=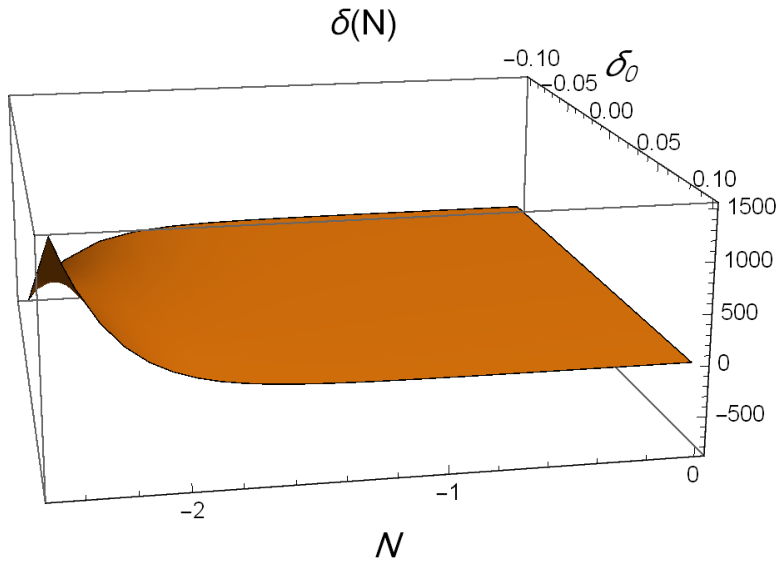, width=.45\linewidth,
height=1.8in} \textbf{(a)} \epsfig{file=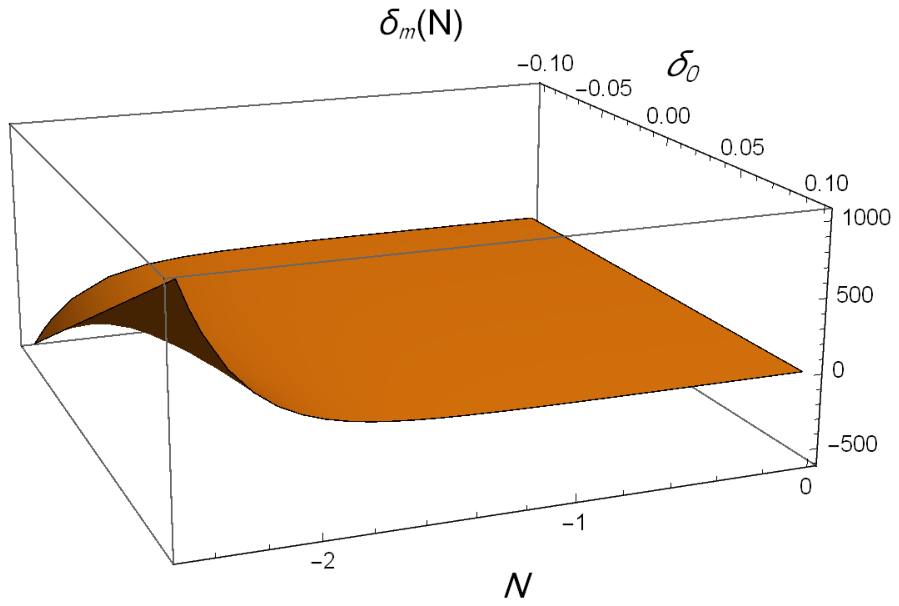, width=.45\linewidth,
height=1.8in}
\centering \epsfig{file=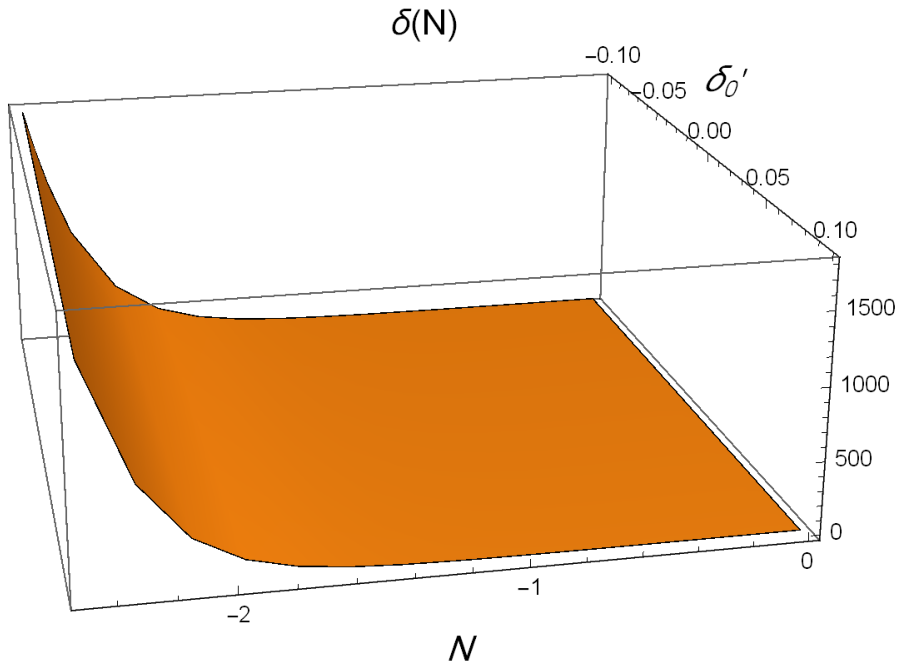, width=.45\linewidth,
height=1.8in} \textbf{(b)} \epsfig{file=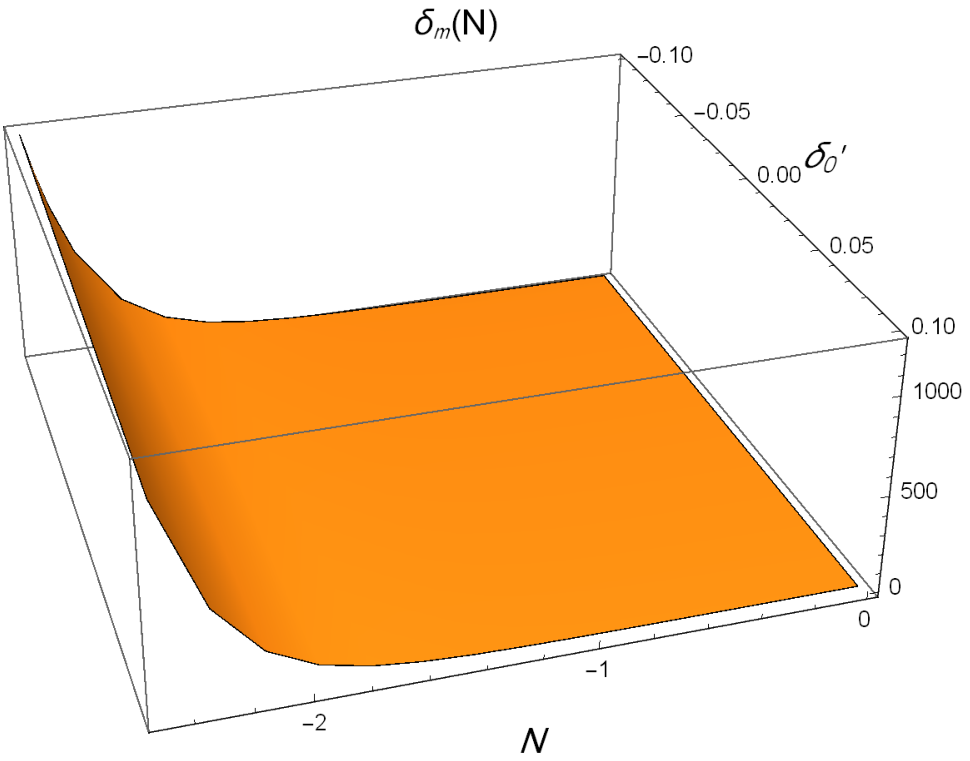, width=.45\linewidth,
height=1.8in}
\centering \epsfig{file=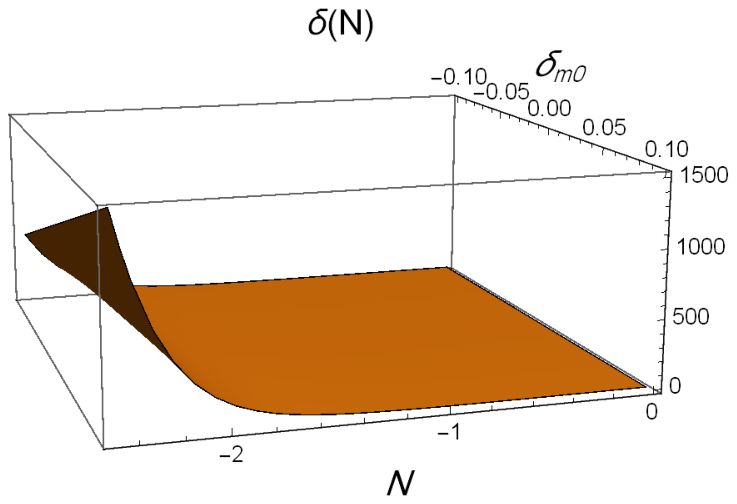, width=.45\linewidth,
height=1.8in} \textbf{(c)} \epsfig{file=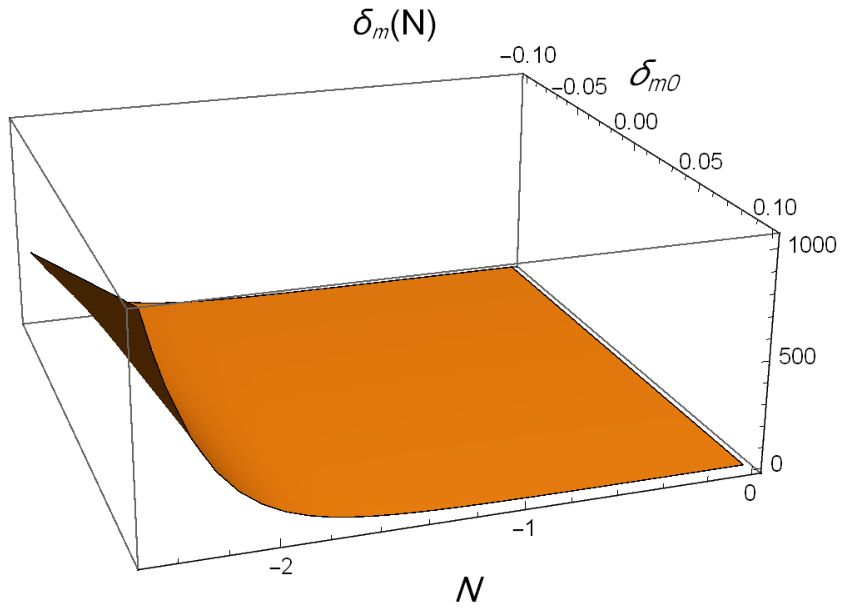, width=.45\linewidth,
height=1.8in}\caption{\label{fig1} This plot shows the stability of exponential model for
$F(\mathcal{R},T)=\mathcal{R}+\beta_1g(\mathcal{R})+\beta_2h(T)$ with $c_i=.2$. Plots for $\delta(N)$ and $\delta_m(N)$
verses N with $\omega=0$ are presented. Thus behavior of such perturbations with defined initial conditions is shown
in Fig.\ref{fig0} which smooth behavior indicates that model is stable near $Z=0$.}
\end{figure}
Consider the exponential evaluation solution with dust matter, the perturbed equation (\ref{64}) takes the form
\begin{eqnarray}\nonumber
&&18H_{\star}^4F_{\mathcal{RR}}^{\star}\delta^{''}+\{-6\rho_{\star}H_{\star}^2F_{\mathcal{R} T}^{\star}+(36H_{\star}^3 H_{\star}^{'}
+54H_{\star}^4)F_{\mathcal{RR}}^{\star}-108H_{\star}^3(H_{\star}^2H_{\star}^{''}+H_{\star}H_{\star}^{'^2}
+4H_{\star}^2H_{\star}^{'})F_{\mathcal{RRR}}^{\star}+18H_{\star}^4\rho_{\star}^{'}f_{RRT}^{\star}\}\delta^{'}\\\nonumber
&&+\{-6H_{\star}^2f_{R}^{\star}-(6\rho_{\star}(H_{\star}H_{\star}^{'}+4H_{\star}^2)+3H_{\star}^2\rho_{\star}^{'})F_{\mathcal{R}T}^{\star}
+18(2H_{\star}^3H_{\star}^{''}+H_{\star}^2H_{\star}^{'^2}+7H_{\star}^3H_{\star}^{'}-4H_{\star}^4)F_{\mathcal{RR}}^{\star}
-108(H_{\star}^3H_{\star}^{''}+H_{\star}^2H_{\star}^{'^2}\\\nonumber &&+ 4H_{\star}^3H_{\star}^{'})(H_{\star}H_{\star}^{'}
+4H_{\star}^2)F_{\mathcal{RRR}}^{\star}+18(H_{\star}^3H_{\star}^{'}+4H_{\star}^4)\rho_{\star}^{'}F_{\mathcal{RR}T}^{\star}\}\delta
+\{k^2\rho_{\star}+\frac{3\rho_{\star}f_{T}^{\star}}{2}+\rho_{\star}^2F_{TT}^{\star}+(\rho_{\star}(3H_{\star}H_{\star}^{'}+H_{\star}^{2})\\\nonumber &&-3H_{\star}^2\rho_{\star}^{'})f_{RT}^{\star}+18\rho_{\star}(H_{\star}^3H_{\star}^{''}+H_{\star}^2H_{\star}^{'^2}
+4H_{\star}^3H_{\star}^{'})F_{\mathcal{RR}T}^{\star}-3H_{\star}^2\rho_{\star}\rho_{\star}^{'}F_{\mathcal{R}TT}^{\star}\}\delta_{m}
-3H_{\star}^2F_{\mathcal{R}T}^{\star}\delta_{m}^{'}=0,
\end{eqnarray}
We consider Eq.(\ref{67}) to examine stability and $F(\mathcal{R},T)$ function which are reconstructed by utilizing the
constraint (\ref{11}). To solve this equation and Eq.(\ref{67}), we use numerical technique for model (\ref{19}).
For this model, above equation yields
\begin{eqnarray}\nonumber
&&18H_{\star}^4F_{\mathcal{R}\mathcal{R}}^{\star}\delta^{''}+\{(36H_{\star}^3 H_{\star}^{'}+54H_{\star}^4)F_{\mathcal{R}\mathcal{R}}^{\star}
-108H_{\star}^3(H_{\star}^2 H_{\star}^{''}+H_{\star}H_{\star}^{'^2}+4H_{\star}^2H_{\star}^{'})F_{\mathcal{R}\mathcal{R}\mathcal{R}}^{\star}\}
\delta^{'}\\\nonumber&&\times\{-6H_{\star}^2f_{R}^{\star}+18(2H_{\star}^3H_{\star}^{''}+H_{\star}^2H_{\star}^{'^2}+7H_{\star}^3H_{\star}^{'}
-4H_{\star}^4)F_{\mathcal{R}\mathcal{R}}^{\star}-108(H_{\star}^3H_{\star}^{''}+H_{\star}^2H_{\star}^{'^2}\\\nonumber
&&+4H_{\star}^3H_{\star}^{'})(H_{\star}H_{\star}^{'}+4H_{\star}^2)F_{\mathcal{R}\mathcal{R}\mathcal{R}}^{\star}\}\delta +\{k^2\rho_{\star}+\frac{3\rho_{\star}f_{T}^{\star}}{2}+\rho_{\star}^2F_{TT}^{\star}\}\delta_{m}=0.
\end{eqnarray}
We set the constants $\beta_1=\beta_2=2$ and numerically solve above equation along with (\ref{67}) for reconstructed
function (\ref{19}) and analyze its stability by plotting graphs as shown in Figure \ref{fig0}. The development of
perturbation $\delta$ and $\delta_m$ regarding to all different initial conditions is presented in Figure \ref{fig0}(a-c),
which indicates that these perturbations incline at small value, i.e., $z=0.010052$ independent of choice of initial
conditions. Hence it can be concluded that reconstructed function exhibit smooth behavior against the introduced
perturbations. The stability of $F(\mathcal{R},T)$ model (\ref{23}) with $\beta_1=\beta_2=1$ is shown in Figure \ref{fig1}
which leads to similar results and hence corresponds to stable behavior. It is worthy to mention here that we cannot analyze
the stability of other reconstructed models like $F(\mathcal{R},T)=\mathcal{R}g(T),~ F(\mathcal{R},T)=\mathcal{R}+Tg(\mathcal{R})$
and $F(\mathcal{R},T)=\mathcal{R}+\mathcal{R}F(T)$ for exponential bouncing as we are unable to find analytical solutions
in these cases.

\subsection{Stability of Oscillatory Solutions}

\begin{figure}
\centering \epsfig{file=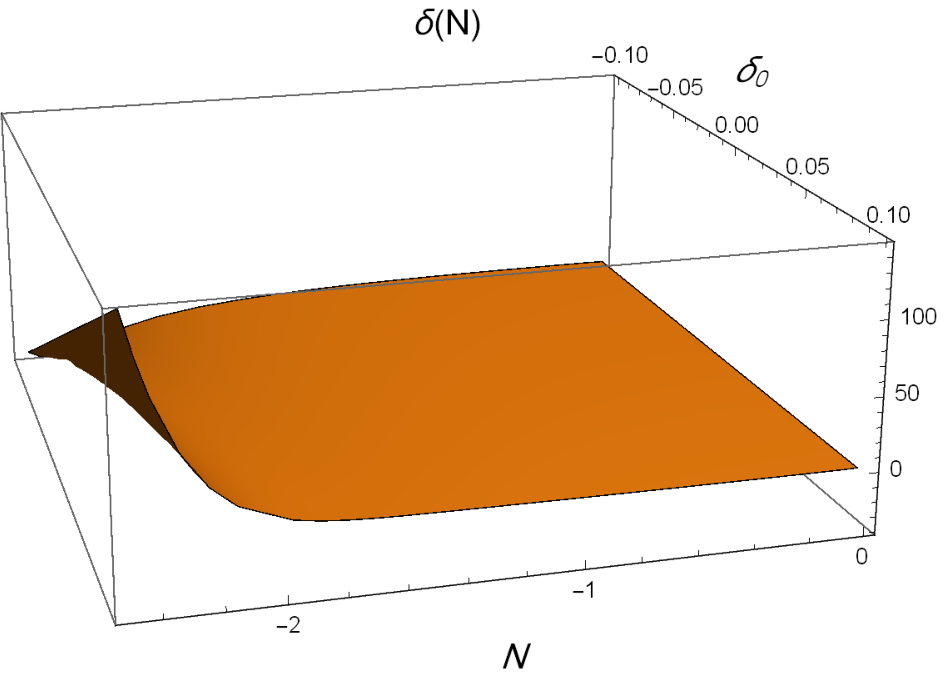, width=.45\linewidth,
height=1.8in} \textbf{(a)} \epsfig{file=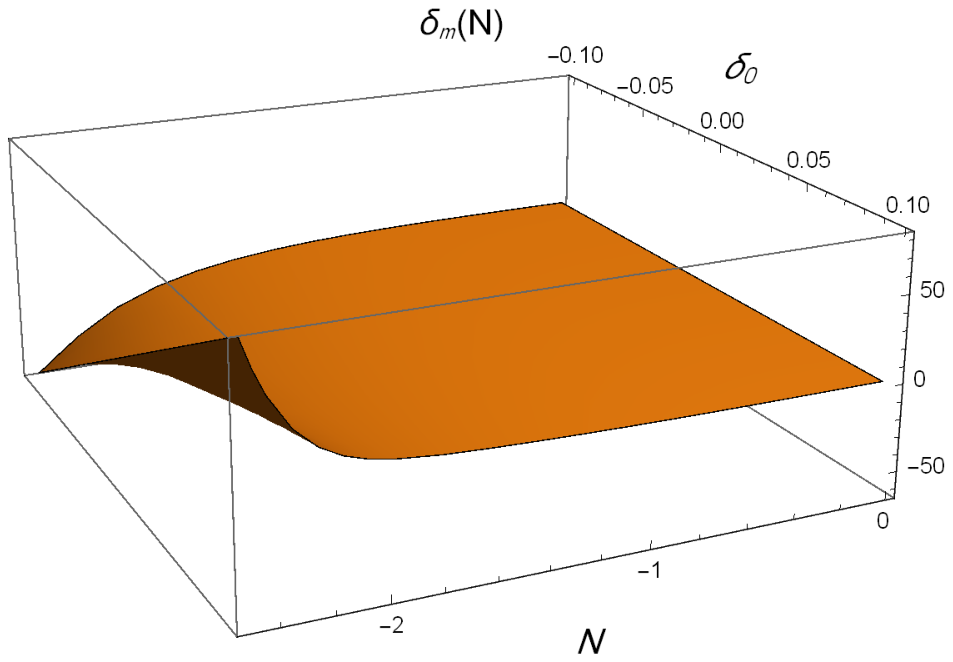, width=.45\linewidth,
height=1.8in}
\centering \epsfig{file=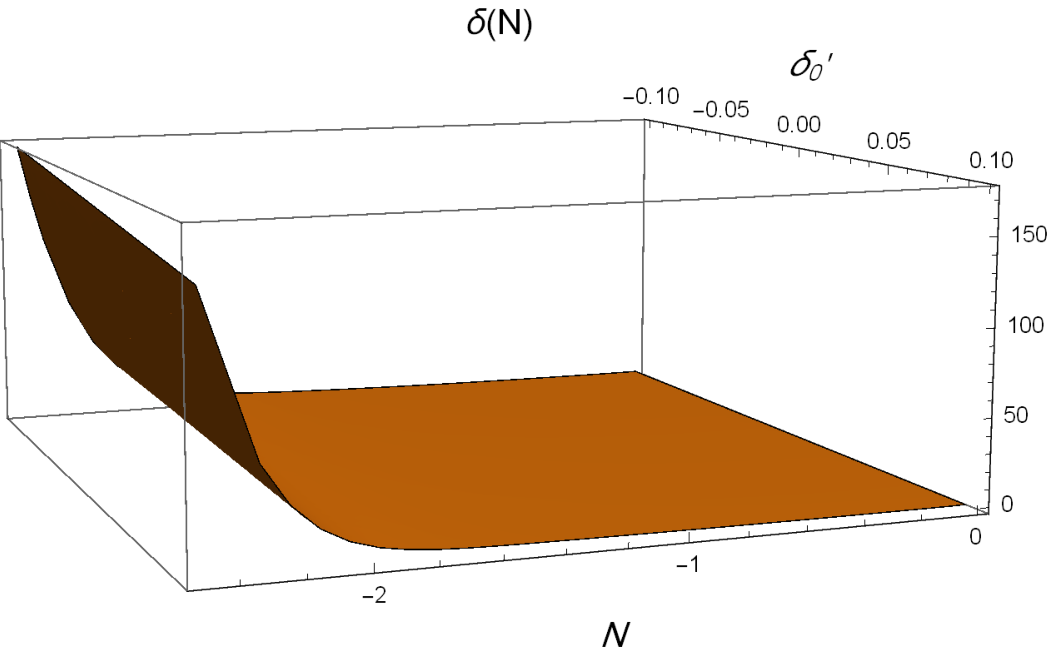, width=.45\linewidth,
height=1.8in} \textbf{(b)} \epsfig{file=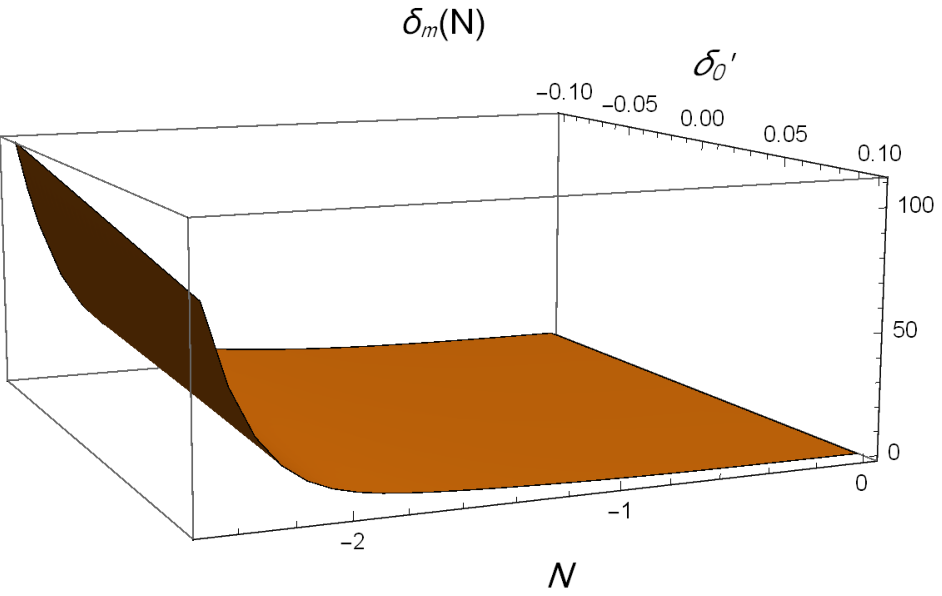, width=.45\linewidth,
height=1.8in}
\centering \epsfig{file=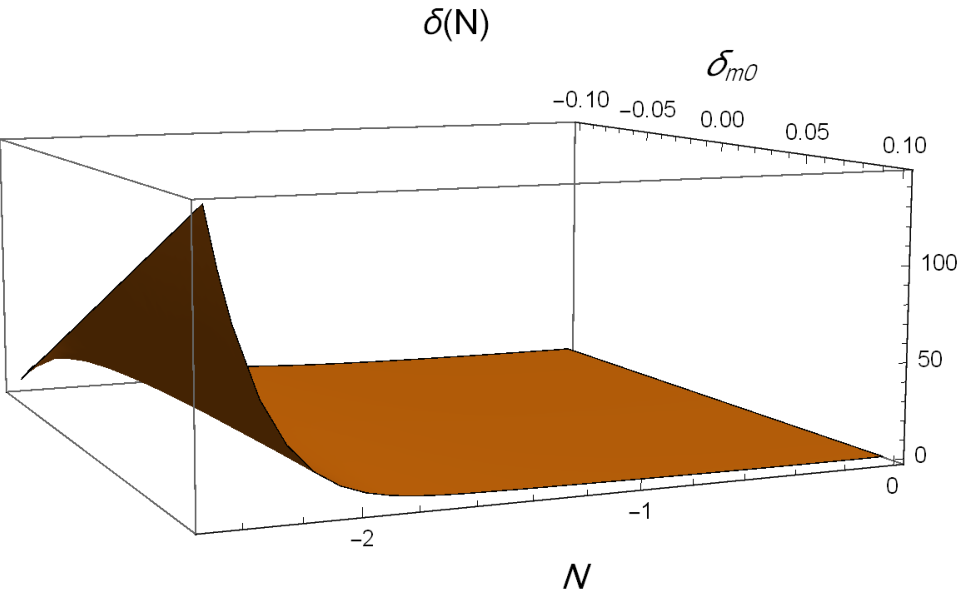, width=.45\linewidth,
height=1.8in} \textbf{(c)} \epsfig{file=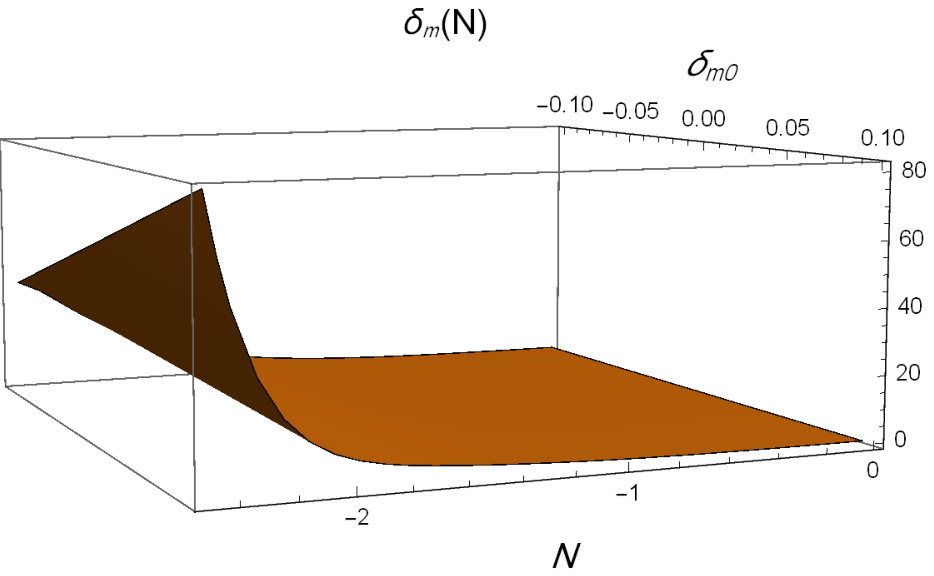, width=.45\linewidth,
height=1.8in}\caption{\label{fig2} Stability of $F(\mathcal{R},T)=\beta_1g(\mathcal{R})+\beta_2h(T)$ model for oscillatory
case with $c_i=.2$. Development of $\delta(N)$ and $\delta_m(N)$ verses $N$ with $\omega=0$ is shown. This plot shows the smooth
behavior near $z=0$, so this model is stable.}
\end{figure}

\begin{figure}
\centering \epsfig{file=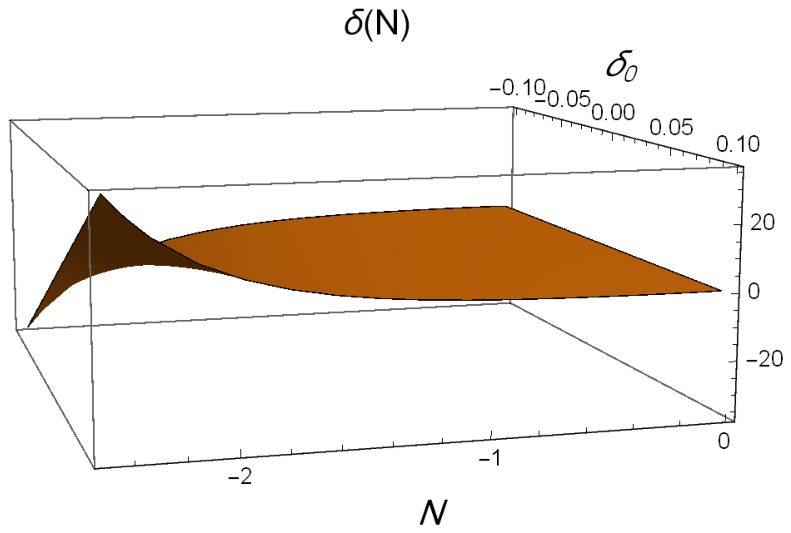, width=.45\linewidth,
height=1.8in} \textbf{(a)} \epsfig{file=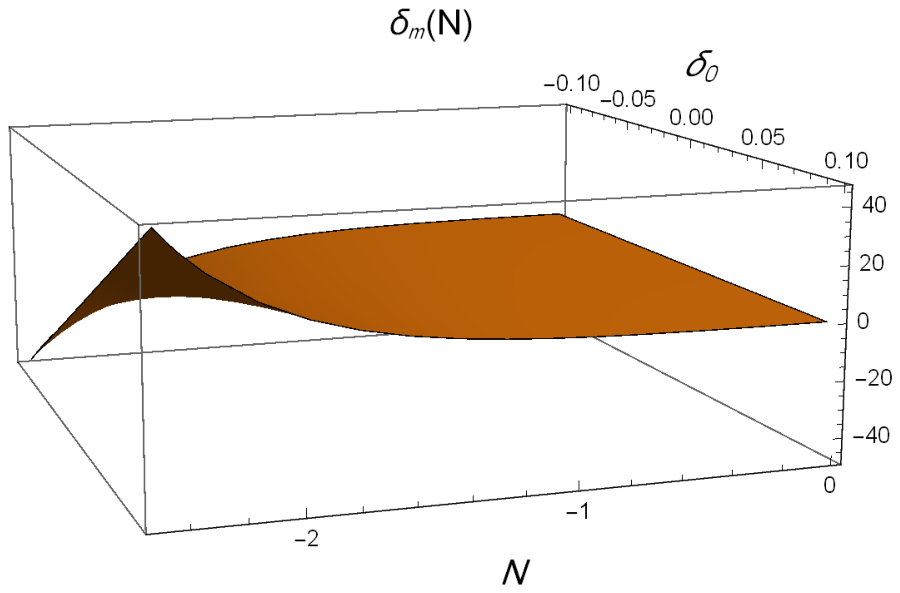, width=.45\linewidth,
height=1.8in}
\centering \epsfig{file=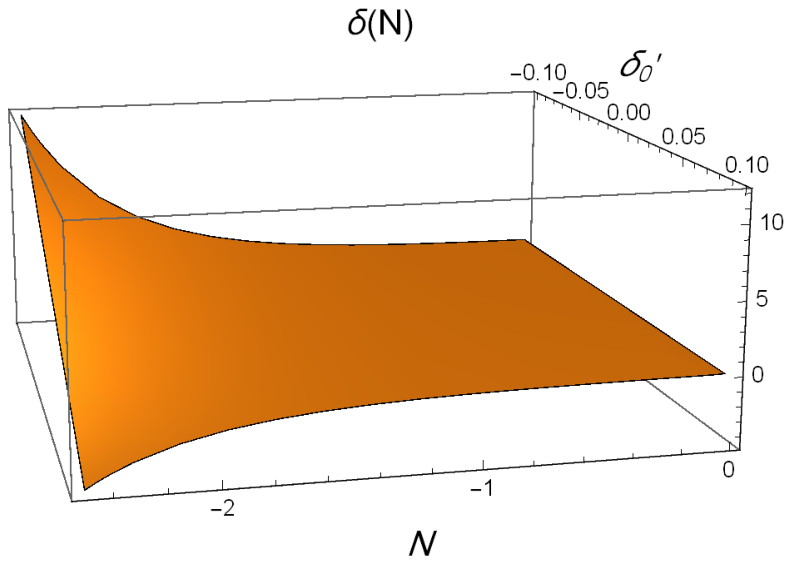, width=.45\linewidth,
height=1.8in} \textbf{(b)} \epsfig{file=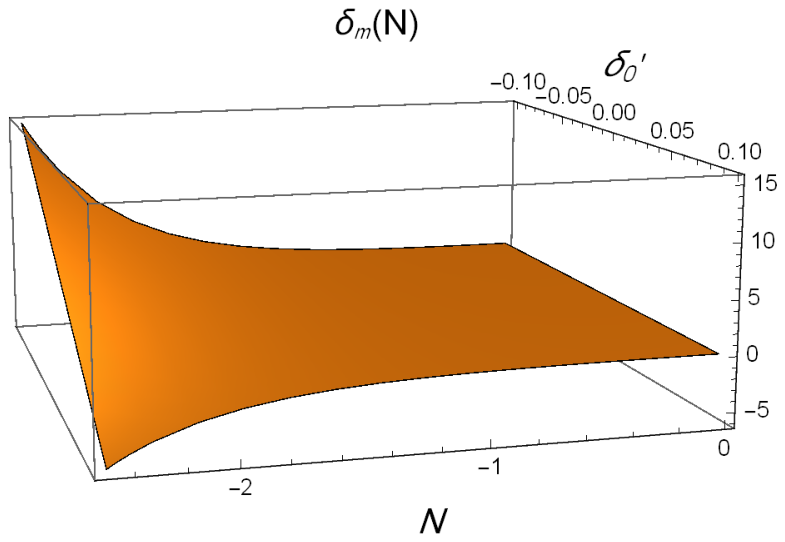, width=.45\linewidth,
height=1.8in}
\centering \epsfig{file=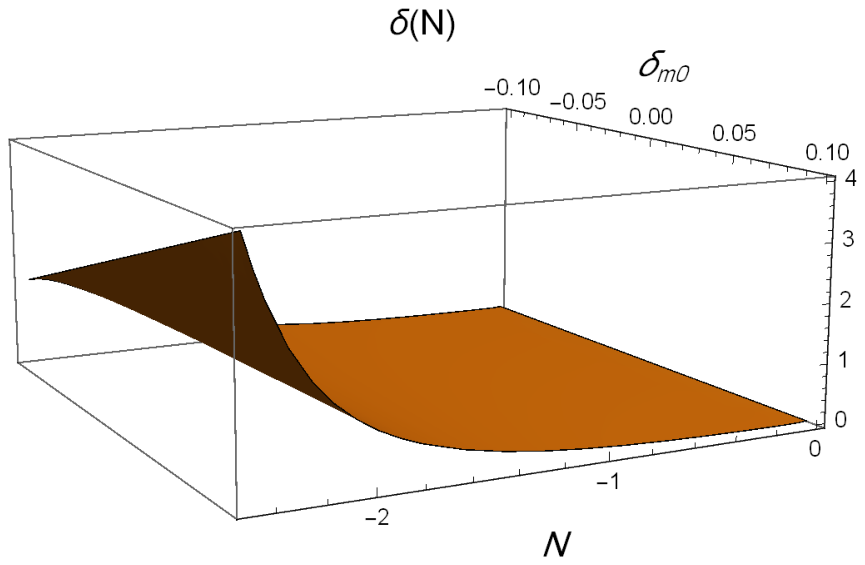, width=.45\linewidth,
height=1.8in} \textbf{(c)} \epsfig{file=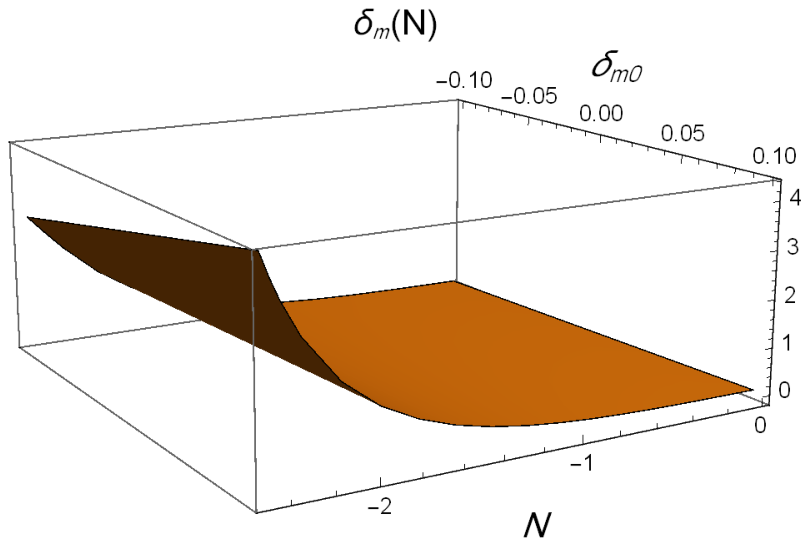, width=.45\linewidth,
height=1.8in}\caption{\label{fig3} Stability of $F(\mathcal{R},T)=\mathcal{R}+\beta_1g(\mathcal{R})+\beta_2h(T)$ form
for oscillatory model with $c_i=.2$. This model is stable due to its smooth behavior near origin.}
\end{figure}

Here we will explore the stability of reconstructed solutions for oscillatory model given in Eqs.(\ref{35}) and (\ref{37}) as
these are the only exact solutions we have found in Section \textbf{III}. For these models, the system of perturbed equations
can be solve numerically by assuming $\beta_1=\beta_2=2$ and applying varied initial conditions. We investigate that
results of these perturbations for all initial conditions are same as provided in the graphs of Figures \ref{fig2} and
\ref{fig3}. These perturbations approach to small value and shows the stable behavior near $z=0$ irrespective of initial
conditions.

\subsection{Stability of Power Law Solutions}

\begin{figure}
\centering \epsfig{file=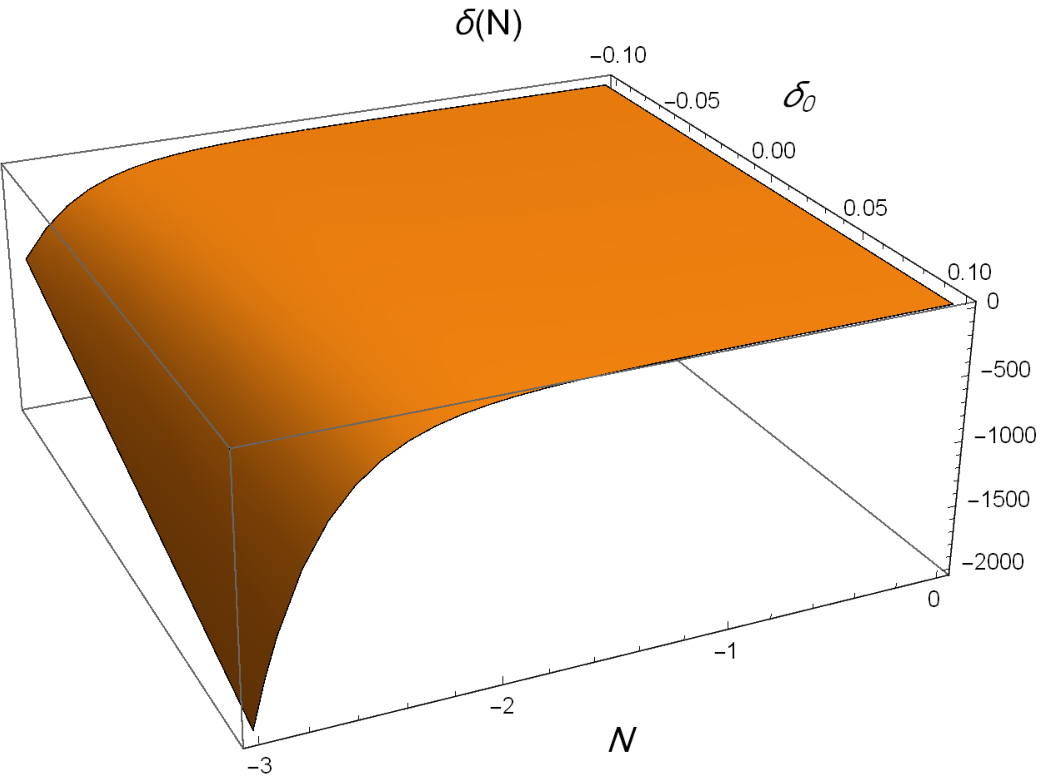, width=.45\linewidth,
height=1.8in} \textbf{(a)} \epsfig{file=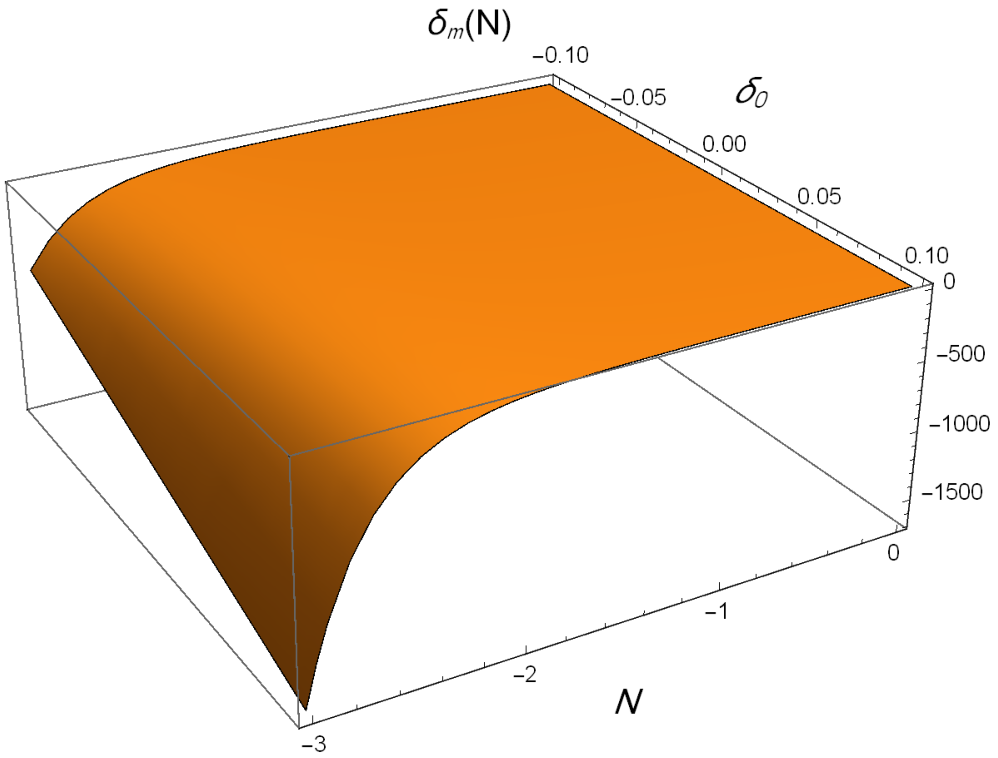, width=.45\linewidth,
height=1.8in}
\centering \epsfig{file=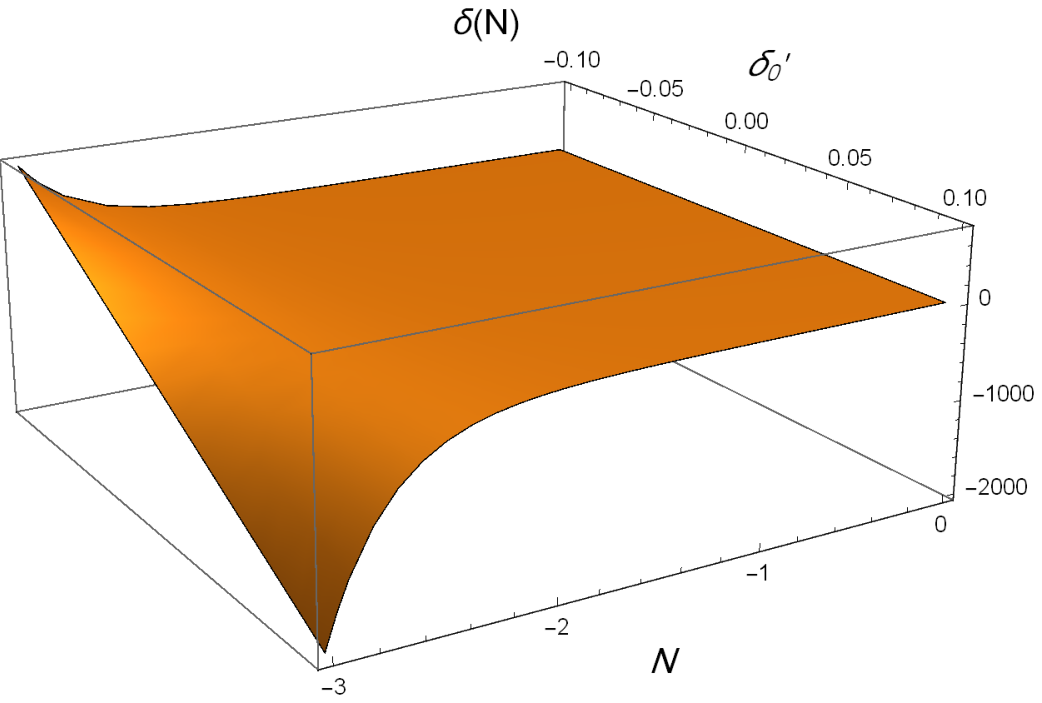, width=.45\linewidth,
height=1.8in} \textbf{(b)} \epsfig{file=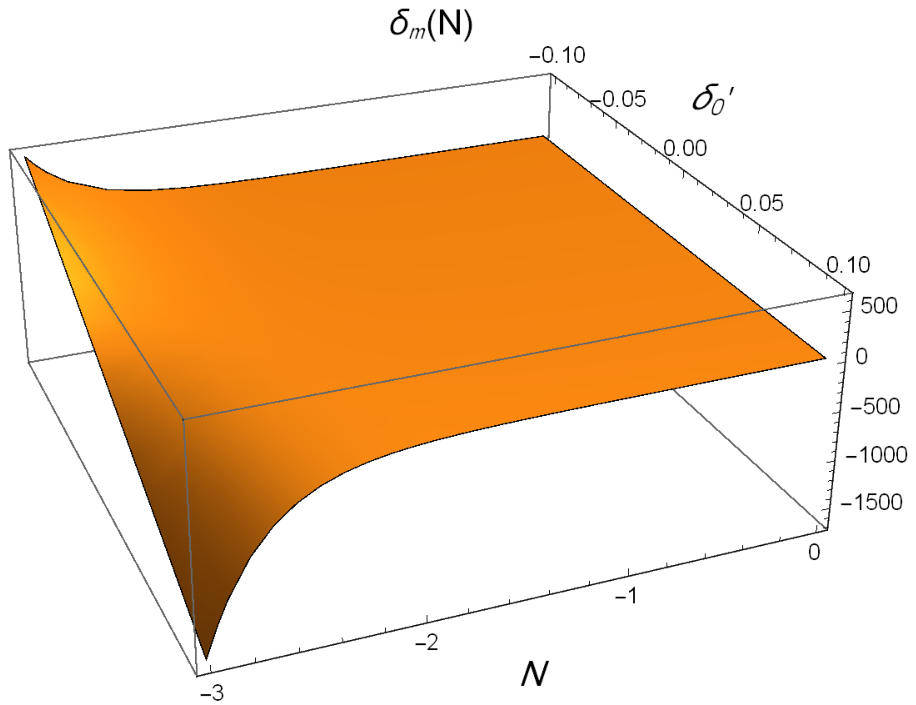, width=.45\linewidth,
height=1.8in}
\centering \epsfig{file=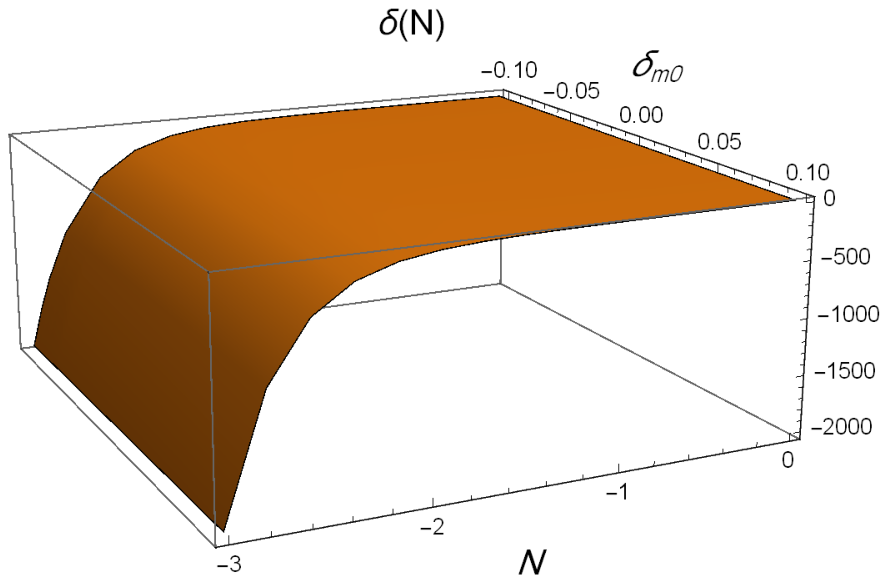, width=.45\linewidth,
height=1.8in} \textbf{(c)} \epsfig{file=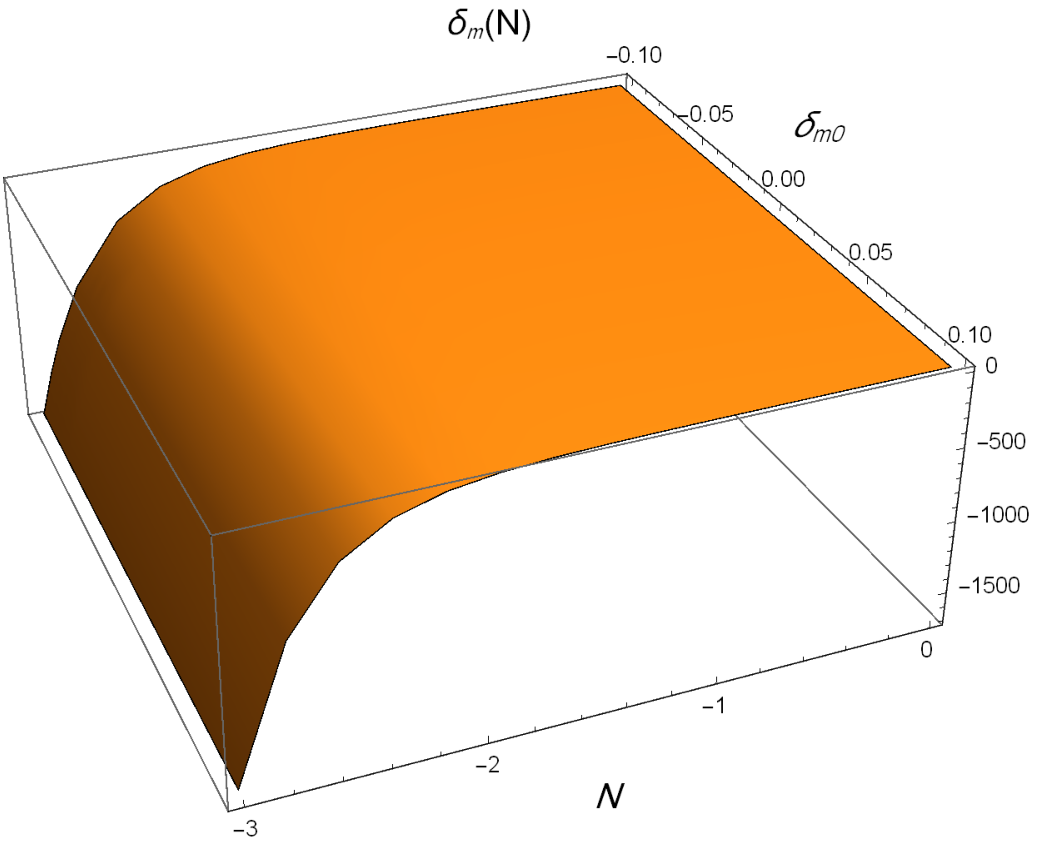, width=.45\linewidth,
height=1.8in}\caption{\label{fig4} Stability of $F(\mathcal{R},T)=\mathcal{R}+\beta_1g(\mathcal{R})+\beta_2h(T)$ model in
power law case with $c_i=.2$.}
\end{figure}

\begin{figure}
\includegraphics[width=0.24\textwidth]{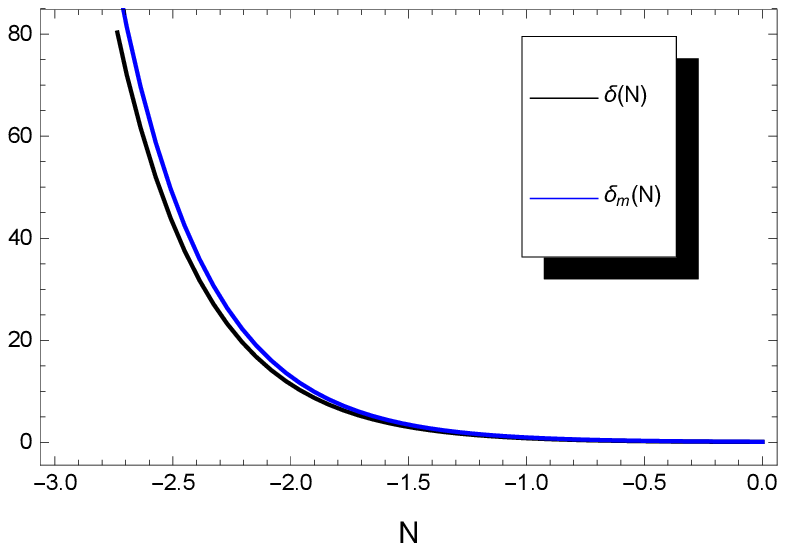}
\includegraphics[width=0.24\textwidth]{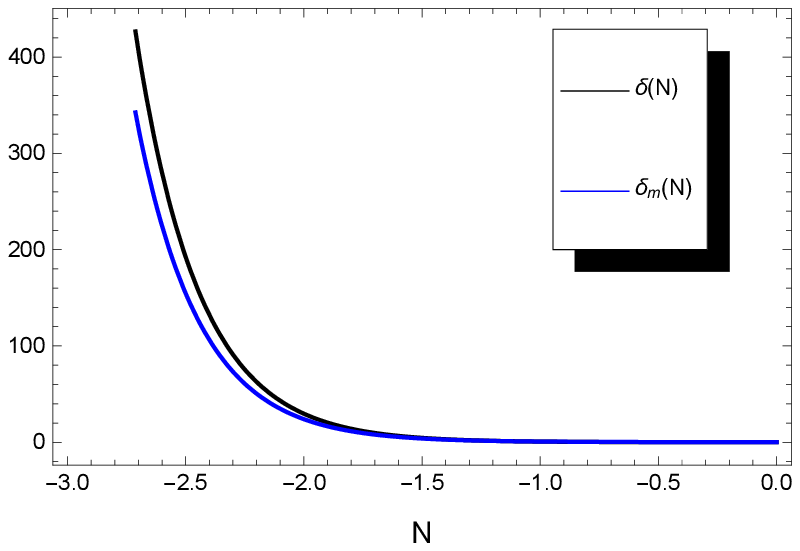}
\includegraphics[width=0.24\textwidth]{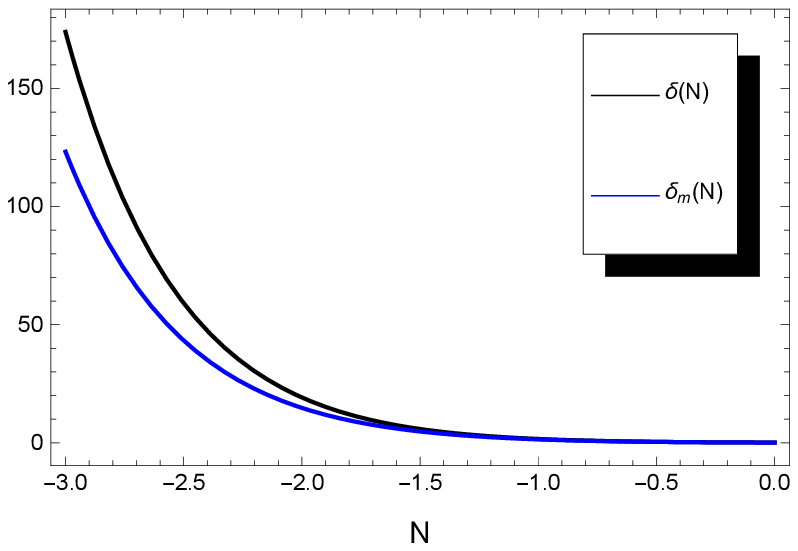}
\includegraphics[width=0.24\textwidth]{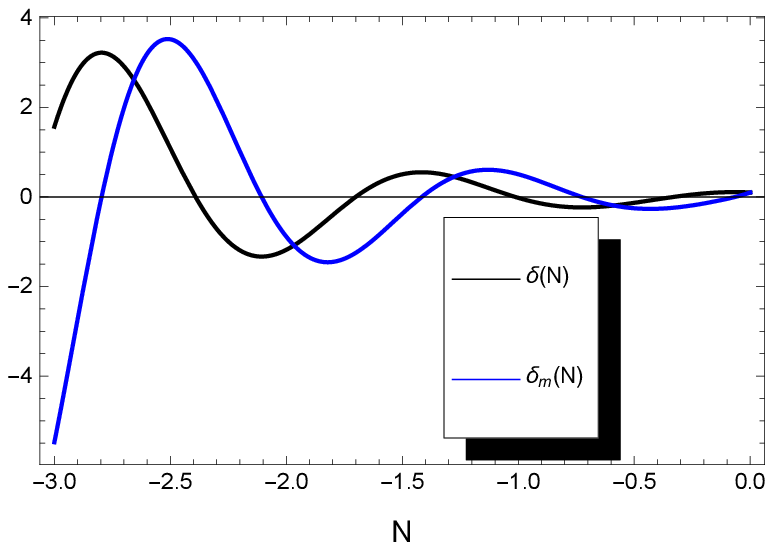}
\caption{\label{fig6} Evolution of $\delta(N)$ and $\delta_m(N)$ for power law model with function (\ref{54}) and (\ref{58}).
First two plots represent the perturbation parameters for function (\ref{54}) with $\alpha=2/3$ and $\alpha=2$ and last two
plots indicate the perturbation parameter for function (\ref{58}) with $\alpha=2/3$ and $\alpha=5/2$. Here, $\omega=0$.}
\end{figure}

In this subsection, we shall present the stability of matter dominated as well as late time epochs of power law solution. The stability of
reconstructed power law model (\ref{47}) has been discussed in Ref. \cite{21}. Now, we first consider the model (\ref{51}) and Eqs.(\ref{65}) and (\ref{67}).
At the first place, we consider the dust dominated era for $\alpha=2/3$ and $\alpha=0$. Figure \ref{fig4} indicates the stability of
perturbations ($\delta(N),\delta_m(N)$) for dust dominated era with $\beta_1=\beta_2=2$.
Similarly, the evolution of perturbation with $\alpha=2$ and $\alpha=0$ for accelerating universe is presented in Figure \ref{fig5}.
Secondly, we analyze the stability of reconstructed solution (\ref{54}) about linear perturbations. Figure \ref{fig6} presents the behavior
of such perturbations in cosmic decelerated phase with $\omega=0$ and $\alpha=2/3$. Here, we choose $c_1=c_2=.2$. It is to be noted that
small oscillations are produced initially and converges to zero so it shows stable behavior.
Figure \ref{fig6} explains the graph of perturbations for model (\ref{54}) in accelerating era with $\alpha=2$. Initially,
it shows the oscillation and then decays in future, so the solution becomes stable.
Thirdly, we explored the stability of reconstructed solution for model $F(\mathcal{R},T)=\mathcal{R} h(T)$ against introduced
perturbations. Here, it is seen that for radiation ($\alpha=2/3$ and $\omega=0$) and matter ($\alpha = 1/2$ and $\omega=0$) dominated
eras, stability cannot be explained for this model as singularity appears which is a non-physical case. Similarly, for cosmic era
with $\alpha=2$ and $\omega=0$, this model can not explain the cosmological evolution because complex terms appear.
Lastly, we investigate the stability of Eq.(\ref{58}) for $\alpha=2/3,~~5/2$ with $\omega=0$. The evolution of perturbations
is presented in graphs of Figure \ref{fig6} by applying different initial conditions. It is easy to check that the perturbations
$(\delta(N), ~\delta_m(N))$ show fluctuating behavior for accelerating universe with $\alpha=5/2$ and even present in late cosmic
epochs, therefore solutions are unstable in this case.

\subsection{Stability of Matter Bounce Solutions}

\begin{figure}
\includegraphics[width=0.35\textwidth]{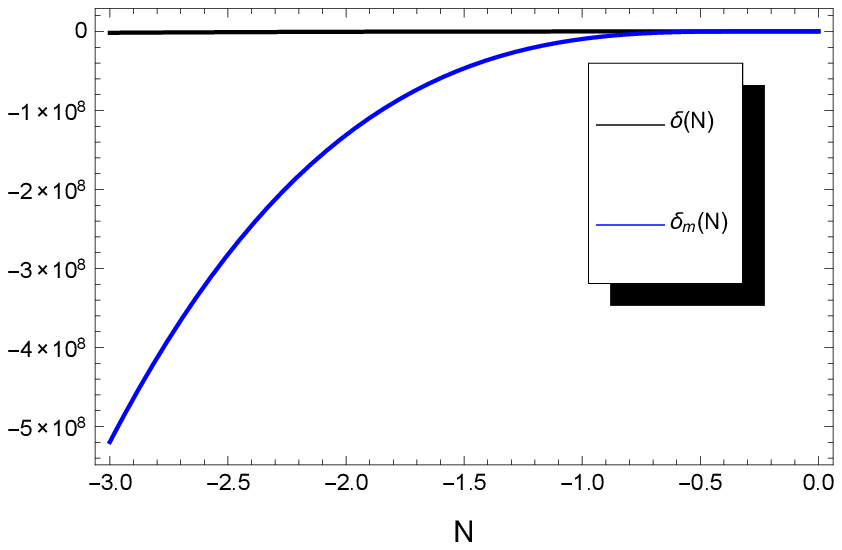}
\caption{\label{Figmb2} Evolution of $\delta(N)$ and $\delta_m(N)$ for matter bounce model with function (\ref{mb2}). Here, $\omega=0$.}
\end{figure}

In this section, we shall analyze the stability of reconstructed matter bounce solution (\ref{mb2}) as shown in Figure \ref{Figmb2}.
One can easily see that it converges to zero so this solution is stable. In case of solutions (\ref{mb1}) and (\ref{mb3}),
it is seen that the stability of reconstructed solutions through perturbation parameters cannot be achieved as singularity or
complex terms appear in these cases. Similarly, stability of perturbations for Lagrangians $(ii),~ (iv), ~(v), ~(vii)$ and $(viii)$
can not be investigated as the analytical solutions were not possible to obtain in these cases (as shown in Section \textbf{III}).

\section{Concluding Remarks}

\begin{figure}
\centering \epsfig{file=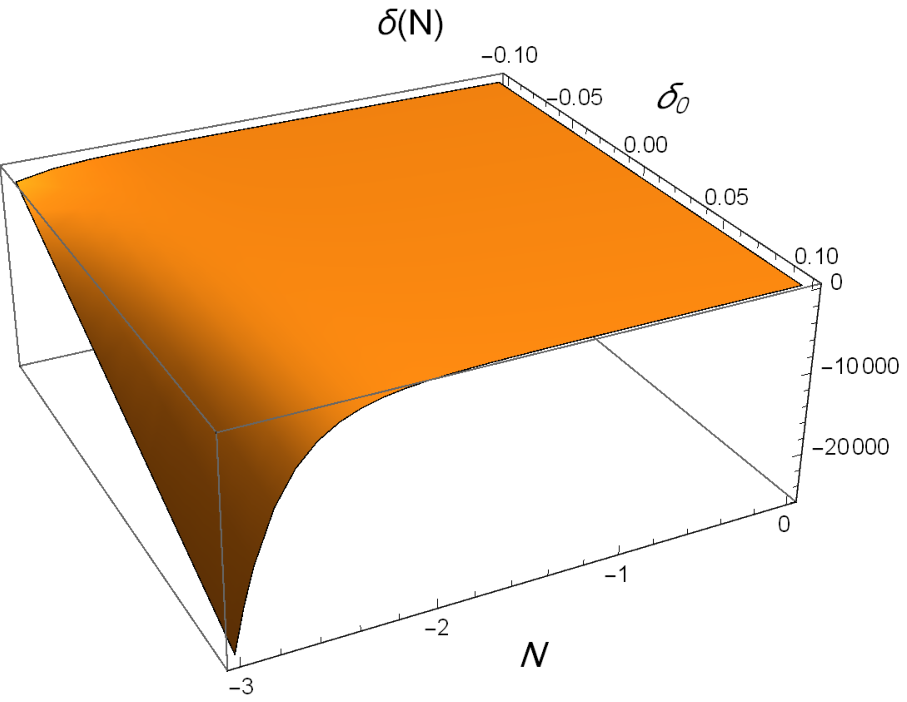, width=.45\linewidth,
height=1.8in} \textbf{(a)} \epsfig{file=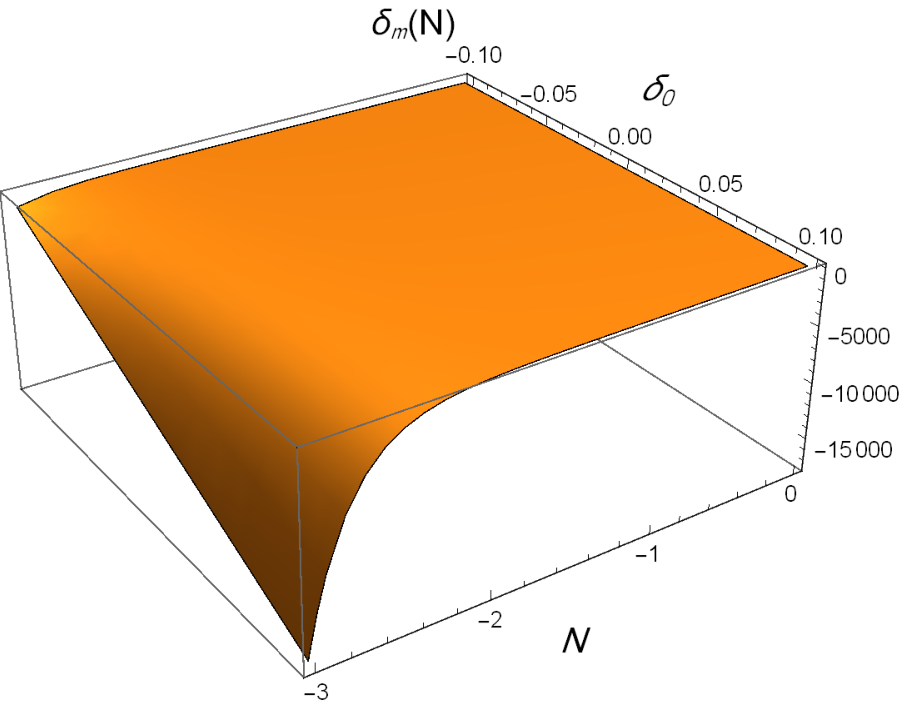, width=.45\linewidth,
height=1.8in}
\centering \epsfig{file=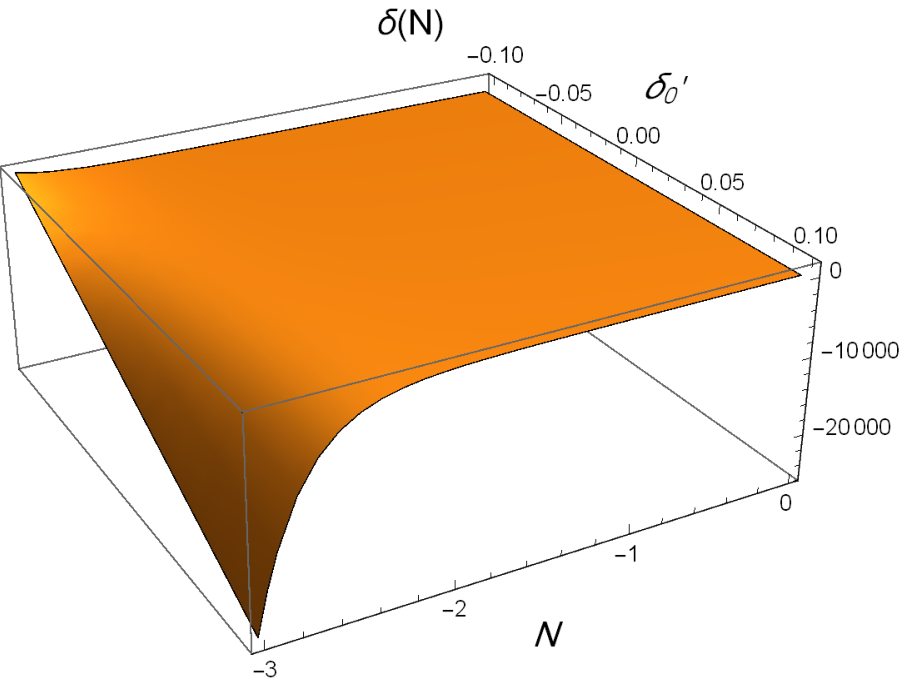, width=.45\linewidth,
height=1.8in} \textbf{(b)} \epsfig{file=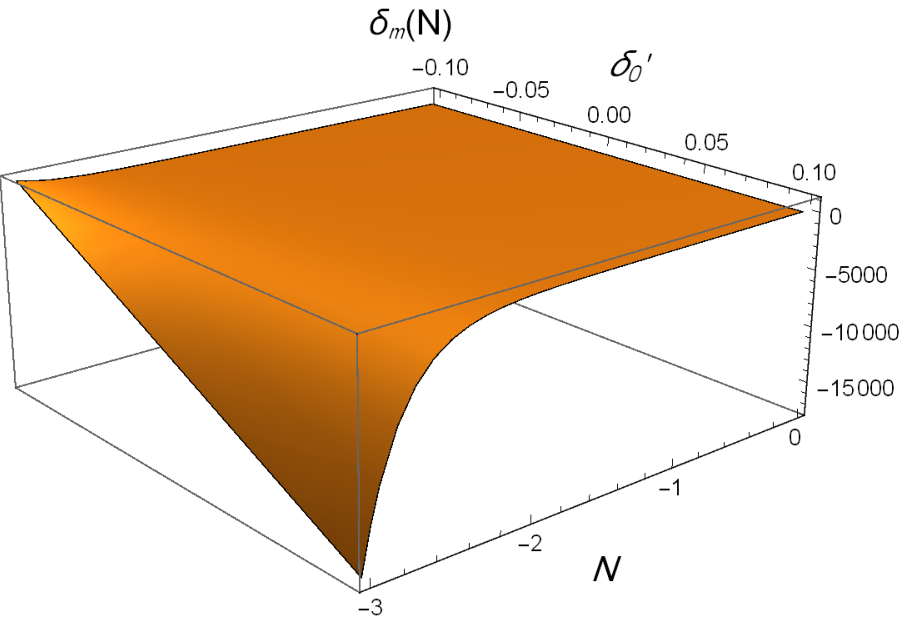, width=.45\linewidth,
height=1.8in}
\centering \epsfig{file=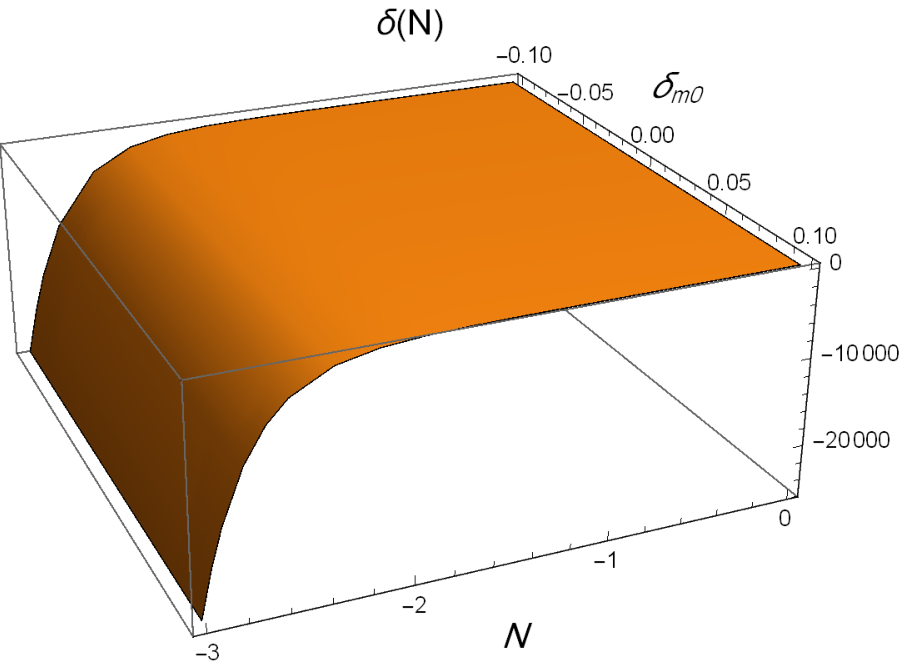, width=.45\linewidth,
height=1.8in} \textbf{(c)} \epsfig{file=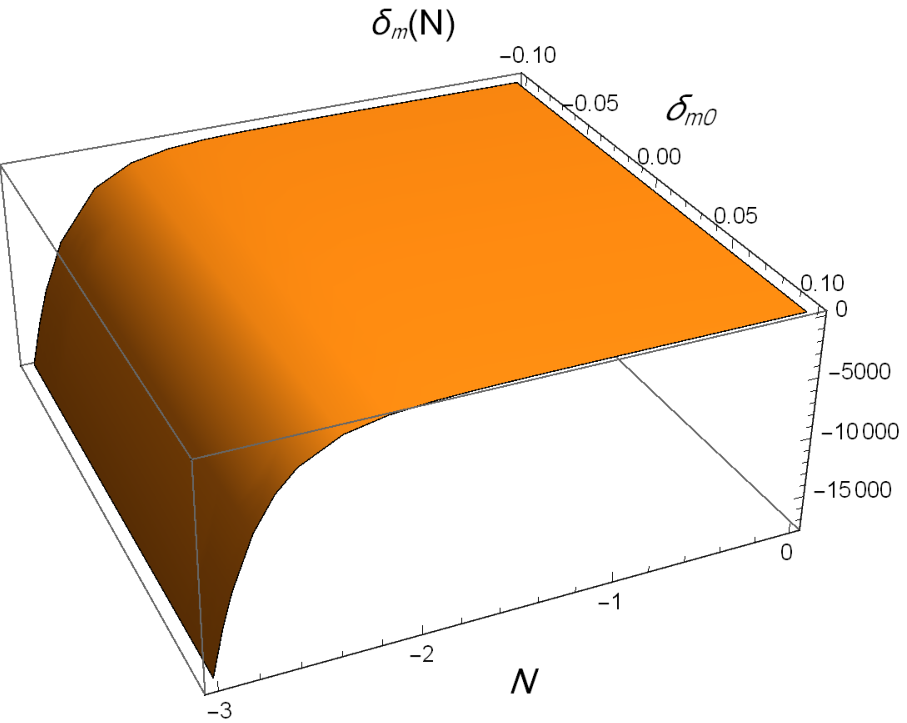, width=.45\linewidth,
height=1.8in}\caption{\label{fig5} Stability of $F(\mathcal{R},T)=\mathcal{R}+\beta_1g(\mathcal{R})+\beta_2h(T)$
function in power law model. The distribution of perturbations for accelerated phase is same as for dust dominated era in Fig. \ref{fig4}}
\end{figure}

\begin{table}
\centering \caption{} \label{Table0}
\begin{tabular}{|c|c|c|c|}
\hline
Models & $F(\mathcal{R},T)$ & Solutions & Vacuum solutions constraints \\\hline
Exponential evaluation &$\beta_{1} g(\mathcal{R}) + \beta_{2} h(T)$ & Analytical solution exist & $32 (a_1)^{\frac{5}{2}}c_1=(6+e^{1/4}\sqrt{\pi}Erf(\frac{1}{2}))c_2$\\
                       &$\mathcal{R}+\beta_{1} g(\mathcal{R}) + \beta_{2} h(T)$& Analytical solution exist & $\frac{d_2}{32(a_1)^{\frac{5}{2}}}(6+\sqrt{\pi}e^{1/4}Erf(\frac{1}{2}))-\frac{27a_1}{128\beta_1}=d_1$\\
                       &$\mathcal{R}g(T)$& No analytical solution exist & Not possible\\
                       &$\mathcal{R}+ Tg(\mathcal{R})$&No analytical solution exist & Not possible\\
                       &$\mathcal{R} + \mathcal{R} F(T)$&No analytical solution exist & Not possible\\
                       &$F_1(\mathcal{R}) + F_2(\mathcal{R}) F_3(T)$&No solution exist & Not possible\\
                       &$\mu \bigg(\frac{R}{R_0}\bigg)^{\beta}\bigg(\frac{T}{T_0}\bigg)^{\gamma}$ & No analytical solution & Not possible \\\hline
Oscillatory           &$\beta_{1} g(\mathcal{R}) + \beta_{2} h(T)$ & Analytical solution exist & $ 2187 a_2^{5/2}c_1=18 c_2\tanh^{-1}(\frac{\iota}{\sqrt{3}})$\\
                       &$\mathcal{R}+\beta_{1} g(\mathcal{R}) + \beta_{2} h(T)$& Analytical solution exist & $4374 a_2^{9/2}\beta_1c_1= -[18a_2^2\tanh^{-1}(\iota/\sqrt{3})$\\
                       & & &$(9a_2^2-2\beta_1c_2)+729 a_2^4(\ln(\frac{-3\sqrt{a_2}+\sqrt{3a_2}\iota}{3\sqrt{a_2}+\sqrt{3a_2}\iota}))]$\\
                       &$\mathcal{R}g(T)$& No analytical solution exist & Not possible\\
                       &$\mathcal{R}+ Tg(\mathcal{R})$&No analytical solution exist & Not possible\\
                       &$\mathcal{R} + \mathcal{R} F(T)$&No analytical solution exist & Not possible\\
                       &$F_1(\mathcal{R}) + F_2(\mathcal{R}) F_3(T)$&No solution exist & Not possible\\
                       &$\mu \bigg(\frac{R}{R_0}\bigg)^{\beta}\bigg(\frac{T}{T_0}\bigg)^{\gamma}$ & No analytical solution & Not possible \\\hline
Power law           &$\beta_{1} g(\mathcal{R}) + \beta_{2} h(T)$ & Analytical solution exist & Always satisfied\\
                       &$\mathcal{R}+\beta_{1} g(\mathcal{R}) + \beta_{2} h(T)$& Analytical solution exist & Always satisfied\\
                       &$\mathcal{R}g(T)$&  Analytical solution exist & Always satisfied\\
                       &$\mathcal{R}+ Tg(\mathcal{R})$& Analytical solution exist & Holds for $\omega=1/3$\\
                       &$\mathcal{R} + \mathcal{R} F(T)$& Analytical solution exist& Always satisfied\\
                       &$F_1(\mathcal{R}) + F_2(\mathcal{R}) F_3(T)$&No solution exist & Not possible\\
                       &$\mu \bigg(\frac{R}{R_0}\bigg)^{\beta}\bigg(\frac{T}{T_0}\bigg)^{\gamma}$ & Analytical solution exist & Holds for $\gamma=0,~3\alpha(1+\omega_i)=2\beta$, $\forall i$ \\\hline
 Matter bounce         & $\beta_{1} g(\mathcal{R}) + \beta_{2} h(T)$ & Analytical solution exist & Not satisfied\\
                       &$\mathcal{R}+\beta_{1} g(\mathcal{R}) + \beta_{2} h(T)$&No Analytical solution exist & Not satisfied\\
                       &$\mathcal{R}g(T)$&  Analytical solution exist & Not satisfied\\
                       &$\mathcal{R}+ Tg(\mathcal{R})$& No solution exist & Not possible\\
                       &$\mathcal{R} + \mathcal{R} F(T)$& Analytical solution exist& Not satisfied\\
                       &$F_1(\mathcal{R}) + F_2(\mathcal{R}) F_3(T)$&No solution exist & Not possible\\
                       &$\mu \bigg(\frac{R}{R_0}\bigg)^{\beta}\bigg(\frac{T}{T_0}\bigg)^{\gamma}$ & No analytical solution exist & Not possible\\\hline
\end{tabular}
\end{table}
The search for a form of Lagrangian which can propose the cosmic evolution in an appropriate way is still under consideration
in the present day cosmology. In this regard, the use of reconstruction schemes has fascinated many researchers due to its
successful applications in explaining different features of cosmology. The $F(\mathcal{R},T)$ gravity plays a vital role to
explain the impact of dark energy on accelerated cosmic expansion. Such modified theory, involving contribution of both
matter Lagrangian and curvature part, has a great importance as a source of dark energy component. Many authors have
explored the cosmological reconstruction in $F(\mathcal{R},T)$ gravity by taking different scenarios into account.
This paper is another valuable addition in this respect by presenting the reconstruction of different isotropic cosmological
bouncing models in $F(\mathcal{R},T)$ gravitational framework. The basic idea of such theory is the inclusion of coupling
between matter and curvature which produces $\bigtriangleup^iT_{ij}\neq0$ and hence in order to get the usual continuity
equation ($\bigtriangleup^iT_{ij}=0$), one needs to impose an additional condition. In the present paper, we have adopted
reconstruction approach to obtain the form of generic function $F(\mathcal{R},T)$ by using four well-known bouncing models
namely exponential, oscillatory, power law and matter bounce models. Due to the complex form of dynamical equations, we have
considered seven simple forms of generic function and explored the corresponding unknown functions. We can summarize our discussion
as follows:
\begin{itemize}
\item In exponential and oscillatory models, only first two ansatz forms of function have analytical solutions while for other
forms, we were unable to find analytical solution. In case of power law model, all assumed forms of function have analytical solution.
For matter bounce, we have obtained analytical solutions for Lagrangian function using three forms of $F(\mathcal{R},T)$. Thus
it can be concluded that for these bouncing models, more general forms of generic function, e.g., $F(\mathcal{R},T)= F(\mathcal{R})g(T)$ and
$F(\mathcal{R},T)=F_1(\mathcal{R})+F_2(\mathcal{R})F_3(T)$ cannot be reconstructed as explained in Table \ref{Table0}.
\item Next we have analyzed the behavior of energy conditions for all reconstructed models graphically which have analytical solution.
Here, we have found that NEC and SEC violate while other energy constraints remain valid. Similarly, in non-phantom region,
the NEC, WEC and SEC are valid but DEC is invalid.
\item We have investigated the stability or instability of different forms of generic function having analytical solutions
for classifying them on physical grounds. Here we have explained the stability of $F(\mathcal{R},T)$ solutions which reproduce
the exponential, oscillatory and power law expansion history. For this purpose, in modified theory,
we have explored a set of perturbation equations and explicit form of coefficients which are presented in Appendix.
On stability perturbations, we impose the initial conditions at $z=.0100502$ and vary the range as $\{-0.1, 0.1\}$.
The results for stability of exponential solutions have been presented in Figures \ref{fig0} and \ref{fig1} which showed the
smooth behavior near $z=0$. Similarly, the stability of oscillatory solutions are provided in Figures \ref{fig2} and \ref{fig3}.
\item For power law solutions, we discussed the perturbations for both cases dust as well as accelerated expansion.
We have chosen integration constants as $c_i$'s=$.2$. The graphs of Figures \ref{fig4}, \ref{fig6} and \ref{fig5} showed the
stability of power law models in $F(\mathcal{R},T)$ gravity. In power law case, we fixed initial conditions at $z=0$ and ranges
vary between \{-.1,.1\}. We concluded that reconstructed solutions $F(\mathcal{R},T)=\mathcal{R}h(T)$ showed unstable behavior
for linear perturbations regarding both accelerating and decelerating eras. Similarly $F(\mathcal{R},T)=\mathcal{R}+\mathcal{R}h(T)$
function presented the unstable behavior for accelerating universe.
\item In case of matter bounce, we have analyzed the stability of solution (\ref{mb2}) and found that it indicates stable behavior.
For solutions (\ref{mb1}) and (\ref{mb3}), singularity or complex terms may appear, so we can not analyze them for stability.
\item Since numerous models can be reconstructed mathematically in different gravity theories, so in order to
distinguish valid reconstructed solutions, an additional constraint can be imposed. For example, one interesting and
viable constraint on such solutions is the requirement that the reconstructed Lagrangian must recover vacuum solutions
such as Minkowski spacetime. In this paper, for all presented solutions, we have checked this constraint and
the corresponding conditions have been obtained and presented in the last column of Table \ref{Table0}.
\end{itemize}
It would be interesting to reconstruct these bouncing models in other modified gravity theories especially in torsion based
formulations.

\section{Appendix}

\begin{eqnarray}\nonumber
b_0&=&-6H_{\star}^2f_{R}^{\star}-(6\rho_{\star}(1+\omega)(H_{\star}H_{\star}^{'}+4H_{\star}^2)
+3(1-3\omega)H_{\star}^2\rho_{\star}^{'})f_{RT}^{\star}+18(2H_{\star}^3H_{\star}^{''}+H_{\star}^2H_{\star}^{'^2}
+7H_{\star}^3H_{\star}^{'}-4H_{\star}^4)f_{RR}^{\star}\\\nonumber
&-&108(H_{\star}^3H_{\star}^{''}+H_{\star}^2H_{\star}^{'^2} + 4H_{\star}^3H_{\star}^{'})(H_{\star}H_{\star}^{'}+4H_{\star}^2)f_{RRR}^{\star}
+18(1-3\omega)(H_{\star}^3H_{\star}^{'}+4H_{\star}^4)\rho_{\star}^{'}f_{RRT}^{\star},\\\nonumber
b_1&=&-6(1+\omega)\rho_{\star}H_{\star}^2f_{R T}^{\star}+(36H_{\star}^3 H_{\star}^{'}+54H_{\star}^4)
f_{RR}^{\star}-108H_{\star}^3(H_{\star}^2 H_{\star}^{''}+H_{\star}H_{\star}^{'^2}+4H_{\star}^2H_{\star}^{'})f_{RRR}^{\star}+ 18(1-3\omega)H_{\star}^4\rho_{\star}^{'}f_{RRT}^{\star},\\\nonumber
b_2&=& 18 H_{\star}^{4}f_{RR}^{\star},\\\nonumber
c_{m1}&=&k^2\rho_{\star}+\frac{(3-\omega)\rho_{\star}f_{T}^{\star}}{2}+(1+\omega)(1-3\omega)\rho_{\star}^2f_{TT}^{\star}+
(1-3\omega)(\rho_{\star}(3H_{\star}H_{\star}^{'}+H_{\star}^{2})- 3H_{\star}^2\rho_{\star}^{'})f_{RT}^{\star}+ 18(1-3\omega)\rho_{\star}\\\nonumber
&\times&(H_{\star}^3H_{\star}^{''}+H_{\star}^2H_{\star}^{'^2}+4H_{\star}^3H_{\star}^{'})f_{RRT}^{\star}-3(1-3\omega)^2H_{\star}^2
\rho_{\star}\rho_{\star}^{'}f_{RTT}^{\star},\\\nonumber
c_{m2}&=&3(1-3\omega)H_{\star}^{2}f_{RT}^{\star},\\\nonumber
d_{1}&=& \rho_{\star}H_{\star}\{k^2+\frac{(3-\omega)f_{T}^{\star}}{2}+3(1+\omega)(1-3\omega)\rho_{\star} H_{\star}f_{TT}^{\star}\},\\\nonumber
d_2&=& (1-3\omega)\rho_{\star}\{(\frac{1}{2}(5+\omega)\rho_{\star}^{'}H_{\star}+3(1+\omega)\rho_{\star}H_{\star})f_{TT}^{\star}
+(1+\omega)(1-3\omega)\rho_{\star}\rho_{\star}^{'}H_{\star}f_{TTT}^{\star}\},\\\nonumber
d_3&=&(3(\omega-3)H_{\star}-18(1+\omega)\rho_{\star}H_{\star}^2)f_{RT}^{\star}-6(1+\omega)(1-3\omega)
\rho_{\star}\rho_{\star}^{'}(H_{\star}^3H_{\star}^{'}+4H_{\star}^{4})f_{RTT}^{\star},\\\nonumber
d_4&=&3(1+\omega)\rho_{\star}H_{\star}(k^2+f_{T}^{\star})-\{3(3-\omega)\rho_{\star}^{'}+18(1+\omega)\rho_{\star}\}
(H_{\star}^2H_{\star}^{'}+4H_{\star}^{3})f_{RT}^{\star}-6(1+\omega)(1-3\omega)\\\nonumber
&\times& \rho_{\star}\rho_{\star}^{'}(H_{\star}^2H_{\star}^{'}+4H_{\star}^{3})f_{RTT}^{\star},\\\nonumber
\hat{b_0}&=&-6H_{\star}^2f_{R}^{\star}+18(2H_{\star}^3H_{\star}^{''}+ H_{\star}^2H_{\star}^{'^2}
+7H_{\star}^3H_{\star}^{'}-4H_{\star}^4)f_{RR}^{\star}-108(H_{\star}^3H_{\star}^{''}+H_{\star}^2H_{\star}^{'^2} + 4H_{\star}^3H_{\star}^{'})(H_{\star}H_{\star}^{'}+4H_{\star}^2)f_{RRR}^{\star},\\\nonumber
\hat{ b_1}&=&(36H_{\star}^3 H_{\star}^{'}+54H_{\star}^4)
f_{RR}^{\star}-108H_{\star}^3(H_{\star}^2 H_{\star}^{''}+H_{\star}H_{\star}^{'^2}+4H_{\star}^2H_{\star}^{'})f_{RRR}^{\star},\\\nonumber
\hat{b_2}&=& 18 H_{\star}^{4}f_{RR}^{\star},\\\nonumber
\hat{c_{m1}}&=&-\bigg[k^2\rho_{\star}+\frac{(3-\omega)\rho_{\star}f_{T}^{\star}}{2}+(1+\omega)(1-3\omega)
\rho_{\star}^2f_{TT}^{\star}\bigg],\\\nonumber\\\nonumber
\hat{ d_{1}}&=& \rho_{\star}H_{\star}\{k^2+\frac{(3-\omega)f_{T}^{\star}}{2}+3(1+\omega)(1-3\omega)\rho_{\star} H_{\star}f_{TT}^{\star}\},\\\nonumber
\hat{d_2}&=& (1-3\omega)\rho_{\star}\{(\frac{1}{2}(5+\omega)\rho_{\star}^{'}H_{\star}+3(1+\omega)\rho_{\star}H_{\star})f_{TT}^{\star}
+(1+\omega)(1-3\omega)\rho_{\star}\rho_{\star}^{'}H_{\star}f_{TTT}^{\star}\},\\\nonumber
\hat{d_4}&=&3(1+\omega)\rho_{\star}H_{\star}(k^2+f_{T}^{\star}),
\end{eqnarray}

\end{document}